\documentclass[11pt]{article}
\usepackage{graphicx,amsmath,amsfonts,amssymb,fullpage,url,natbib}
\usepackage{amsmath}
\usepackage{bm}
\usepackage{xcolor}
\usepackage{subfigure}
\usepackage{microtype}
\usepackage{authblk}

\def\h{\bm h}
\def\W{\bm W}
\def\w{\bm w}
\def\x{\bm x}
\def\g{\bm g}

\def\M{\bm M}

\def\v{\bm v}

\def\u{\bm u}
\def\A{\bm A}
\def\mR{\mathcal R}
\def\mH{\mathcal H}
\def\mG{\mathcal G}
\def\mC{\mathcal C}

\begin{document}
\title{\textbf{Disordered Dynamics in High Dimensions: Connections to Random Matrices and Machine Learning}}

\author[1]{Blake Bordelon\thanks{ blake@cmsa.fas.harvard.edu}}
\affil[1]{Center of Mathematical Sciences and Applications}

\author[2]{Cengiz Pehlevan\thanks{cpehlevan@seas.harvard.edu}}
\affil[2]{John A. Paulson School of Engineering and Applied Sciences, Center for Brain Science, Kempner Institute for the Study of Natural and Artificial Intelligence  }
\affil[1,2]{Harvard University, Cambridge, MA, USA}

\maketitle

\begin{abstract}
    We provide an overview of high dimensional dynamical systems driven by random matrices, focusing on applications to simple models of learning and generalization in machine learning theory. Using both cavity method arguments and path integrals, we review how the behavior of a coupled infinite dimensional system can be characterized as a stochastic process for each single site of the system. We provide a pedagogical treatment of dynamical mean field theory (DMFT), a framework that can be flexibly applied to these settings. The DMFT single site stochastic process is fully characterized by a set of (two-time) correlation and response functions. For linear time-invariant systems, we illustrate connections between random matrix resolvents and the DMFT response. We demonstrate applications of these ideas to machine learning models such as gradient flow, stochastic gradient descent on random feature models and deep linear networks in the feature learning regime trained on random data. We demonstrate how bias and variance decompositions (analysis of ensembling/bagging etc) can be computed by averaging over subsets of the DMFT noise variables. From our formalism we also investigate how linear systems driven with random non-Hermitian matrices (such as random feature models) can exhibit non-monotonic loss curves with training time, while Hermitian matrices with the matching spectra do not, highlighting a different mechanism for non-monotonicity than small eigenvalues causing instability to label noise. Lastly, we provide asymptotic descriptions of the training and test loss dynamics for randomly initialized deep linear neural networks trained in the feature learning regime with high-dimensional random data. In this case, the time translation invariance structure is lost and the hidden layer weights are characterized as spiked random matrices. 
\end{abstract}

\pagebreak
\tableofcontents
\pagebreak

\section{Introduction}

In many areas of physics and applied mathematics, one encounters high dimensional dynamical systems which depend on some source of randomness. In statistics or machine learning theory, the randomness could come from randomly sampled data points \cite{krogh1990dynamics, advani2020high, montanari2025dynamical, agoritsas2018out}, stochastic gradient noise \cite{mignacco2020dynamical, mignacco2022effective}, or from the initialization of model parameters \cite{mei2022generalization, adlam2020neural}. In physics, randomness in the interactions of a high dimensional system, such as a spin glass, can generate rich high dimensional dynamics \cite{de1978dynamics, sompolinsky1981dynamic, berthier2011theoretical}. This idea has also been pursued in theoretical neuroscience, where randomly connected recurrent neural networks have been analyzed for decades as simple solvable models of high dimensional chaotic dynamics \cite{sompolinsky1988chaos, vogels2005neural, helias2020statistical}. Many works in theoretical ecology have analyzed the population dynamics of interacting species with complex cross-species interactions \cite{may1972will, cui2021diverse, blumenthal2024phase, cui2024houches}.


Despite the diversity of contexts where disordered high dimensional dynamical systems appear, these systems often share a common mathematical structure in the large system size limit. In this note, we examine a commonly used tool, known as dynamical mean field theory (DMFT) which can be used to analyze low dimensional summary statistics of the system. Of key interest are the correlation and response functions of the system, which both describe how long perturbations are remembered in a system.

\subsection*{Plan for this Note}
We aim to present a high level overview of DMFT by providing simple examples relevant to machine learning theory, where the limiting dynamics are sufficiently simple to retain Gaussian process structure. These include random feature models, kernel methods, recurrent neural networks, and deep linear networks in the feature-learning (non-lazy) training regime. Concretely, we will examine the following examples:
\begin{itemize}
    \item \textbf{Linear Dynamics with GOE Matrix Interactions.} We will start with a simple example involving linear dynamics with an interaction matrix sampled from the Gaussian Orthogonal Ensemble (GOE) and recover the Wigner Semi-Circle Law from the response function. We illustrate both the cavity and path integral approaches for this problem. 
    \item \textbf{Gradient Descent/Flow Dynamics of Linear Regression.} We will consider linear regression on random covariates with $P$ data points in $D$ dimensions with $P = \alpha D$ \cite{krogh1990dynamics}. For isotropic covariates, the DMFT equations will encode the Marchenko-Pastur Law \cite{advani2020high}. 
    \item \textbf{Gradient Descent for Kernel Regression.} We next examine gradient flow on structured regression problems such as kernel methods. As an example, we will describe ``stage like" multiple descent structures in problems with large numbers of degenerate eigenvalues \cite{lu2025equivalence, xiao2022precise, hu2024asymptotics} as well as powerlaw behavior for powerlaw eigenspectra \cite{bordelon2020spectrum, paquette20244+, bordelon2024dynamical}.
    \item \textbf{Random Feature Models.} Instead of one random matrix from the randomly sampled training data, we then introduce another random matrix representing a feature projection \cite{mei2022generalization}. In this case, we show that the spectrum of the matrix which governs the linear dynamics of the test loss is insufficient to characterize the loss dynamics. Rather the DMFT two-point correlation function can capture interesting non-monotonic behavior in the test loss near the interpolation threshold where parameters and data are equal.
    \item \textbf{Generic Free Products.} We then generalize the analysis beyond matrices with independent entries to allow the analysis of generic free products, which is a product of the form $\bm O \bm B \bm O^\top \bm A$ (asymmetric) or $\bm A^{1/2} \bm O \bm B \bm O^\top \bm A^{1/2}$ (symmetrized) where $\bm O$ is a random orthogonal matrix and $\bm A, \bm B$ have known spectra. We compare and contrast asymmetric and symmetrized free products, which have identical spectra but generate distinct dynamics at the level of the correlation function. We show how the dynamics in both cases can be handled within DMFT.
    \item \textbf{Non-Hermitian Examples.} We provide a simple DMFT recipe to compute spectra for non-Hermitian matrices with complex spectra by studying long-time behavior of a Hermitianized flow. We use this to derive the circular law \cite{girko1985circular, feinberg9703118non, cui2024elementary} \& and the spectrum of a diagonally modulated asymmetric random matrix, which is related to the Jacobian of random recurrent neural networks \cite{krishnamurthy2022theory}. 
    \item \textbf{Beyond Linear Dynamics.} We provide simple examples of systems which are not described by a linear dynamics where a (Gaussian) DMFT still provides the exact asymptotic description of the system. We will first illustrate a toy example of (Anti-)Hebbian dynamics on an initially GOE recurrent network before using this theory to describe multilayer linear networks in a proportional scaling limit where data, input dimension, and width are all comparable \cite{bordelon_deep_linear}. 
\end{itemize}

\paragraph{Notation.} We will often use the shorthand $\text{tr} \ \bm M \equiv \frac{1}{N} \text{Tr} \ \bm M$ for the normalized trace of a $N \times N$ matrix $\bm M$ and use $\left<  \cdot \right>$ to denote an average over a random variable. We also use the notation $u(t) \sim \mathcal{GP}(0,C(t,t'))$ to represent a mean-zero Gaussian process with covariance $\left< u(t) u(t') \right> = C(t,t')$. The set of natural numbers less than or equal to $n$ will be denoted as $[ n ] = \{ 1, ..., n \}$. Path integrals (with appropriate normalization in discretized time under Ito convention, see Appendix \ref{app:path_integral}) will be denoted as $\int \mathcal Df ...$ for single variable function $f(t)$ or $\int \mathcal D C$ for two variable function $C(t,t')$. We use callographic font for a Fourier transform $\mathcal R(\omega) = \int d\tau e^{- i\omega \tau} R(\tau)$ of function $R(\tau)$; the inverse transform $R(\tau) = \frac{1}{2\pi}\int d\omega e^{i\omega \tau} \mR(\omega)$ will contain the factor of $\frac{1}{2\pi}$.

\section{Disordered Linear Dynamical Systems}

In the first sections of this note, we will are focus on linear dynamical systems of the form
\begin{align}
    \frac{d}{dt} \h(t) = - \M \h(t) + \bm j(t),
\end{align}
where the matrix $\bm M \in \mathbb{R}^{N\times N}$ is a fixed random matrix that depends on the specific problem details (we will show several examples in the coming sections). The two objects that we wish to track are 
\begin{align}
    C(t,t') = \frac{1}{N} \h(t) \cdot \h(t')  \ , \ R(t,t') = \frac{1}{N}   \text{Tr} \frac{\delta \h(t)}{\delta \bm j(t')^\top},
\end{align}
which intuitively measure the cross correlation between the state variables $\h(t)$ and the response of the variables to perturbations to their dynamics. The functions $C(t,t')$ and $R(t,t')$ are central objects to describe high dimensional disordered systems. For linear systems, the response function $R(t,t')$ for matrix $\bm M$ actually encodes its spectral density $\rho(\lambda) = \frac{1}{N} \sum_{i=1}^N \delta(\lambda-\lambda_i)$, since 
\begin{align}
    R(t,t') = \frac{1}{N} \text{Tr} \frac{\delta \h(t)}{\delta \bm j(t')^\top} =  \frac{1}{N} \text{Tr} \exp\left( - \bm M (t-t') \right) \Theta(t-t') = \int d\lambda \rho(\lambda) e^{-\lambda (t-t')} \Theta(t-t')
\end{align}
where $\Theta(t-t') = \mathbf{1}_{t>t'}$ is the Heaviside step function which is $1$ for all $t > t'$ and zero otherwise.

For linear systems, the response function $R(t,t')$ also encodes the trace of the \textit{resolvent} matrix for $\bm M$ through a Fourier or Laplace transform
\begin{align}
    \mR(\omega) = \int d\tau R(t+\tau, t) e^{-i\omega \tau} = \frac{1}{N} \text{Tr} \left[ i\omega + \bm M \right]^{-1}  = \int d\lambda \  \frac{\rho(\lambda)}{i\omega + \lambda}  . 
\end{align}
The trace of the resolvent, $\mR(\omega)=\frac{1}{N} \text{Tr} \left[ i\omega + \bm M  \right]^{-1}$ is often termed the Stieltjes transform, is a central object in the study of random matrices \cite{potters2020first}. For matrices $\bm M$ with real spectra, the eigenvalue density $\rho(\lambda)$ can be obtained from the Sokhotski–Plemelj formula
\begin{align}
    \rho(\lambda) = \lim_{\epsilon \to 0} \frac{1}{\pi}\Im \ \mR(\omega)|_{i\omega= -\lambda - i\epsilon} .
\end{align}
We will be focused on matrices with real spectra for the first sections of this paper, before moving onto matrices whose spectra extend into the complex plane in Section \ref{sec:complex_spectra}.

\section{Warmup Problem: GOE (Wigner) Matrix} 

To first see how the DMFT description of the high dimensional limit works, we start with the simplest example of a random matrix, the symmetric Gaussian (GOE/Wigner) matrix $\bm M = \frac{1}{\sqrt N} \bm A$ where $\bm A = \bm A^\top$ with $A_{ij} \sim \mathcal{N}(0, 1)$. We consider the linear dynamical system
\begin{align}
    \frac{d}{dt} \h(t) = - \frac{1}{\sqrt N} \bm A \h(t) + \bm j(t) . 
\end{align}
We are interested in the behavior of this dynamical system as the system size diverges $N \to \infty$. In this limit, we will find that each of the variables effectively decouple from one another, enabling a simple description of the full dynamics in terms of a one-dimensional stochastic process. 

\subsection{Cavity Derivation}

\begin{figure}[h]
    \centering
    \includegraphics[width=0.9\linewidth]{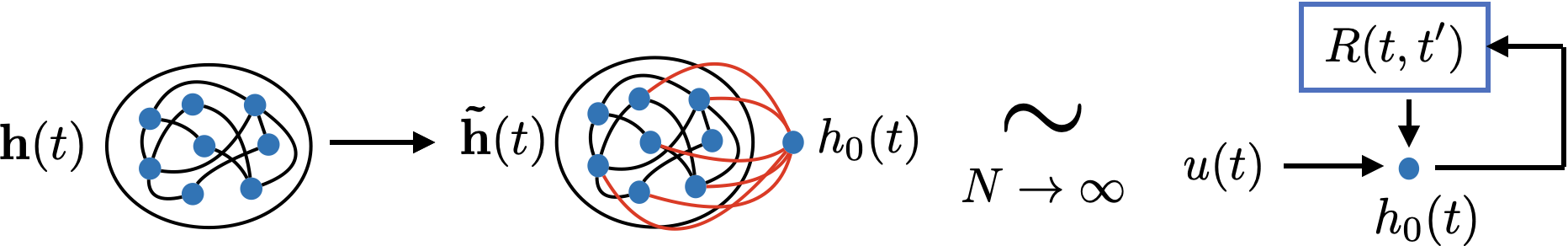}
    \caption{Cavity derivation of the marginal dynamics for a single site of the system as $N \to \infty$. Adding a new site to the system comes with $N$ reciprocal couplings $\bm a_0 \in \mathbb{R}^N$ to the original variables which are now perturbed $\h(t) \to \tilde{\h}(t)$. In the large system size limit $N \to \infty$ the system can be viewed as a single-variable stochastic process driven by a colored noise process with a delayed feedback through response function $R(t,t')$. }
    \label{fig:cavity_GOE}
\end{figure}

To characterize the dynamics in the large system size limit $N \to \infty$, we will first illustrate the cavity method. In this method, we consider adding a new site $h_0(t)$ to the system, resulting in a $N+1$ site system. We illustrate this procedure in Figure \ref{fig:cavity_GOE}.

\paragraph{Adding a New Site}
Upon the addition of a new site $h_0$ the other $N$ variables will experience a $\mathcal{O}(N^{-1/2})$ perturbation to their original dynamics which we denote as $\h(t) \to \tilde{\h}(t)$ where 
\begin{align}
    \tilde{\h}(t) = \h(t) - \frac{1}{\sqrt{N}} \int_{-\infty}^t dt' \frac{\partial \h(t)}{\partial \bm j(t')^\top} \bm a_0 \ h_0(t')   + \mathcal{O}(N^{-1})
\end{align}
where $\bm a_0 \in \mathbb{R}^N$ are the added weights between the $N+1$st neuron and the original $N$ neuron system (the red lines in Figure \ref{fig:cavity_GOE}). This expression results from considering that the added term to the right hand side can be considered a perturbation to the source $\bm j(t) \to \bm j(t) - \frac{1}{\sqrt {N}} \bm a_0 \  h_0(t)$, allowing us to expand the dynamics by differentiating in the source at all earlier times $t' < t$. 

\paragraph{Computing Correction to Dynamics from Feedback}
These perturbations to the original $N$ variables feedback into the dynamics for the added site $h_0(t)$. Computing these corrections at leading order, we find
\begin{align}
    \frac{d}{dt} h_0(t) &\sim - \frac{1}{\sqrt{N}} \bm a_0 \cdot \tilde{\h}(t)   + j_0(t) \nonumber
    \\
    &\sim - \underbrace{\frac{1}{\sqrt{N}} \bm a_0 \cdot {\h}(t)}_{\text{Noise Term}}  +  \underbrace{\frac{1}{N} \int_{-\infty}^t dt' \bm a_0^\top \frac{\partial \h(t)}{\partial \bm j(t')^\top }  \bm a_0 h_0(t')}_{\text{Response Term}}  + j_0(t) \nonumber
\end{align}
To proceed, we next work out the statistics of the noise term and the response term. 
\paragraph{Statistics of The Noise Term} The noise term $u_0(t) \equiv - \frac{1}{\sqrt N} \bm a_0 \cdot \bm h(t)$ is a random variable due to the random $\bm m_0$ vector. By construction, the $\bm h(t)$ variables are statistically independent of $\bm m_0$. As a consequence, $u_0(t)$ will behave as a Gaussian process with covariance
\begin{align}
    \left< u_0(t) u_0(t') \right> = \frac{1}{N} \sum_{i=1}^N h_i(t) h_i(t') \equiv C(t,t')
\end{align}
which is computed as a population average over the $N$ original sites. 
\paragraph{Statistics of the Response Term} Next, we note that the response term concentrates (over random draws of the $\bm m_0$ vector) as $N \to \infty$ by the law of large numbers. Computing the mean and standard deviation of this term, we find that 
\begin{align}
     \frac{1}{N} \int_{-\infty}^t dt' \bm a_0^\top \frac{\partial \h(t)}{\partial \bm j(t')^\top }  \bm a_0 \  h_0(t') = \int_{-\infty}^t dt' \underbrace{\left[ \frac{1}{N} \sum_{i=1}^N \frac{\partial h_i(t)}{\partial j_i(t')} \right]}_{\equiv R(t,t')} h_0(t') + \mathcal{O}(N^{-1/2}) 
\end{align}
where we introduced the \textit{response function} $R(t,t') \equiv \frac{1}{N} \sum_{i=1}^N \frac{\partial h_i(t)}{\partial j_i(t')}$ which is another population average over the $N$ original sites. This quantity asks about how a perturbation to the dynamics for the $i$-th site impacts the later value of the $i$-th site at a later time. 

\paragraph{Decoupled Dynamics} Combining the two computations above, we find that the dynamics of the variable $h_0(t)$ in the $N \to \infty$ limit take the form
\begin{align}
    \frac{d}{dt}  h_0(t) = u_0(t) + \int_{-\infty}^t dt' R(t,t') h_0(t') + j_0(t) , \quad  u(t) \sim \mathcal{GP}(0, C(t,t'))
\end{align}
which only depends \textit{on its own dynamics} except through self-averaging functions $R(t,t')$ and $C(t,t')$. Thus, if $C$ and $R$ become non-random in the $N \to \infty$ limit, the variable $h_0$ will be effectively decoupled from the other $N$ variables in the model. 

\paragraph{Each Variable Will Behave Like $h_0$} In this model, as $N \to \infty$, all sites $h_i(t)$ become statistically equivalent. Therefore, in the $N \to \infty$ limit, they should each behave identically and independently. Averages over the population of $N$ sites can be replaced with averages over the Gaussian noise $u(t)$ which generates the variance. 
\begin{align}
    \frac{\partial}{\partial t} h(t) = u(t) + \int dt' R(t,t') h(t') + j(t)
\end{align}

\paragraph{Response Function Dynamics} We can differentiate with respect to the source $j(t)$ to find the following equation for the linear response function
\begin{align}
    \frac{\partial}{\partial t} R(t,t') = \delta(t-t') + \int_0^t dt'' R(t,t'') R(t'',t') 
\end{align}
The solution is clearly time translation invariant so that $R(t,t') = R(\tau)$ where $\tau = t-t'$ is the time-lag. Taking a Fourier transform gives
\begin{align}
    \mR(\omega) &= \int_{-\infty}^\infty d\tau R(\tau) e^{-i\omega \tau} \implies i\omega \mR(\omega) = 1 + \mR(\omega)^2 \implies \mR(\omega) = \frac{1}{2} \left[ i \omega + \sqrt{ (i\omega)^2 - 4 } \right]
\end{align}
We can compute the inverse transform $R(\tau) = \frac{1}{2\pi} \int d\omega e^{i\omega \tau } \mR(\omega)$ to obtain the 

\paragraph{Eigenvalue Distribution (Semi-circle Law)} We note that the response function $R(\tau)$ has a straightforward connection to the eigenvalue density $\rho(\lambda) = \frac{1}{N}\sum_{i=1}^N \delta(\lambda - \lambda_i)$. 
\begin{align}
    R(\tau) &= \frac{1}{N} \text{Tr} \exp\left( - \bm M \tau \right) = \int d\lambda \rho(\lambda) e^{-\lambda \tau}  \implies \mR(\omega) = \int d\lambda \ \frac{\rho(\lambda)}{i\omega + \lambda} \nonumber
    \\
    &\implies \rho(\lambda) =  \lim_{\epsilon \to 0}  \frac{1}{\pi} \ \text{Im} \ \mR( i \lambda - \epsilon ) = \frac{1}{2\pi} \sqrt{ \left[4-\lambda^2 \right]_+ } .
\end{align}
This recovers Wigner's semicircle distribution which has support for $\lambda \in [-2,2]$.

\paragraph{Connection between Correlation and Response for this System} For this problem, since $\bm M = \bm M^\top$, the correlation function satisfies
\begin{align}
    C(t,t') = \text{tr} \ \exp\left( - \bm M t' \right)^\top \exp\left( - \bm M t \right) = \text{tr} \exp\left( - \bm M (t + t')  \right) = R(t+t') .
\end{align}
Thus, the response function $R(\tau)$ completely characterizes the correlation function in this setting. We stress that this correspondence between correlation and response will fail in almost all of the subsequent settings of this note, especially in cases where $\bm M$ is asymmetric or if the correlation function involves traces against other matrices which do not commute with $\exp( \bm M t)$. 

\begin{figure}[h]
    \centering
    \subfigure[GOE Eigenvalue Density]{\includegraphics[width=0.425\linewidth]{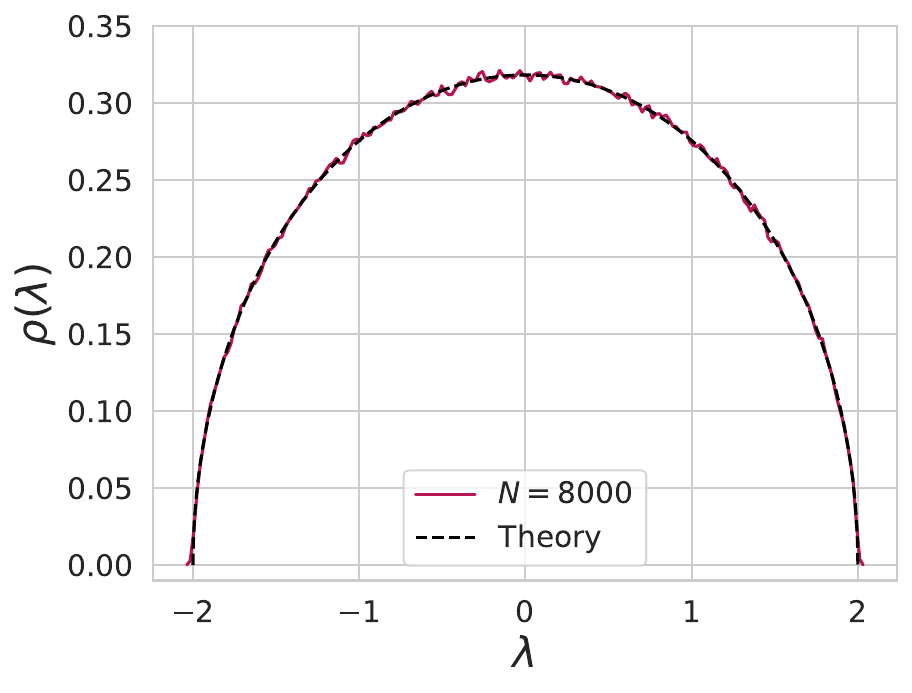}}
    \subfigure[Response Function]{\includegraphics[width=0.425\linewidth]{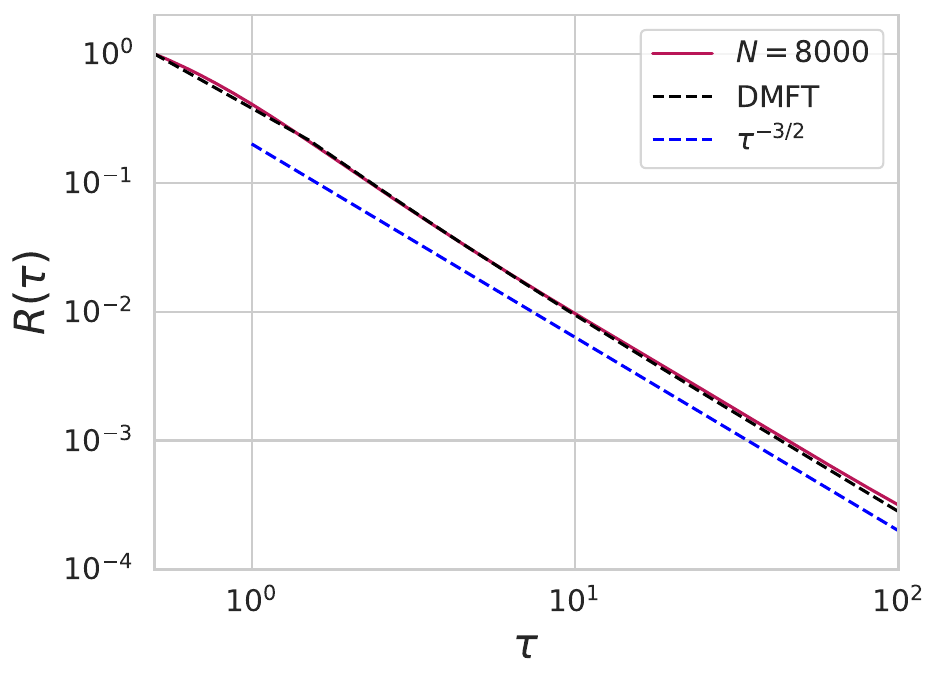}}
    \caption{The DMFT response function for the random Wigner matrix encodes the semicircle eigenvalue law. (a) The spectral distribution $\rho(\lambda)$ for a randomly sampled $N = 8000$ Wigner matrix (red) is compared to the asymptotic theoretical density $\rho(\lambda) = \frac{1}{2\pi} \sqrt{4-\lambda^2}$ (dashed black lines). (b) The response function $R(\tau)$ as a function of the time lag $\tau$ for the dynamics $\frac{d}{dt} \h(t) = - \bm M \h(t) - z \h(t)$ where $\bm M$ is a Wigner matrix and $z=2$. For this system, the relaxation rate at $z=2$ is powerlaw with $R(\tau) = \frac{1}{2\pi} \int e^{-( \lambda + 2)\tau}\sqrt{4-\lambda^2 } \propto \tau^{-3/2}$ for large $\tau$.  }
    \label{fig:enter-label}
\end{figure}

\subsection{Path Integral Approach}

In this section, we provide an additional method to compute the DMFT equations, known as the path integral approach. A reader less interested in the formal approaches used to derive the equations and more interested in applications could skip this section and proceed to Section \ref{sec:lin_reg_isotropic}.

\paragraph{Moment Generating Function } An alternative approach to derive the DMFT equations in the limit of $N \to \infty$ is to analyze a path integral / moment generating function for the random variables $\h(t)$
\begin{align}
    Z[\bm \zeta] = \left< \exp\left( \int dt \  \bm{\zeta}(t) \cdot \h(t)  \right) \right>_{\bm h}
\end{align}
where the average is taken over the distribution of the random variables $\h(t)$ induced by the random couplings $\bm M$. While moments of the distribution for $\h$ can be obtained through derivatives of $Z$ with respect to $\bm \zeta$, it often suffices to analyze $Z$ formally for $\bm \zeta=0$, where $Z = 1$ \cite{de1978dynamics, helias2020statistical, crisanti2018path}, which we explain in Appendix \ref{app:path_integral}.  
Starting from an integral representation of the Dirac delta function $\delta( z ) = \int \frac{d\hat z}{2\pi} \ \exp(i \hat z z )$, we can write a formal path integral that enforces the dynamics for $\h(t)$, yielding 
\begin{align}
    Z &= \int \mathcal D \h \mathcal D\hat{\h} \left< \exp\left( i \int dt \ \hat{\h}(t) \cdot \left( \partial_t \h(t) + \bm M \h(t) - \bm j(t) + \delta(t) \h_0  \right) \right)  \right>_{\bm M } \nonumber
    \\
    &= \int \mathcal D C \mathcal D \hat C \mathcal D R \mathcal D \hat{R} \ \exp\left( - N \mathcal S\left[C,\hat C, R, \hat R \right] \right)  .
\end{align}
By averaging over the Gaussian random matrix $\bm M$ we were able to re-express $Z$ in terms of correlation and response variables (often referred to as \textit{order-parameters} for the theory) 
\begin{align}
    C(t,t') = \frac{1}{N} \h(t) \cdot \h(t') \ , \ R(t,t') = - \frac{i}{N} \h(t) \cdot \hat{\h}(t') . 
\end{align}
These definitions are enforced with conjugate variables $\hat{C}(t,t'), \hat R(t,t')$. The mean-field action $\mathcal S$ has the following form
\begin{align}
    \mathcal S = - \frac{1}{2} \int dt dt' &\hat C(t,t') C(t,t')  + \frac{1}{2} \int dt dt' R(t,t') \hat{R}(t',t) - \ln \mathcal Z_h  \nonumber
    \\
    \mathcal Z_h = \int \mathcal Dh \mathcal D\hat{h} \  &\exp\left( - \frac{1}{2} \int dt dt' \  \hat{C}(t,t') h(t) h(t') - \frac{1}{2} \int dt dt' \ C(t,t') \hat{h}(t) \hat{h}(t)  \right) \nonumber
    \\
    &\exp\left(  i \int dt \ \hat{h}(t)\left [ \partial_t h(t) - \frac{1}{2} \int dt' \left(R(t,t') + \hat{R}(t,t') \right) h(t') + \delta(t)   \right  ]  \right)
\end{align}
The object $\mathcal Z_h$ is often known as the single site generating function for $h(t)$, reflecting the fact that, after averaging over $\bm M$, all entries in $\h(t)$ are identically distributed. As $N \to \infty$ the dominant contribution comes from the saddle point, where the following conditions are satsified
\begin{align}
    \frac{\partial \mathcal S}{\partial \hat C(t,t')} &= - \frac{1}{2} C(t,t') + \frac{1}{2} \left< h(t) h(t') \right> = 0 \nonumber
    \\
    \frac{\partial \mathcal S}{\partial C(t,t')} &= - \frac{1}{2} \hat C(t,t') + \frac{1}{2} \left< \hat h(t) \hat h(t') \right> = 0  \nonumber
    \\
    \frac{\partial \mathcal S}{\partial \hat R(t',t)} &= \frac{1}{2} R(t,t') + \frac{i}{2} \left< h(t)  \hat h(t' ) \right>  = 0 \nonumber
    \\
    \frac{\partial \mathcal S}{\partial R(t',t)} &= \frac{1}{2} \hat R(t,t') + \frac{i}{2}\left< h(t)  \hat h(t' ) \right>  = 0
\end{align}
where $\left< \cdot \right>$ represents an average over the stochastic process for $h(t)$ defined by the single-site moment generating function $\mathcal Z_h$. The above saddle point equations imply that $\hat{C}(t,t') = 0$ and $\hat{R}(t,t') = R(t,t') = - i\left<h(t) \hat{h}(t') \right>$. Lastly, we can introduce a Gaussian variable $u(t)$ to linearize the $\hat{h}(t)$ terms
\begin{align}
    \exp\left( - \frac{1}{2} \int dt dt' C(t,t') \hat{h}(t) \hat{h}(t') \right) = \left< \exp\left( - i \int dt \ u(t) \hat{h}(t) \right) \right>_{u(t) \sim\mathcal{GP}(0, C(t,t'))}
\end{align}
After introducing this variable, it becomes possible to exchange correlation functions of $\hat{h}$ with derivatives with respect to the variable $u(t)$, resulting in the following formula for the response function
\begin{align}
    R(t,t') = \left< \frac{\partial h(t)}{\partial u(t')} \right> ,
\end{align}
which agrees with the cavity formula. Lastly, performing the integral over the $\hat{h}(t)$ variables, we recover the original DMFT equations for the GOE matrix
\begin{align}
    \frac{\partial}{\partial t} \ h(t) = u(t) + \int dt' R(t,t') h(t') + \delta(t)  \ , \ u(t) \sim \mathcal{GP}(0, C(t,t'))  .
\end{align}
Under this dynamics, we compute the correlation $C(t,t') = \left< h(t) h(t') \right>$. 


\subsection{Anti-Symmetrized Version} 

The oscillatory version of this system, where $\frac{d}{dt} \h(t) = \bm M \h(t)$ where $\bm M= \frac{1}{\sqrt{2N}} ( \bm A - \bm A^\top )$ where $A_{ij} \sim \mathcal{N}(0,1)$ was analyzed with DMFT in another recent note \cite{blumenthal2025building}. In this case, one arrives at very similar DMFT equations modified by flipping the sign of the term involving the response function 
\begin{align}
    \frac{d}{dt} h(t) = u(t) - \int dt' R(t,t') h(t')  ,
\end{align}
where the response function can be computed from its Fourier transform $i\omega \mR(\omega) = 1 - \mR(\omega)^2$. Geometrically, the eigenvalues of $\bm M$ follow the semicircle distribution on the imaginary axis in the complex plane from $z = - 2i$ to $z= 2i$. Consequently, the response function can also be viewed as a linear combination of oscillatory modes (rather than exponential decay timescales)
\begin{align}
    R(\tau) = \int d\lambda \rho(\lambda) e^{ - i \lambda \tau} = \frac{1}{\pi} \int_0^2 d\lambda \sqrt{4-\lambda^2} \cos(\lambda\tau) .
\end{align}
In this setting, the correlation function is a function of the \textit{time-lag} $\tau = t-t'$ 
\begin{align}
    C(t,t') = \text{tr} \exp\left( \bm M t' \right)^\top \exp\left( \bm M t \right) = \int d\lambda \rho(\lambda) e^{i \lambda (t-t')} = R(t-t') .
\end{align}
While the correlation function in the symmetric case depended on $t+t'$, we see that the anti-symmetric case depends on the value of the response function evaluated at the time-lag $t-t'$. In this system, the correlation function does not monotonically decay with $t-t'$, but rather exhibits oscillations. 

\section{Linear Regression with Isotropic Covariates}\label{sec:lin_reg_isotropic}

In this section we consider gradient flow on a linear regression problem. Let $\bm \Psi \in \mathbb{R}^{N \times P}$ represent the $P$ samples $\{ \bm\psi_\mu \}_{\mu=1}^P$ of $N$ dimensional isotropic covariates $\left< \bm\psi \bm\psi^\top \right> = \bm I$ which will be used in the learning problem. We will be interested in the proportional limit where 
\begin{align}
    P,N \to \infty \quad  \text{with} \quad  \frac{P}{N } \equiv \alpha
\end{align}
The model $f$ and the target function $y$ are both linear in the features $\bm\psi$
\begin{align}
    f = \frac{1}{\sqrt N} \w  \cdot \bm\psi \ , \ y = \frac{1}{\sqrt N} \bm \beta_\star \cdot \bm\psi + \epsilon \ , \  \left< \epsilon^2 \right> = \sigma^2 
\end{align}
The noise $\epsilon$ is zero mean with variance $\sigma^2$ and the target weights $\bm\beta_\star$ are normalized $\frac{1}{N} |\bm \beta_\star|^2 = 1$ \footnote{This scaling would occur automatically for randomly sampled $\bm\beta_\star \sim \mathcal{N}(0,\bm I)$ as $N \to \infty$}. 
We will optimize the weights $\w(t)$ using a gradient flow algorithm and aim to track the dynamics of the training loss $\hat{\mathcal L}(t)$ and the test loss $\mathcal L(t)$ which are defined as
\begin{align}
    &\underbrace{\hat{\mathcal L}(t)}_{\text{Train Loss}} =  \underbrace{\frac{1}{P} \sum_{\mu=1}^P   \left( \frac{1}{\sqrt N} \bm\psi_\mu \cdot \w(t) -  y_\mu \right)^2}_{\text{Empirical Average}} = \frac{1}{P} \left| \frac{1}{\sqrt N} \bm\Psi \w(t) - \bm y \right|^2  \nonumber
    \\
    &\underbrace{\mathcal L(t)}_{\text{Test Loss}} =  \underbrace{\left< \left( \frac{1}{\sqrt N} \bm\psi_\mu \cdot \w(t) -  y_\mu \right)^2 \right>_{\bm \psi, y}}_{\text{Population Average}} =  \frac{1}{N} |\w(t) - \bm\beta_\star|^2 + \sigma^2
\end{align}
The weights $\w(t)$ are optimized with gradient flow (with learning rate $\propto N$ to ensure the learning dynamics occur on timescales $t = \Theta(1)$ \footnote{This scaling of the learning rate with $\propto N$ is sensible under this parameterization as the first term has $\mathcal{O}(1)$ mean and the second term has $\mathcal{O}(1)$ variance in the proportional scaling limit $P,N \to \infty$ with $P=\alpha N$, leading to changes in $\h(t)$ in $\mathcal O(1)$ time.}) on the \textit{training loss} $\hat{\mathcal L}(t)$ 
\begin{align}
    \frac{d}{dt} \w(t) = - N \ \nabla \hat{\mathcal L}(t) \ = \frac{\sqrt N}{P} \bm \Psi^\top \left( \bm y - \frac{1}{\sqrt N} \bm\Psi \w(t)  \right) =  \left( \frac{1}{P} \bm\Psi^\top \bm\Psi \right) (\bm\beta_\star - \w(t))  +  \frac{\sqrt N}{P} \bm\Psi^\top \bm\epsilon   .
\end{align} 
The dynamics are thus governed by the empirical covariance matrix, known as a Wishart Matrix
\begin{align}
    \bm M = \frac{1}{P} \bm \Psi^\top \bm \Psi \in \mathbb{R}^{N \times N}.
\end{align}
We will now proceed to compute the typical case loss dynamics in the proportional scaling limit.

\paragraph{Decomposing the Dynamics} To derive the asymptotic limit, we will find it more convenient to work with the dynamics on the weight vector \textit{residual error} $\h(t) \equiv \bm\beta_\star - \w(t)$ which determines the test loss $\mathcal L(t) = \frac{1}{N} |\h(t)|^2 + \sigma^2$ and evolves as
\begin{align}
    &\frac{d}{dt} \h(t) = - \left( \frac{1}{P} \bm\Psi^\top \bm\Psi \right) \h(t) - \frac{\sqrt N}{P} \bm\Psi^\top \bm \epsilon + \bm  j_h(t)  = - \frac{1}{\alpha \sqrt N} \bm\Psi^\top \bm\Delta(t) + \bm j_h(t) \nonumber
    \\
    &\bm \Delta(t) \equiv \frac{1}{\sqrt N} \bm\Psi \bm h(t) + \bm\epsilon + \bm j_\Delta(t) 
\end{align}
In the above expressions, we introduced the variables $\bm\Delta(t)$ and added source variables $\bm j_h(t)$ and $\bm j_\Delta(t)$ which will be set to zero after the computation. The benefit of introducing the intermediate variables $\bm\Delta(t)$ is that each of the variable definitions are now separately \textit{linear} in the random matrix $\bm\Psi$. The above equation for $\h(t)$ is integrated from the initial condition $\bm h(0) = \bm\beta_\star$. Following the same idea in the GOE example, we can set up both a simple cavity analysis and a path integral analysis for this system. 

\subsection{Bipartite Cavity Analysis} 
\begin{figure}[h]
    \centering
    \includegraphics[width=0.75\linewidth]{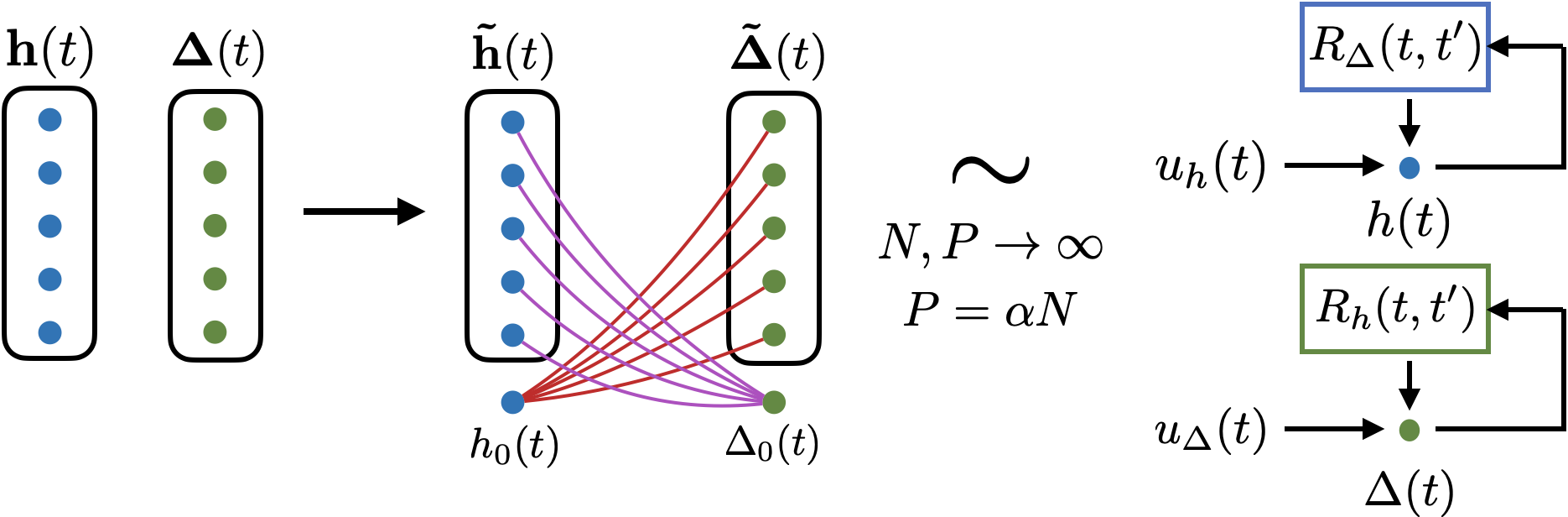}
    \caption{The cavity method for the linear regression problem can proceed in two steps. First, a computation of the marginals for the weight discrepancy $h_0 = w_0 - w_0^*$ when a new $N+1$st feature is added requires considering feedback through the perturbed training errors $\tilde{\Delta}_\mu(t)$. Second, the training error made on an added $P+1$st data point $\Delta_0(t)$ requires considering feedback through the perturbed weight discrepancies $\tilde{\h}(t)$. Under the joint limit $N,P \to \infty$ with $P = \alpha N$, the loss  }
    \label{fig:cavity_lin_reg}
\end{figure}

For the regression problem, we introduced two sets of variables $\{\h(t), \bm\Delta(t)\}$ which are both linearly related through the matrix $\bm\Psi$. We can therefore perform two-steps of a cavity argument, one for each of these variables (see \cite{clark2025simplified} for application of this idea for several static settings). This proceeds in two steps which are visualized in Figure \ref{fig:cavity_lin_reg}. First, we consider adding a $(N+1)$st feature dimension which leads to a new variable $h_0(t)$, which leads to new couplings to the original $P$ traning errors (red lines). Next, we consider adding the $(P + 1)$st data point, which couples to all $N$ original features (purple lines).

\paragraph{Adding a Feature} First, let's consider the addition of a new feature. This leads to a perturbation in the $P$ training error variables $\Delta_\mu(t)$. We again try to express the perturbed dynamics $\tilde\Delta_\mu(t)$ in terms of the unperturbed $N$-feature system $\Delta_\mu(t)$ which can be viewed as having a shifted source ${j}_\mu(t) \to j_\mu(t) + \frac{1}{\sqrt N} \psi_{\mu 0} h_0(t)$. We can therefore consider a Taylor series expansion in these variables, keeping the leading term
\begin{align}
    \tilde{\Delta}_\mu(t) = \Delta_\mu(t) + \frac{1}{\sqrt N} \int  dt' \frac{\partial \Delta_\mu(t)}{\partial j_\nu(t')} \psi_{\nu 0} \ h_0(t') + \mathcal{O}(N^{-1}) .
\end{align}
Plugging this perturbation back into the new feature $h_0(t)$'s differential equation, we find
\begin{align}
    \frac{d}{dt} h_0(t) &= - \frac{1}{\alpha\sqrt N} \sum_{\mu=1}^P \tilde{\Delta}_\mu(t) \psi_{\mu 0} + j_{0}^h(t) \nonumber
    \\
    &\sim - \frac{1}{\alpha\sqrt N} \sum_{\mu=1}^P {\Delta}_\mu(t) \psi_{\mu 0}  - \frac{1}{P} \int  dt'  \sum_{\mu \nu} \psi_{\mu 0} \frac{\partial \Delta_\mu(t)}{\partial j_\nu(t')} \psi_{\nu 0} h_0(t') + j^h_0(t) \nonumber
    \\
    &\sim \underbrace{u_{h,0}(t)}_{\text{Gaussian Process}} - \int dt'  \underbrace{\left[ \frac{1}{P} \sum_{\mu=1}^P \frac{\partial \Delta_\mu(t)}{\partial j_\mu(t')} \right]}_{\text{Response Function}  \   R_\Delta(t,t')} h_0(t') 
\end{align}
In the last line, we noted that all $\Delta_\mu(t)$ are independent of the new features $\psi_{\mu 0}$ and thus invoked a central limit theorem. The second term concentrates. The Gaussian process $u_{h,0}(t)$ has covariance
\begin{align}
    \left< u_{h,0}(t) u_{h,0}(t') \right> = \frac{1}{\alpha } \underbrace{\left[ \frac{1}{P} \sum_{\mu=1}^P \Delta_\mu(t) \Delta_\mu(t')  \right]}_{\text{Correlation Function} \ C_\Delta(t,t')} \equiv \frac{1}{\alpha} C_{\Delta}(t,t') 
\end{align}
We note that the correlation function $C_\Delta(t,t)$ exactly gives the the training loss dynamics $\hat{\mathcal L}(t)$. This first cavity argument describes the effective stochastic process for $h_0(t)$. Next we need to describe the behavior of the $\Delta(t)$ variables. 

\paragraph{Adding a Data Point} We can now add a data point which leads to a small perturbation of all of the $N$ features 
\begin{align}
   \tilde \h(t) = \h(t) - \frac{1}{\alpha \sqrt N} \int dt' \frac{\partial \h(t)}{\partial \bm j^h(t')^\top} \ \bm\psi_0 \  \Delta_0(t') + \mathcal{O}(N^{-1})
\end{align}
Plugging this perturbation into the dynamics for the added training error, we find
\begin{align}
    \Delta_0(t) &= \frac{1}{\sqrt N} \bm\psi_0 \cdot \tilde{\h}(t) + \epsilon_0 \nonumber
    \\
    &= \frac{1}{\sqrt N} \bm\psi_0 \cdot {\h}(t) - \frac{1}{\alpha N} \int  dt' \  \bm\psi_0^\top \frac{\partial \h(t)}{\partial \bm j^h(t')^\top} \bm\psi_0  \ \Delta_0(t')  + \epsilon_0  \nonumber
    \\
    &\sim \underbrace{u_{\Delta,0}(t)}_{\text{Gaussian Process}} - \frac{1}{\alpha} \int_0^t dt' \underbrace{\left[ \frac{1}{N} \sum_{i=1}^N \frac{\partial h_i(t)}{\partial j^h_i(t')} \right]}_{\text{Response Function} \ \mR_h(t,s)} \Delta_0(t')  + \epsilon_0
\end{align}
The Gaussian process $u_{\Delta,0}(t)$ has covariance structure
\begin{align}
    \left< u_{\Delta,0}(t) u_{\Delta,0}(t') \right> =  \frac{1}{N} \sum_{i=1}^N h_i(t) h_i(t') \equiv C_h(t,t') 
\end{align}
while the label noise term $\epsilon_0 \sim \mathcal{N}(0,\sigma^2)$ is constant across time. We again note that the test loss can be expressed as $\mathcal L(t) = C_h(t,t) + \sigma^2$. 

\paragraph{Closing the Equations} The final step is to acknowledge that in the $N \to \infty$ limit there is nothing special about the additional feature or data point. Rather, all $N$ features and $P$ training errors will behave as iid random variables. Thus, we can describe the stochastic process for a typical variable 
\begin{align}
    &\frac{d}{dt} h(t) = u_h(t) - \int_0^t dt' R_\Delta(t,t') h(t')   \nonumber
    \\
    &\Delta(t) = u_\Delta(t) - \frac{1}{\alpha} \int_0^t dt' \mR_h(t,t') \Delta(t') + \epsilon  \nonumber
    \\
    &u_h(t) \sim \mathcal{GP}(0, C_\Delta(t,t') ) \ , \ u_\Delta(t) \sim \mathcal{GP}(0, C(t,t') ) \ , \ \epsilon \sim \mathcal{N}(0, \sigma^2) 
\end{align}
We also removed the sources $j$ as we see that we can alternatively differentiate with respect to the variable $u_h(t')$ or $u_\Delta(t')$ which has the same effect (i.e. $R_h(t,t') = \frac{\partial h(t)}{\partial u_h(t')}$). 

\subsection{Path Integral Approach}
The above DMFT equations can also be derived from the moment generating function (path-integral) perspective as well as the cavity method. In the path integral approach, we introduce our variables $\h(t), \bm\Delta(t)$ and enforce their dynamics with conjugate variables $\hat{\h}(t), \hat{\bm\Delta}(t)$ and average over the random data matrix $\bm\Psi \in \mathbb{R}^{N \times P}$ and noise vector $\bm\epsilon \in \mathbb{R}^P$
\begin{align}
    Z &= \int \mathcal D \h \mathcal D \hat{\h} \mathcal D\bm\Delta \mathcal D \hat{\bm\Delta} \nonumber
    \\
    &\times \left<  \exp\left( i \int dt \left[\hat{\h}(t) \cdot \left(\partial_t \h(t) + \frac{1}{\alpha \sqrt N} \bm\Psi^\top \bm\Delta(t)  \right) +  \hat{\bm\Delta}(t) \cdot \left( \bm\Delta(t) - \frac{1}{\sqrt N} \bm\Psi \h(t) + \bm\epsilon \right)  \right] \right)  \right>_{\bm\Psi,\bm\epsilon} \nonumber
    \\
    &= \int \mathcal D C_h \mathcal D C_\Delta \mathcal D R_h \mathcal D R_\Delta \exp\left( - \frac{N}{2} \mathcal S[C_h, C_\Delta, R_h, R_\Delta] \right) .
\end{align}
as we outline in Appendix \ref{app:lin_reg_path_integral}. By averaging over the random matrix $\bm\Psi$, we are able to rewrite the integral over $\{\h, \bm \Delta, \hat{\h}, \bm{\hat\Delta}\}$ in terms of the overlap order parameters
\begin{align}
    &C_h(t,t') \equiv \frac{1}{N} \h(t) \cdot \h(t') \ , \ C_\Delta(t,t') \equiv \frac{1}{P} \bm\Delta(t) \cdot \bm\Delta(t')  \nonumber
    \\
    &R_h(t,t') \equiv -\frac{i}{N} \h(t) \cdot \hat{\h}(t') \ , \ R_\Delta(t,t') \equiv - \frac{i}{P} \bm\Delta(t) \cdot \bm{\hat\Delta}(t')
\end{align}
As $N \to \infty$, this integral is dominated by a single value for these order parameters which are determined by the saddle point. Taking the saddle point equations for the above action $\mathcal S$ generates equations 
\begin{align}
    \frac{\partial \mathcal S}{\partial C_h(t,t')} = \frac{\partial \mathcal S}{\partial C_\Delta(t,t')} = \frac{\partial \mathcal S}{\partial R_h(t,t')} = \frac{\partial \mathcal S}{\partial R_\Delta(t,t')} = 0
\end{align} 
As we show in detail in Appendix \ref{app:lin_reg_path_integral} equations reproduce the formulas obtained with the bipartite cavity method. 

\paragraph{Fourier Transform} We can recognize that the response functions in the above system will have time-translation invariant structure so that $R(t,t') = R(t-t')$. We can therefore take a Fourier transform of these equations, which gives
\begin{align}
    h(\omega) = \mR_h(\omega) \left[ \beta_\star + u_h(\omega) \right] \ , \ i \omega \ \mR_h(\omega) = 1 - \frac{\mR_h(\omega)}{1+\alpha^{-1} \mR_h(\omega)}
\end{align}

\begin{figure}
    \centering
    \subfigure[Response Functions]{\includegraphics[width=0.32\linewidth]{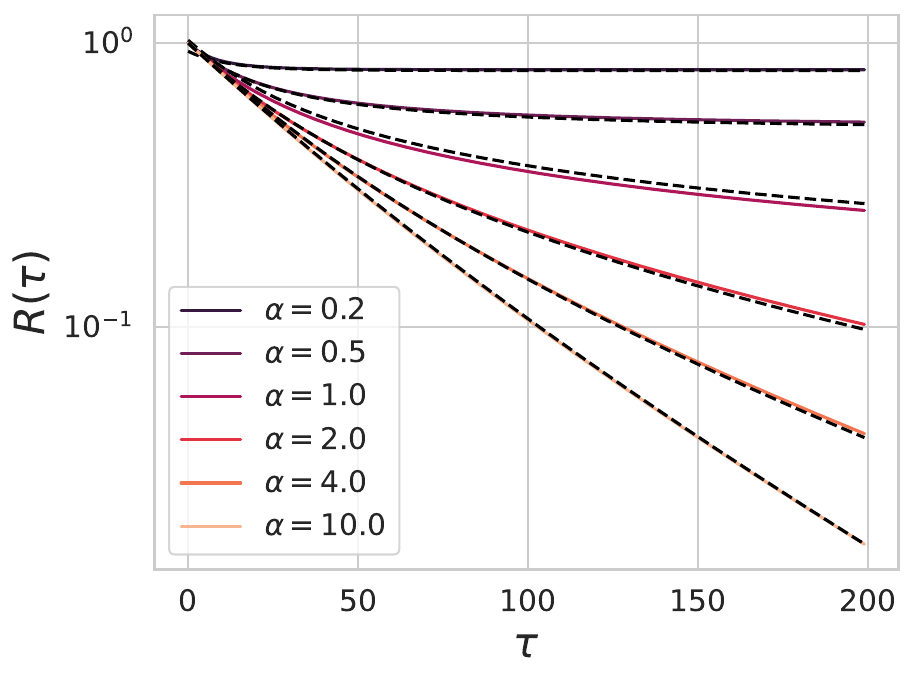}}
    \subfigure[Theoretical Eigenvalue Density]{\includegraphics[width=0.32\linewidth]{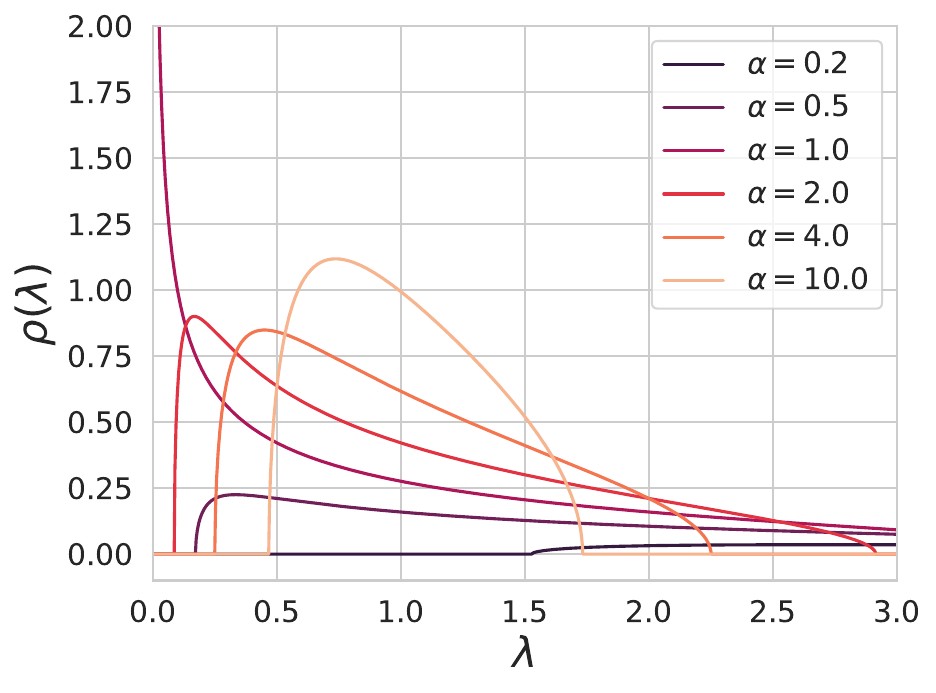}}
    \subfigure[Test Loss Dynamics]{\includegraphics[width=0.32\linewidth]{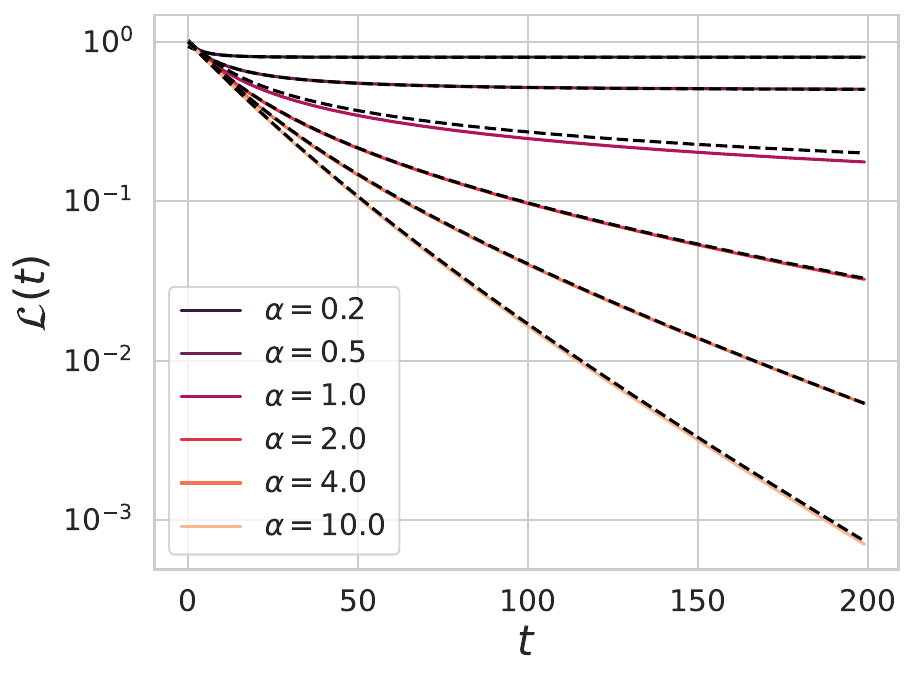}}
    \caption{The dynamics of linear regression with random dataset of side $P = \alpha N$ are governed by the DMFT response function $R(\tau)$ which encodes the spectral properties of a Wishart matrix. Experiments with $N=1000$ are shown in solid lines while the DMFT is plotted in black dashed lines. (a) For $\alpha < 1$, the response function $R(\tau)$ saturates to $1-\alpha$ at large time lag $\tau \to \infty$. Alternatively if $\alpha > 1$, the response function at large $\tau$ relaxes exponentially with timescale set by the minimum eigenvalue $R(\tau) \sim \exp\left(- \left[1 -  \alpha^{-1/2} \right]^2 \tau \right)$. (b) From the Fourier transform of the response $\mR(\omega)$, we can recover the eigenvalue density $\rho(\lambda) = \frac{1}{\pi} \ \text{Im} \mR( i\lambda -\epsilon)$ which we plot without the singularity at $\lambda= 0$. (c) The dynamics of gradient flow accurately describe the effect of subsampled data in the proportional regime $P/ N = \alpha$.   }
    \label{fig:lin_regression_example}
\end{figure}

\paragraph{Eigenvalue Distribution (Marchenko-Pastur Law)} Solving explicitly for the response $\mR_h(\omega)$, we have
\begin{align}
    \mR_h(\omega) = \frac{\alpha}{2(i\omega)} \left[ -(1 + i \omega - \alpha^{-1}) + \sqrt{ (1 + i \omega - \alpha^{-1} )^2 + 4 (i\omega) \alpha^{-1} }  \right] 
\end{align}
Again evaluating at $i \omega = - \lambda - i\epsilon$, taking the $\epsilon \to 0$ limit and using the fact that, the eigenvalue density $\rho(\lambda)$ is recovered
\begin{align}
    \rho(\lambda)= \frac{1}{\pi} \Im  \ \mR(\omega)|_{i\omega=-\lambda - i\epsilon} = \frac{1}{2\pi \lambda} \sqrt{ \left[ 4 \lambda \alpha^{-1} - ( \alpha^{-1}- 1 + \lambda )^2 \right]_+ } + [1-\alpha]_+ \delta(\lambda)
\end{align}
where $\left[ z \right]_+ = \max(z,0)$.  This bulk density has support over $\lambda \in [ (1 - \alpha^{-1/2})^2 , (1+\alpha^{-1/2})^2 ]$. We plot the bulk portion of the density in Figure \ref{fig:lin_regression_example}.

\paragraph{Test Loss} The test loss is governed by the two-frequency correlation function $C(\omega,\omega') = \left< h(\omega) h(\omega')  \right>$ which can be expressed in terms of the response functions $\mR_h(\omega)$ and $\mR_\Delta(\omega)$ 
\begin{align}
    &\mC(\omega,\omega') \equiv \left< h(\omega) h(\omega') \right> = \frac{\mR_h(\omega) \mR_h(\omega') \left[\left< \beta_\star^2  \right> + \frac{\sigma^2}{\alpha (i\omega)(i\omega') } \mR_\Delta(\omega) \mR_\Delta(\omega')  \right]}{1 - \frac{1}{\alpha} \mR_h(\omega) \mR_h(\omega') R_\Delta(\omega) R_\Delta(\omega) }  \nonumber
\end{align}
The real-time test loss can be recovered from an inverse Fourier transform
\begin{align}
    \mathcal L(t) - \sigma^2  = \frac{1}{(2\pi)^2} \int d\omega d\omega' \mC(\omega,\omega') e^{i (\omega + \omega') t} . 
\end{align}
We show an example of the response function $\mR_h(\tau)$, the Marchenko-Pastur eigenvalue densities, and the test loss dynamics in Figure \ref{fig:lin_regression_example}.

\subsubsection{Bias / Variance Decomposition}\label{sec:bias_var_linear} One may also be interested in the separate contributions to the loss dynamics from the bias and the variance induced by random sampling of the dataset. We define the bias as the loss for the \textit{dataset averaged} weight vector $\left< \bm w \right>$ 
\begin{align}
    B(t) = \frac{1}{N} \left| \left< \bm w(t) \right> - \bm w_\star \right|^2  
\end{align}
and the variance is the remaining error $V(t) = \mathcal L(t) - B(t)$. Both the bias $B(t)$ and the variance $V(t)$ can be easily accessed from the DMFT equations. To illustrate this, consider a \textit{bagging} operation where the predictions $E$ separate weight vectors $\bm w_e(t)$ each on their own $\{ \bm \Psi_e , \bm\epsilon_e \}_{e=1}^E$  
\begin{align}
    \bar{\bm w}(t) = \frac{1}{E} \sum_{e=1}^E \bm w_e(t)  \ , \ \frac{d}{dt} \bm w_e(t) = \left( \frac{1}{P} \bm\Psi_e^\top \bm \Psi_e \right) (\bm\beta_\star - \bm w_e(t)) + \frac{\sqrt N}{P} \bm\Psi_e^\top \bm\epsilon_e  \ , \ e \in [E]
\end{align}
The loss of $\bar{\w}(t)$ represents the error of averaging learned models over these $E$ random draws. We can consider the DMFT equations for this $E$-fold replicated system in terms of the error variables $\bm h_e(t) = \bm\beta_\star - \bm w_e(t)$ from the following
\begin{align}
    &\frac{d}{dt} h_e(t) = u_{h,e}(t) - \int dt' R_\Delta(t,t') h_e(t')   \ , \ \left< u_{h,e}(t) u_{h,e'}(t') \right> =  \delta_{e,e'} \alpha^{-1} C_{\Delta,e}(t,t') 
    \\
    &\Delta_e(t) = u_{\Delta, e}(t) - \alpha^{-1} \int R_h(t,t') \Delta_e(t') + \epsilon_e   \ , \ \left< u_{\Delta,e}(t) u_{\Delta,e'}(t') \right> =  \delta_{e,e'} C_{h,e}(t,t') 
\end{align}
We see that the Gaussian processes $u_h(t)$ and $u_\Delta(t)$ are uncorrelated across different copies $e \neq e'$ of the system. Further, each copy will have identical within-system correlations $C_{h,e}(t,t') = C_h(t,t')$ for all $e \in [E]$. Thus, our averaged system $\bar{h}(t) = \frac{1}{E}\sum_{e=1}^E h_e(t)$ has the dynamics
\begin{align}
    &\frac{d}{dt} \bar h(t) = \bar u_h(t)  - \int dt' R_\Delta(t,t') \bar h(t')  
    \\
    &\bar u_h(t) = \frac{1}{E} \sum_{e=1}^E u_{h,e}(t) \sim \mathcal{GP}\left( 0 , \frac{1}{\alpha E} C_\Delta(t,t')  \right)
\end{align}
Thus, bagging over $E$ independent datasets effectively reduces the variance of the Gaussian process driving the right hand side of the system. The exact bias can be computed as the limit as $E \to \infty$. 

The bias $B(\omega,\omega')$ can be computed from the average of the dynamics over many random draws of datasets. Concretely, the average predictor has Fourier transform 

\begin{align}
    &\mathcal B(\omega,\omega') \equiv \left< \left< h(\omega) \right>_{u_h} \left< h(\omega') \right>_{u_h} \right> = \mR_h(\omega) \mR_h(\omega') \left< \beta_\star^2  \right> \nonumber
    \\
    &\mathcal V(\omega,\omega') \equiv C(\omega,\omega') - B(\omega,\omega') = \frac{\frac{1}{\alpha} \mR_h(\omega) \mR_h(\omega') \mR_\Delta(\omega) \mR_\Delta(\omega) B(\omega,\omega')}{1- \frac{1}{\alpha} \mR_h(\omega) \mR_h(\omega') R_\Delta(\omega) R_\Delta(\omega)}
\end{align}
We therefore see that the response function $\mR_h(\omega)$ completely determines both the bias and the variance components

\subsection{Hermiticity \& Monotonicity in Noise Free Setting}
 We note that the response function and correlation function are linked for linear dynamical systems defined by a Hermitian matrix $\bm M = \bm M^\top$
\begin{align}
    C(t,t') = \text{tr} \left[ \exp\left( - \bm M t \right) \right]^\top \left[ \exp\left( - \bm M t' \right) \right] = \text{tr} \exp\left( -   \bm M (t+t')  \right) = R( t+ t' ) 
\end{align} 
As a consequence, the loss function $\mathcal L(t) = C(t,t) + \sigma^2$ can in fact be computed from the \textit{one-point function}/resolvent/response function $R(t)$. However, this is not always the case if $\bm M  \neq \bm M^\top$. We will return to this point explicitly in Section \ref{sec:random_features}.

\paragraph{Monotonic Convergence in Noise Free Setting} The autonomous (noise-free $\sigma^2=0$) linear dynamics for this regression problem leads to monotonic decrease in the test loss since
\begin{align}
    \frac{d}{dt}\mathcal L(t) = - \frac{2}{N} \h(t)^\top \left( \frac{1}{P} \bm\Psi^\top \bm\Psi \right) \h(t) = - \frac{2}{N P} |\bm\Psi \h(t)|^2 = - 2\int_0^\infty d\lambda \rho(\lambda) \lambda e^{-2\lambda t} \leq 0
\end{align}
This is a consequence of the fact that the matrix $\left( \frac{1}{P} \bm\Psi^\top \bm\Psi \right)$ is Hermitian and positive semidefinite. In the next section, however, we show how that a simple extension of this model to a random feature projection can be characterized as a linear dynamical system with \textit{non-Hermitian} random matrix, allowing for non-monotonic training dynamics for certain values of the parameters.

\paragraph{Final Loss} We can access the final value of the loss by examining the low-frequency $\omega,\omega'\to 0$ limit of the correlation function (e.g. final value theorem). To start, we need the 
\begin{align}
    \lim_{\tau \to \infty} R_h(\tau) = \lim_{\omega\to 0} (i\omega) \mR_h(\omega) = 
    \begin{cases}
        1-\alpha & \alpha < 1
        \\
        0 & \alpha > 1
    \end{cases}
\end{align}
From this final value for the response function, we can access the final value of the correlation function 
\begin{align}
    \lim_{t\to\infty} \mathcal L(t) - \sigma^2 = \lim_{t \to \infty} C(t,t) = \lim_{\omega,\omega' \to 0} (i\omega)(i\omega') \mC(\omega,\omega') = \begin{cases}
        1-\alpha + \frac{\sigma^2 \alpha}{1-\alpha } & \alpha < 1
        \\
        \frac{\sigma^2}{\alpha - 1} & \alpha > 1
    \end{cases}
\end{align}
We note that in the presence of noise $\sigma^2 > 0$, the loss curve exhibits an overfitting peak at $\alpha = 1$. The final bias and variance can be similarly deduced
\begin{align}
    \lim_{t \to \infty} B(t) = 
    \begin{cases}
        (1-\alpha)^2   & \alpha < 1
        \\
        0 & \alpha > 1
    \end{cases} 
    \ , \ \lim_{t \to \infty} V(t) = 
    \begin{cases}
        \alpha (1- \alpha) + \frac{\sigma^2 \alpha}{1-\alpha} & \alpha < 1
        \\
        \frac{\sigma^2}{\alpha-1} & \alpha > 1
    \end{cases} .
\end{align}
While the bias monotonically decreases with $\alpha$, we see that the variance can be non-monotonic and can even diverge at $\alpha = 1$. 

\section{Structured Covariates and Kernel Methods}

\begin{figure}
    \centering
    \subfigure[Effective Density for $k$-th mode]{\includegraphics[width=0.4\linewidth]{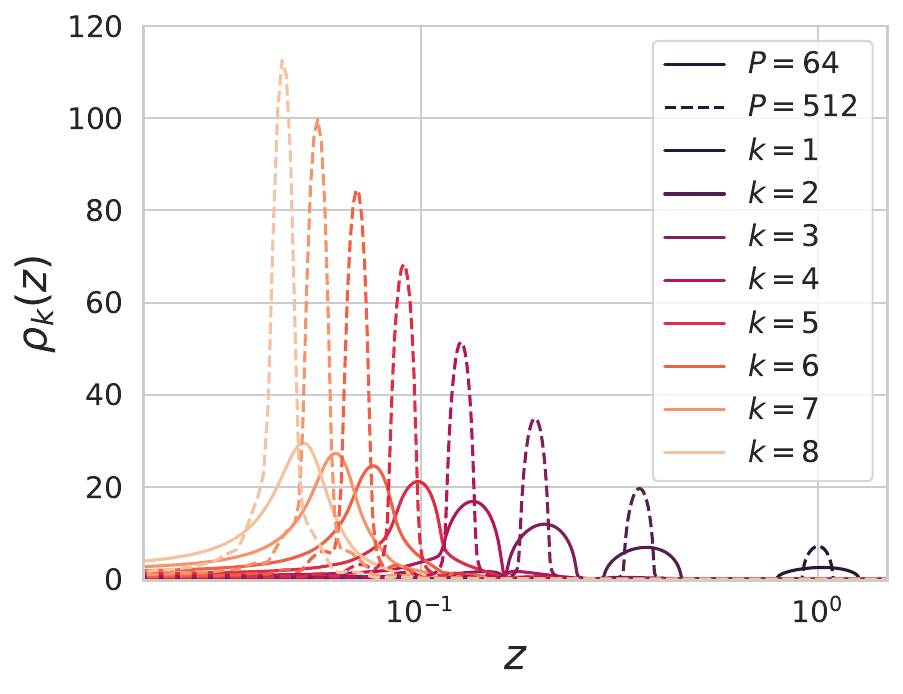}}
    \subfigure[Response Functions vs $\tau$]{\includegraphics[width=0.4\linewidth]{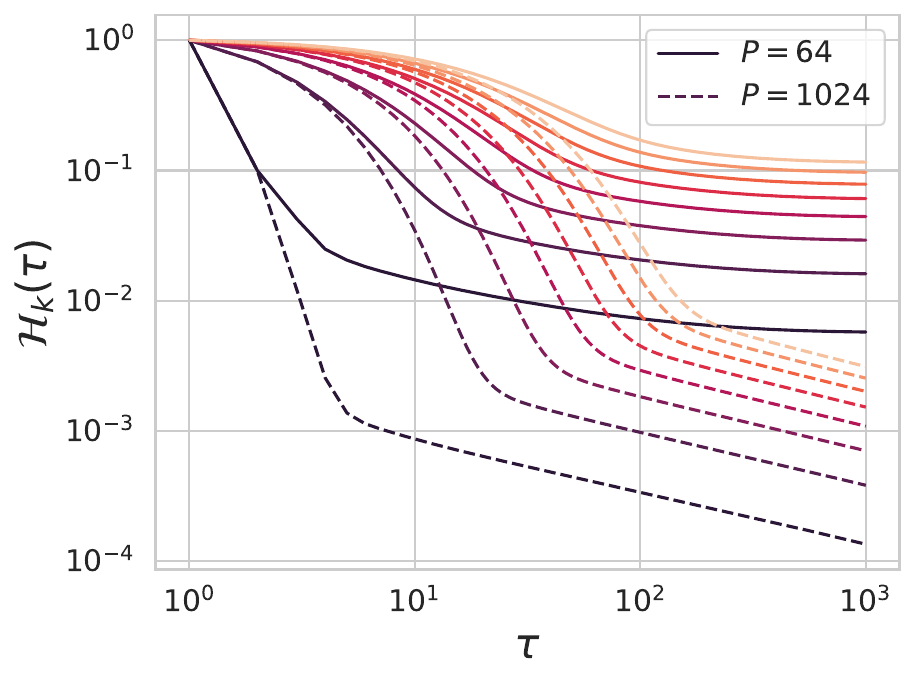}}
    
    \caption{The effective densities $\rho_k(z)$ and corresponding response functions $\mathcal H_k(\tau)$ for structured kernel regression with eigenvalues $\lambda_k = k^{-b}$ with $b=1.25$. We plot the first $8$ eigemodes $k \in [10]$ for $P \in \{64, 512\}$. (a) The effective densities $\rho_k(z)$ exhibit larger spread for smaller $P$. As $P \to \infty$ these converge to Dirac masses $\rho_k(z) \to \delta(z-\lambda_k)$. (b) The response functions $\mathcal H_k(\tau)= \int dz e^{-z\tau} \rho_k(z) $ as a function of the timelag $\tau$. For small $\tau$, these functions scale as $\mH_k(\tau) \sim e^{-\lambda_k \tau}$, while for large $\tau$ they relax to a constant.  }
    \label{fig:kernel_theory}
\end{figure}

So far, our mean field equations have resulted in a description of the system where each individual site becomes statistically identical as a stochastic process. We can generalize this to a structured random matrix by adding a specific set of eigenvalues $\lambda_k$ for each feature $\psi_k$  
\begin{align}
    \frac{d}{dt} \h(t) = - \left( \frac{1}{P} \bm\Psi^\top \bm\Psi \right) \h(t) + \frac{1}{P} \bm\Psi^\top \bm\epsilon \quad  , \quad \left< \Psi_{\mu k} \Psi_{\nu \ell} \right> = \delta_{\mu \nu} \delta_{k \ell} \lambda_k  \quad , \quad y = \sum_k \beta^\star_k \psi_k + \epsilon 
\end{align}
The values $\lambda_k$ represent the \textit{population covariance eigenvalues} while the coefficients $\beta_k$ represent the decomposition of the target function in these eigenfeatures $\psi_k$ \footnote{Note that if the population feature covariance was instead a non-diagonal matrix $\bm\Sigma$ we are free to perform a change of basis to render it diagonal.}. 
In what follows, we will take the number of eigenvalues to infinity first and assume trace class (dimension-free) structure 
\begin{align}
    \sum_{k=1}^\infty \lambda_k < \infty \quad , \quad  \left< y^2 \right> = \sum_{k=1}^\infty \lambda_k (\beta^\star_k)^2  + \sigma^2 < \infty
\end{align}
In this case, the mean field dynamics provide a decoupled set of stochastic processes for each population eigenmode, though we note that the stochastic processes $h_k(t)$ are not exchangeable in the limit and follow distinct dynamics depending on $\lambda_k$\footnote{We note that the mean field limit is no longer exact in the trace class regression setting, but becomes more accurate as $P$ increases. Analyzing leading order fluctuations around this DMFT is possible through study of higher order derivatives of the DMFT action. 
}.
\begin{align}
    &\frac{d}{dt} h_k(t) =  u_k(t) - \lambda_k \int dt'  R_{\Delta}(t,t')  h_k(t') \ , \ u_k(t) \sim \mathcal{GP}\left(0, \frac{\lambda_k}{P} C_\Delta(t,t') \right)  \nonumber
    \\
    &\Delta(t) = u_\Delta(t) - \frac{1}{P} \int dt' \sum_{k=1}^\infty \lambda_k H_{k}(t,t')  \Delta(t') + \epsilon \ , \ u_\Delta(t) \sim \mathcal{GP}(0, C(t,t')) \ , \ \epsilon \sim \mathcal{N}(0,\sigma^2) \nonumber
\end{align}
where the $h_k(t)$ dynamics are integrated from the initial condition $h_k(0) = \beta^\star_k$. We see from the above DMFT equations that the $h_k(t)$ variables are independent but not identically distributed (unlike the previous examples). The correlation functions $C(t,t'), C_\Delta(t,t')$ and response functions $\{ H_k(t,t') \}_{k=1}^\infty, R_\Delta(t,t')$ are defined as
\begin{align}
    &C(t,t') =  \sum_{k=1}^\infty \lambda_k  \left< h_k(t) h_k(t') \right>  \ , \ C_\Delta(t,t') = \left< \Delta(t) \Delta(t') \right>
    \\
    &H_k(t,t') = \frac{\partial h_k(t)}{\partial u_k(t')} \ , \ R_\Delta(t,t') =  \frac{\partial\Delta(t)}{\partial u_\Delta(t')}
\end{align}
where $\left< \right>$ above denotes averages over the $u_k(t), u_\Delta(t), \epsilon$ random variables. The response functions can be solved for directly in Fourier space
\begin{align}
    \mH_k(\omega) = \frac{1}{i\omega + \lambda_k \mR_\Delta(\omega)} \ , \ \mR_\Delta(\omega) = 1 - \frac{1}{P} \sum_{k=1}^\infty \frac{\lambda_k \mR_\Delta(\omega)}{i\omega + \lambda_k \mR_\Delta(\omega) }  . 
\end{align}
In analogy to the relationship between the response function and a density of timescales (eigenvalues), we can define a collection of effective densities $\rho_k(z)$ which characterize the spread in time constants for each eigenmode $k$ due to finite $P$ 
\begin{align}
    \rho_k(z) \equiv \lim_{\epsilon \to 0} \  \frac{1}{\pi} \Im \ \mH_k( i z - \epsilon)  \ , \ \mH_k(\tau) = \int dz \rho_k(z) e^{-z\tau }
\end{align}
We visualize these effective densities $\rho_k(z)$ for the first few eigenmodes in Figure \ref{fig:kernel_theory}. These describe the spread of time-constants along each population eigendirection. From this solution to the response functions $\mH_k(\omega)$, we can 
\begin{align}
    &\mC(\omega,\omega') = \frac{1}{1-\Gamma(\omega,\omega')} \left[ \sum_k \lambda_k (\beta^\star_k)^2 \mH_k(\omega) \mH_k(\omega')  + \frac{\sigma^2}{(i\omega)(i\omega')} \Gamma(\omega,\omega')  \right] \nonumber
    \\
    &\Gamma(\omega,\omega') = \frac{1}{P} \sum_k \lambda_k^2 \ \mH_k(\omega) \mH_k(\omega') \mR_\Delta(\omega) \mR_\Delta(\omega')
\end{align}
We see that the correlation function depends on $P$ through the response functions $\mH_k(\omega)$ and $\mR_\Delta(\omega)$, which control the bias, as well as the function $\Gamma(\omega,\omega')$ which controls the variance component of the loss. 

\paragraph{Final Loss}
In the long time limit (equivalently low frequency $i\omega\to 0$ limit), we have $\mR_\Delta(\omega) \sim i \omega r_\Delta$ so that the final loss 
\begin{align}
   \lim_{t \to \infty } \mathcal L(t) = \lim_{t \to \infty} C(t,t) + \sigma^2 =  \frac{1}{1-\Gamma} \left[ \sum_{k=1}^\infty   \frac{\lambda_k (\beta^\star_k)^2 }{( 1 + \lambda_k r_\Delta )^2  } + \sigma^2  \right]
\end{align}
where $r_\Delta$ and $\Gamma$ are defined by the low-frequency limit of the DMFT equations
\begin{align}
    1 = \frac{1}{P} \sum_{k=1}^\infty \frac{\lambda_k r_\Delta}{1 + \lambda_k r_\Delta} \ , \ \Gamma = \frac{(r_\Delta)^2}{P}  \sum_{k=1}^\infty \frac{\lambda_k^2}{(1 + \lambda_k r_\Delta )^2} , 
\end{align}
The large time limit of these DMFT equations recover static computations for the test loss of the final predictor \cite{bordelon2020spectrum, canatar2021spectral, loureiro2021learning}.

\subsection{Gapped and Degenerate Spectra: Multiple Stages in Time and Data} 

\begin{figure}
    \centering
\subfigure[Finite $D$ Theoretical Loss Dynamics]{\includegraphics[width=0.38\linewidth]{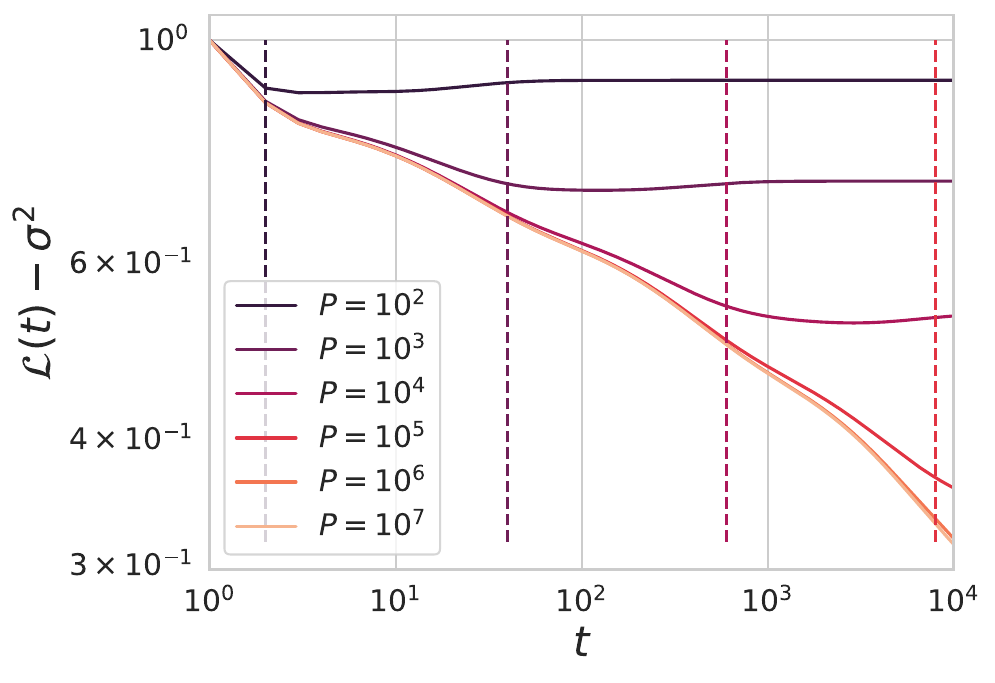}}
\subfigure[Finite $D$ Theoretical Final Losses]{\includegraphics[width=0.38\linewidth]{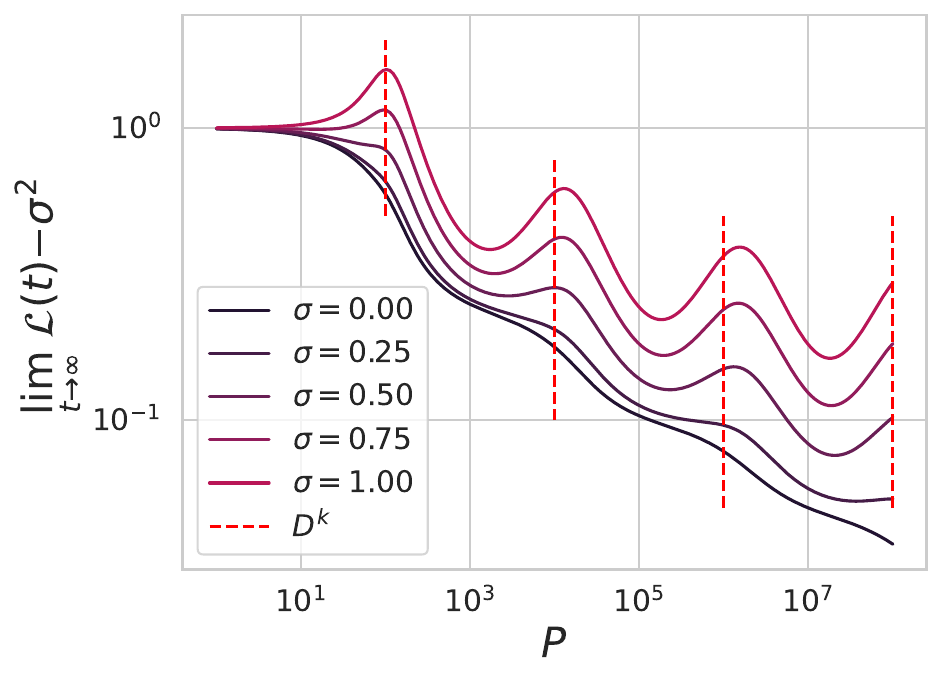}}
\subfigure[$k$-th Stage Dynamics $D \to \infty$]{\includegraphics[width=0.38\linewidth]{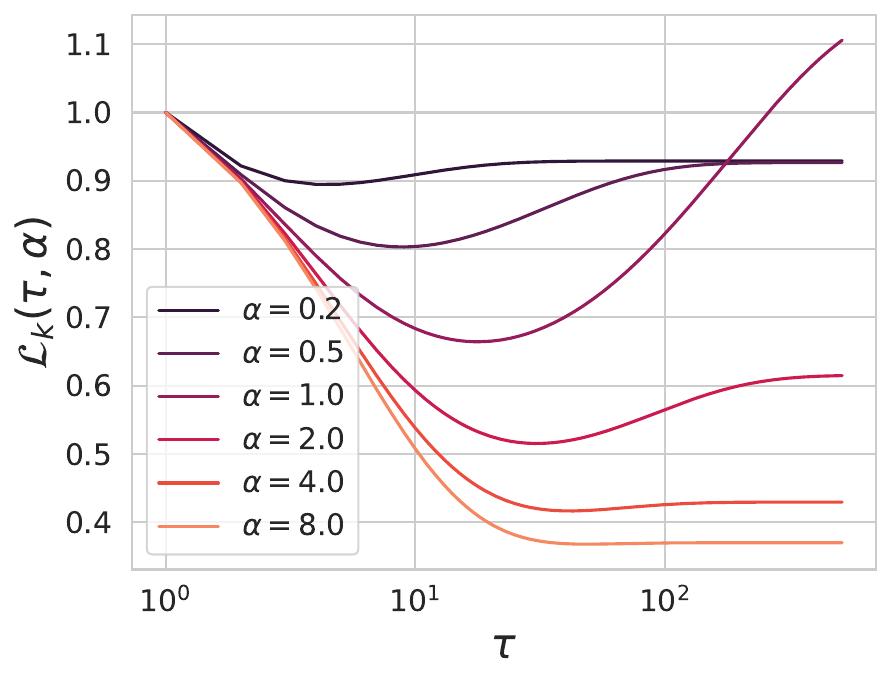}}
\subfigure[Development of $k$-th Overfitting Peak]{\includegraphics[width=0.38\linewidth]{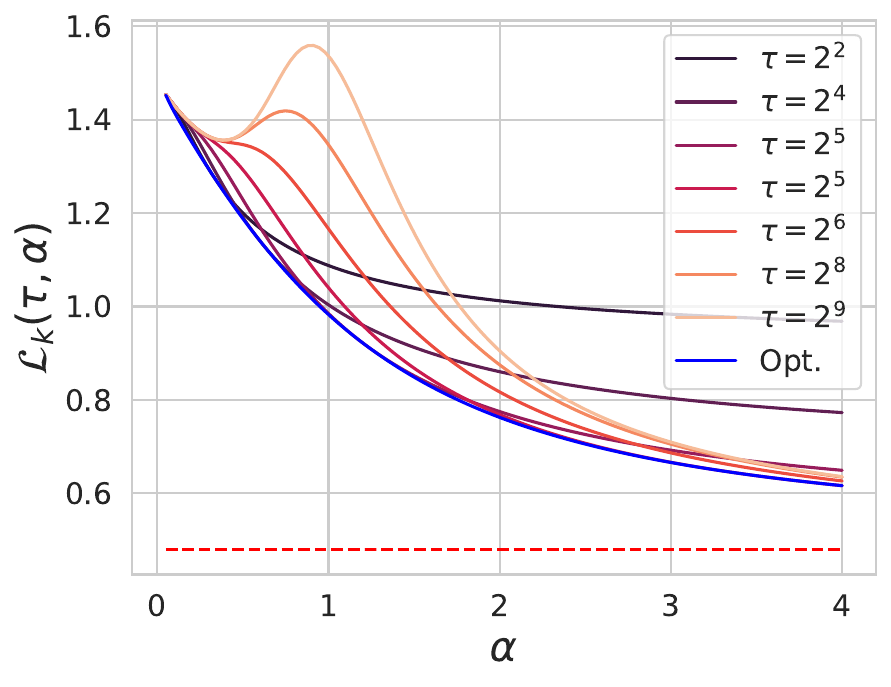}}
    \caption{Multistage learning curves for gapped degenerate spectra at finite $D$ as well as a degree-$k$ scaling limit. (a) Theoretical loss dynamics across various $P$ exhibit multiscale behavior with transitions at $\eta^{-1} \lambda_k^{-1}$ (dashed vertical lines). (b) The final value of the loss $\lim_{t\to\infty} \mathcal L(t) - \sigma^2$ across varying $P$ and noise levels $\sigma$. (c) The $k$-th stage limit $\mathcal L_k(\tau,\alpha)$ exhibits dynamics in the rescaled time $\tau$ and rescaled data $\alpha$. (d) The loss exhibits non-monotonicity in $\alpha$ and $\tau$ around the $k$-th stage with an optimal early stopping time $\tau_\star(\alpha)$ (blue). The loss relaxes as $\tau,\alpha \to \infty$ to the unlearnable variance $\sum_{\ell>k} \eta_\ell \beta_\ell^2 + \sigma^2$ (dashed red line), which can be interpreted as an effective noise level.  }
    \label{fig:multiple_stage}
\end{figure}

In this section, we explore the case of a spectrum that consists of stages of a large number of degenerate eigenvalues. One commonly studied setting which generates this is dot-product kernels for spherically symmetric data in $D \gg 1$ dimensions \cite{mei2022generalization, bordelon2020spectrum, canatar2021spectral}. In this case, eigenvalues $\lambda_k$ are all equal for the orthogonal degree $k$ polynomials (Hermite polynomials or spherical harmonics) which carry multiplicity $\mathcal N_k$ which scales polynomially with the dimension $D$ in the following way (for $D \gg k$)
\begin{align}
    \mathcal N_k \sim \mathcal{O}( D^{k} ) \ , \ \lambda_k \sim  \mathcal{O}( D^{-k} )  \ \ \text{for} \ D \gg k
\end{align}
Under this degenerate spectrum, the formula for the response function $\mR_\Delta(\omega)$ takes the form
\begin{align}
     \mR_\Delta(\omega) = 1 - \frac{1}{P} \sum_{k=1}^\infty \frac{\lambda_k \mR_\Delta(\omega) \mathcal N_k}{i\omega + \lambda_k \mR_\Delta(\omega)}
\end{align}
This equation can be solved exactly for any collection of $\{ \mathcal N_k , \lambda_k, \omega \}$ at finite $D,P$ as we show in Figure \ref{fig:multiple_stage} (a)-(b). These equations exhibit multiple stages of learning in both time $t$ and data $P$. To gain additional insight into these stages, we can take the following high dimensional $D \to \infty$ limit where timescales $t = 1/(i\omega) \propto D^k$ and data is $P \propto D^k$ both scale polynomially in $D$ 
\begin{align}
   k\text{-th stage limit: } \quad  \lim_{\substack{D,t,P \to \infty \\ t=\tau D^k , P = \alpha D^k}} \mathcal L(t,D,P) \equiv \mathcal L_k(\tau, \alpha)
\end{align}
Under this scaling limit, we define the following limiting response functions 
\begin{align}
    \mathfrak h_{k,\ell}(\Omega) = \lim_{D \to\infty } \  D^{-k} \ \mathcal H_\ell(\Omega) = 
    \begin{cases}
        0 & \ell < k
        \\
        \left( i\Omega + \eta_k R_\Delta(\Omega) \right)^{-1} & \ell = k
        \\
        (i\Omega)^{-1} & \ell > k
    \end{cases}
\end{align}
which implies that in this scaling limit, all of the modes $\ell < k$ have been perfectly learned, the mode $k$ is currently being learned (note the competition between the $i\Omega$ and the response function $R_\Delta(\omega)$ and all modes $\ell > k$ are unlearnable at these timescales. Under this limit, the response function $\mR_\Delta(\Omega)$ satisfies the following equation
\begin{align}
    \mR_\Delta(\Omega) = 1 - \frac{1}{\alpha} \frac{\eta_k n_k \mR_\Delta(\Omega)}{i\Omega + \eta_k \mR_\Delta(\Omega)} - \frac{\mR_\Delta(\Omega)}{\alpha (i\Omega)} \sum_{\ell > k} \eta_\ell n_\ell 
\end{align}
Defining $\mathfrak{h}_k(\Omega) = \lim_{D \to\infty} D^{-k} \mH_k(\Omega)= \left( i\Omega + \eta_k \mR_\Delta(\Omega) \right)^{-1}$, the loss takes the form 
\begin{align}
    \mathcal C_k(\Omega,\Omega') = \frac{1}{1-\Gamma(\Omega,\Omega')} \left[ \underbrace{\eta_k (\beta_k^\star)^2 \mathfrak h_k(\Omega) \mathfrak h_k(\Omega')}_{\text{Mode} \ k \ \text{learning curve}} + \underbrace{\frac{1}{i\Omega i\Omega'} \sum_{\ell > k} \eta_\ell (\beta_\ell)^2 +  \frac{\sigma^2}{i\Omega i\Omega} \Gamma(\Omega,\Omega')}_{\text{Effective noise}}  \right]
\end{align}
where we have identified the contribution from the learnable component (the $k$-th stage) and the components which act as effective noise (all higher stages $\ell > k$). The $k$-th stage loss $\mathcal L_k(\tau,\alpha)$ in rescaled time $\tau$ can be accessed again as a two-variable Fourier transform
\begin{align}
  \mathcal L_k(\tau, \alpha) = \int \frac{d\Omega d\Omega'}{(2\pi)^2} \left[ \mathcal C_k(\Omega,\Omega') + \frac{\sigma^2}{i\Omega i \Omega'}  \right] e^{i (\Omega+\Omega') \tau} .
\end{align}
We plot the function $\mathcal L_k(\tau,\alpha)$ in Figure \ref{fig:multiple_stage} (c)-(d). This function exhibits non-monotonicity in both $\tau$ and $\alpha$, with an optimal early stopping time (blue) and a potential overfitting peak at late times near $\alpha \approx 1$. The large $\tau$ limit in stage-$k$ takes the form
\begin{align}
    \lim_{\tau \to\infty} \mathcal L_k(\tau,\alpha) &= \frac{1}{1-\gamma} \left[ \underbrace{\frac{\eta_k (\beta_k^\star)^2  }{(1 + \eta_k r)^2}}_{\text{Mode} \ k \ \text{learning curve}}  +   \underbrace{\sum_{\ell > k} \eta_\ell (\beta_\ell)^2  + \sigma^2}_{\text{Effective noise}}  \right] 
    \\
    1  &= \frac{r_\Delta}{\alpha} \left[ \frac{\eta_k n_k  }{1 + \eta_k r_\Delta} + \sum_{\ell > k} \eta_\ell n_\ell \right] \ , \ \Gamma = \frac{r^2_\Delta}{\alpha} \frac{\eta_k^2 n_k}{(1+\eta_k r_\Delta)^2}
\end{align}
We note that $\lim_{\tau,\alpha \to \infty} \mathcal L_k(\tau,\alpha) = \sum_{\ell>k} \eta_\ell (\beta_\ell)^2 + \sigma^2$ is larger than the $\lim_{D \to \infty} \lim_{t,P \to\infty} \mathcal L(t,D,P) = \sigma^2$ since the former reflects the best predictor possible at scales $t \sim D^k$ and $P \sim D^k$  
\begin{align}
    \underbrace{\lim_{\tau,\alpha \to \infty} \mathcal L_k(\tau,\alpha)}_{\text{Final Value for $P=\alpha D^k$}} = \sum_{\ell>k} \eta_\ell (\beta_\ell)^2 + \sigma^2 >  \lim_{D\to\infty} \underbrace{\lim_{t,P \to \infty} \mathcal L(t,D,P)}_{\text{Large Time \& Data, Fixed $D$}} = \sigma^2 
\end{align}

\subsection{Power Laws} We can also analyze the case of powerlaw features for which
\begin{align}
    \lambda_k \sim k^{-\nu} \ , \ \sum_{\ell < k} \lambda_\ell (\beta_\ell)^2 \sim k^{-\nu\chi } . 
\end{align}
where the exponents $\chi\nu$ and $\nu$ are known as the source and capacity exponents respectively \cite{cui2021generalization, bordelon2024dynamical}. The solution to $\mR_\Delta(\omega)$ can be approximated by an early time (high frequency) and late time (low frequency) expansion in $P (i\omega)^\nu$ 
\begin{align}
    \mathcal R_\Delta(\omega) = 1 - \frac{1}{P } \sum_{k=1}^\infty \frac{\lambda_k \mR_\Delta(\omega)}{i\omega + \lambda_k \mR_\Delta(\omega)} \approx 
    \begin{cases}
    1 & \omega P^\nu \gg 1
    \\
    i \omega P^{\nu}  & \omega P^{\nu} \ll 1
    \end{cases} , 
\end{align} 
which leads to the following approximations for the response functions $\mH_k(\omega)$
\begin{align}
    \mH_k(\omega) \approx 
    \begin{cases}
        \frac{1}{i\omega + \lambda_k } + \frac{1}{P (i\omega)^{\frac{1}{\nu}}} \frac{\lambda_k}{(i\omega + \lambda_k)^2}   + ... &  \omega P^{\nu} \gg 1
        \\
        \frac{1}{i\omega} \frac{1}{\left( 1 + (P / k )^\nu \right) } + ... & \omega P^\nu \ll 1
    \end{cases}
\end{align}
These equations indicate that the early stage of the dynamics, the loss dynamics along the $k$-th population eigenmode obey $\mathcal H_k(\tau) \approx e^{-\lambda_k \tau} + \mathcal{O}(P^{-1})$ for small times $\tau$ and converge to $\lim_{\tau \to \infty} \mH_k(\tau) \approx \frac{1}{1 + (P/k)^\nu }$ at late time. This indicates the intuitive fact that if $P \gg k$ this eigenmode will be learned effectively after sufficient training, while for $P \ll k$ this eigenmode cannot be resolved at this sample size $P$. As a consequence the loss can be roughly approximated as a combination of powerlaws (motivated by \cite{hoffmann2022training}) 
\begin{align}
    \mathcal L(t,P) \approx c_t \  t^{- \chi} + c_P \ P^{- \nu \chi} .
\end{align}
in the sense that this expression captures the $\lim_{t \to \infty} \mathcal L \sim P^{-\nu\chi}$ and $\lim_{N \to \infty} \mathcal L \sim t^{-\chi}$. This is the same functional form explored in ``Chinchilla" neural scaling laws \cite{hoffmann2022training, paquette20244+}. Additional terms can be incorporated which capture the dynamical effect of variance from sampling the random dataset (which are mixed terms involving both finite $t,P$).

\section{Random Feature Model}\label{sec:random_features}

We can consider a simple random feature model of the form
\begin{align}
    f =  \frac{1}{N_1} \w(t)^\top \bm A \bm\psi \ , \ y = \bm\beta^\star \cdot \bm\psi  \ , \  \left< \psi_k \psi_\ell \right> = \lambda_k \delta_{k\ell} \ , \ \left< \epsilon^2 \right> = 1
\end{align}
where $\bm A \in \mathbb{R}^{N_1 \times N_0}$ is a frozen random matrix and $\w(t) \in \mathbb{R}^{N_1}$ is trained with gradient flow. The important variable to track is the discrepancy between the target weights $\bm\beta_\star$ and the effective model weights $\frac{1}{N_1} \bm A^\top \bm w(t)$ which gives
\begin{align}
    \h_0(t) \equiv  \bm\beta_\star -  \frac{1}{N_1} \bm A^\top  \w(t) \in \mathbb{R}^{N_0}
\end{align}
The test loss is simply $\mathcal L(t) = \bm h_0(t)^\top \bm\Lambda \bm h_0(t)$. Gradient flow on $\w(t)$ induces the following dynamics on this error variable $\h_0(t)$  
\begin{align}
    \frac{d}{dt} \h_0(t) = - \left( \frac{1}{N_1} \bm A^\top \bm A \right) \left( \frac{1}{P} \bm\Psi^\top \bm \Psi \right) \h_0(t) .
\end{align}
Before setting out to describe the DMFT equations for this model for random $\bm A$ matrix and random data $\bm\Psi$ matrix, we first note that this matrix is asymmetric and non-normal.

\begin{figure}
    \centering
     \includegraphics[width=0.9\linewidth]{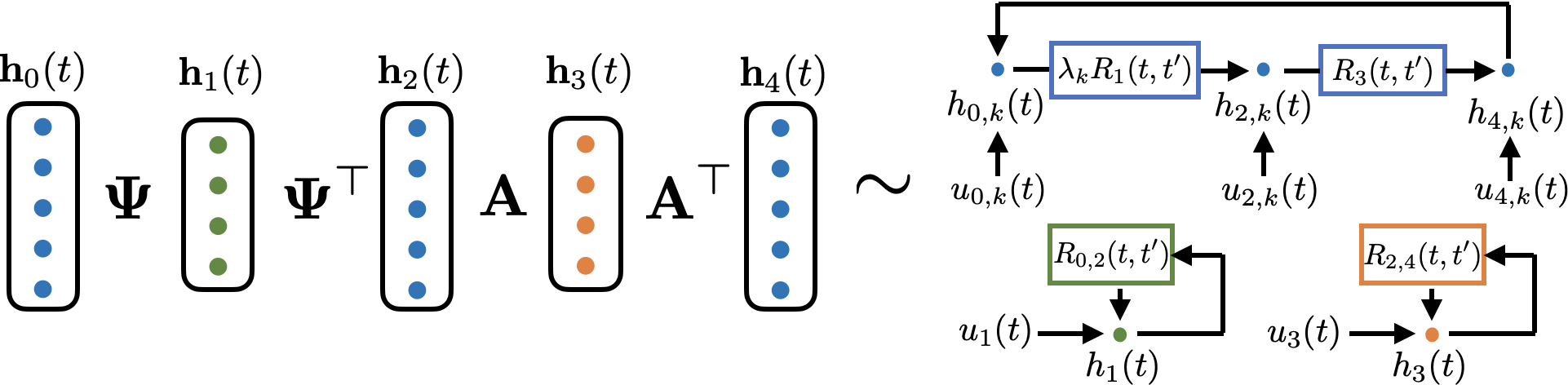}
    \caption{Visualization of the decomposition of the dynamics into four separate components. The $M$ dimensional base features are represented in blue. The $P$ dimensional space for the training predictions $\h_1(t)$ are green and the projection of $\h_2$ into an $N$ dimensional random feature space is represented in orange. In the proportional limit, the variables $\{ \h_0(t) , \h_2(t) , \h_4(t) \}$ are coupled and statistically dependent but the variables $\h_1(t)$ and $\h_3(t)$ evolve independently. The variable $\h_1(t)$ is related to the training error statistics while $\h_0(t)$ encodes test error statistics.   }
    \label{fig:visualize_random_features}
\end{figure}

\paragraph{Possibility of Non-Normal Overfitting} From this above equation, we identify the matrix which drives the dynamics as $\bm M = \left( \frac{1}{N_1} \bm A^\top \bm A \right) \left( \frac{1}{P} \bm\Psi^\top \bm \Psi \right)$, we can again try writing the dynamics for the test loss 
\begin{align}
    \frac{d}{dt} \mathcal L(t) &= - 2 \ 
    \h_0(t)^\top \ \bm\Lambda \  \bm M  \ \h_0(t) \nonumber = - 2 \ \bm h_0(t)^\top \bm\Lambda \left[ \left( \frac{1}{N_1} \bm A^\top \bm A \right)  \right]\left[\left( \frac{1}{P} \bm\Psi^\top \bm \Psi \right) \right] \h_0(t)  
\end{align}
which is not-necessarily negative since the vectors $\bm\Lambda \left( \frac{1}{N} \bm A^\top \bm A \right) \h_0(t)$ and $\left( \frac{1}{P} \bm\Psi^\top \bm \Psi \right) \h_0(t)$ do not necessarily have positive dot-product. However, as $N_1 \to \infty$ (or $P \to \infty$) then $\frac{1}{N_1} \bm A^\top \bm A \to \bm I$ (respectively $\frac{1}{P} \bm\Psi^\top \bm\Psi \to \bm \Lambda$) and then monotonicity of the loss dynamics is recovered as $\bm\Lambda \bm M$ recovers Hermitian positive-semidefinite structure. We will see that for $N = P$ finite, the non-normal dynamics can lead to a divergence in the loss as $t \to \infty$. 





\paragraph{Decomposing the Dynamics} As we did in the linear regression problem, we can break up the matrix vector product $\left( \frac{1}{N} \bm A^\top \bm A \right) \left( \frac{1}{P} \bm\Psi^\top \bm \Psi \right) \h(t)$ into four components, each of which are linear in one of the random matrices. To do so, we first introduce the ratios $\alpha = P/M$ and $\nu = N / M$, allowing us to rewrite the dynamics as 
\begin{align}
    &\h_1(t) =  \bm\Psi \h_0(t) \ , \ \h_2(t) = \frac{1}{P} \bm\Psi^\top \h_1(t) \nonumber
    \\
    &\h_3(t) = \bm A \h_2(t) \ , \ \h_4(t) = \frac{1}{N_1} \A^\top \h_3(t) \nonumber
    \\
    &\frac{d}{dt} \h_0(t) = - \h_4(t) .
\end{align}
We visualize this decomposition in Figure \ref{fig:visualize_random_features} which shows how each of these four vectors are computed graphically. In the mean field limit, each single site (represented as a dot) evolves as an independent stochastic process. We now describe this limit.

\paragraph{Mean Field Description} By averaging over the random matrices $\bm\Psi, \bm A$, we derive the following stochastic differential equations for each typical entry in the vectors $\h_{\ell}$ for $\ell \in \{0 ,... ,4\}$. Following a cavity argument like what was presented previously, we find the limiting stochastic process to be
\begin{align}
    &h_1(t) = u_1(t) + \frac{1}{P} \int dt' R_{0,2}(t,t') h_1(t') \ , \ u_1(t) \sim \mathcal{GP}(0, C_0(t,t') ) \nonumber
    \\
    &h_{2,k}(t) = u_{2,k}(t) + \lambda_k \int dt' R_{1}(t,t') h_{0,k}(t')  \ , \ u_2(t) \sim \mathcal{GP}\left(0, \frac{1}{P} \lambda_k C_1(t,t') \right) \nonumber
    \\
    &h_3(t) = u_3(t) + \frac{1}{N_1} \int dt' R_{2,4}(t,t') h_3(t')  \ , \ u_3(t) \sim \mathcal{GP}\left(0, C_2(t,t') \right) \nonumber
    \\
    &h_{4,k}(t) = u_{4,k}(t) + \int dt' R_{3}(t,t') h_{2,k}(t')  \ , \ u_4(t) \sim \mathcal{GP}\left(0, \frac{1}{N_1} C_3(t,t') \right) \nonumber
    \\
    &\frac{d}{dt} h_{0,k}(t) = - h_{4,k}(t)
\end{align}
where the correlation and response functions are defined as 
\begin{align}
    &C_0(t,t') = \sum_{k} \lambda_k \left< h_{0,k}(t) h_{0,k}(t')\right> \ , \ C_2(t,t') =  \sum_{k} \left< h_{2,k}(t) h_{2,k}(t')\right>
    \ , \ C_\ell(t,t') = \left< h_\ell(t) h_\ell(t') \right> \ \ell \in \{1,3\}\nonumber
    \\ 
    &R_{0,2}(t,t') = \sum_{k} \lambda_k \left< \frac{\partial h_{0,k}(t)}{\partial u_{2,k}(t')} \right> \ , \ R_{2,4}(t,t') = \sum_{k} \left< \frac{\partial h_{2,k}(t)}{\partial u_{4,k}(t')} \right> \ , \ R_{\ell}(t,t') = \left< \frac{\partial h_\ell(t)}{\partial u_\ell(t')} \right> \ \ell \in \{ 1,3 \} .
\end{align}
We note that $h_1(t), h_3(t)$ variables do not carry an index as each of the variables are identical and exchangeable. However the variables $h_{0,k}, h_{2,k}, h_{4,k}$ obey dynamics that are distinct across different population eigenvalues $\lambda_k$. The test and train losses are simply
\begin{align}
    \mathcal L(t) = C_0(t,t) \ , \ \mathcal{\hat L}(t) = C_1(t,t) .  
\end{align}
To gain additional insight into these equations, we wil first analyze them for isotropic covariates where the response functions can be written explicitly.

\subsubsection{Isotropic Features} 
For isotropic features where the covariance of the $N_0$ dimensional features is $\bm\Lambda = \frac{1}{N_0} \bm I \in \mathbb{R}^{N_0 \times N_0}$, the DMFT is exact under a proportional scaling
\begin{align}
    N_0, N_1, P \to \infty \ , \   \frac{N_1}{N_0} = \nu  \ , \ \frac{P}{N_0} = \alpha .
\end{align} 
By the TTI property of linear systems, the response functions can be solved directly after Fourier transformation. Further, by the symmetry of this covariance eigenvalues $\lambda_k$, all of the $h_{0,k}, h_{2,k}, h_{4,k}$ variables are identically distributed across $k$. The self-response for $h_0(\omega)$ which we again denote as $\mH(\omega) \equiv - \frac{\partial h_0(\omega)}{\partial u_4(\omega)}$ satisfies the following cubic equation
\begin{align}
    &\mH(\omega) = \frac{1}{i\omega + (1 - \alpha^{-1} + i\omega \alpha^{-1}  \mH(\omega) )  (1 - \nu^{-1} + i\omega \nu^{-1} \mH(\omega) ) } . \nonumber
\end{align}
From this solution, the response functions for the $h_1$ and $h_3$ variables can be directly computed
\begin{align}
    \mR_1(\omega) = 1 - \alpha^{-1} + \alpha^{-1}  i\omega  \mH(\omega) \ , \ \mR_3(\omega) = 1 - \nu^{-1} + \nu^{-1} i\omega \mH(\omega). 
\end{align}
Similarly, the Fourier transformed correlation functions satisfy
\begin{align}
    &\mC_0(\omega,\omega') = \frac{1}{1-\Gamma(\omega,\omega')} \  \mH(\omega) \mH(\omega')  \nonumber
    \\
    &\Gamma(\omega,\omega') = (1-i\omega\mH(\omega)) (1 - i\omega'\mH(\omega')) \left[ \frac{1}{\nu} + \frac{1}{\alpha} - \frac{1}{\nu\alpha} +  \frac{1}{\nu\alpha} ( i\omega \ \mH(\omega) + i\omega' \ \mH(\omega') ) \right]
\end{align}
The function $\Gamma(\omega,\omega')$ captures the multiplicative noise induced by the random processes $\{u_2, u_4\}$. Under ensembling matrices $\bm A$ and bagging over datasets $\bm \Psi$ the factor $\frac{1}{1-\Gamma(\omega,\omega')}$ will disappear.  The train loss $\hat{\mathcal L}(t)$ can be accessed from the correlation function $\mC_1(\omega,\omega') = \mR_1(\omega) \mR_1(\omega') \mC_0(\omega,\omega')$.

\begin{figure}[h]
    \centering
    \subfigure[Test Losses $\alpha=0.75$]{\includegraphics[width=0.32\linewidth]{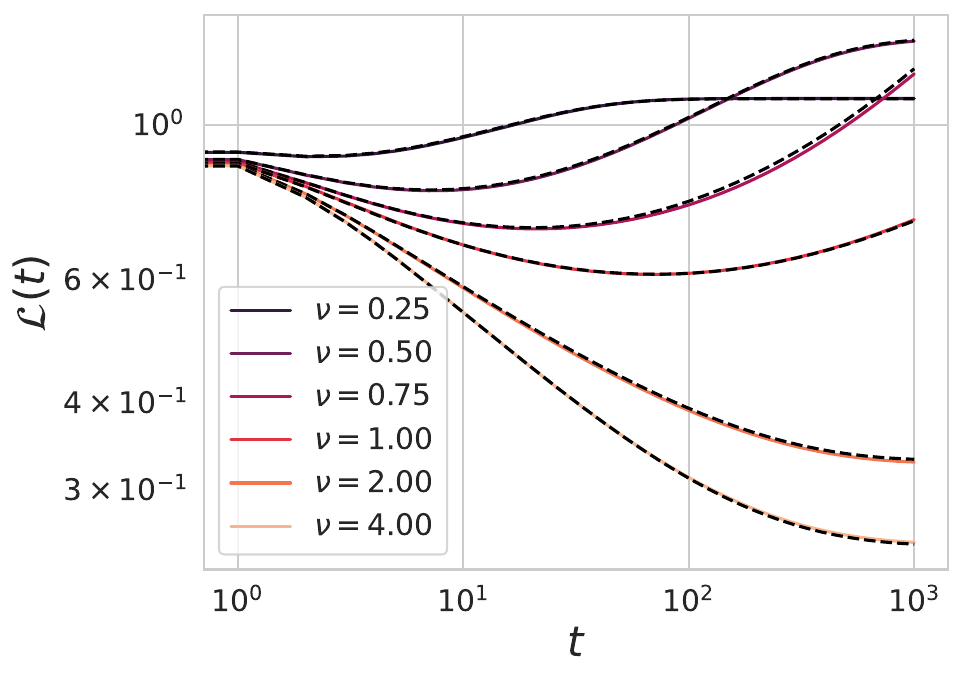}}
    \subfigure[Train Losses]{\includegraphics[width=0.32\linewidth]{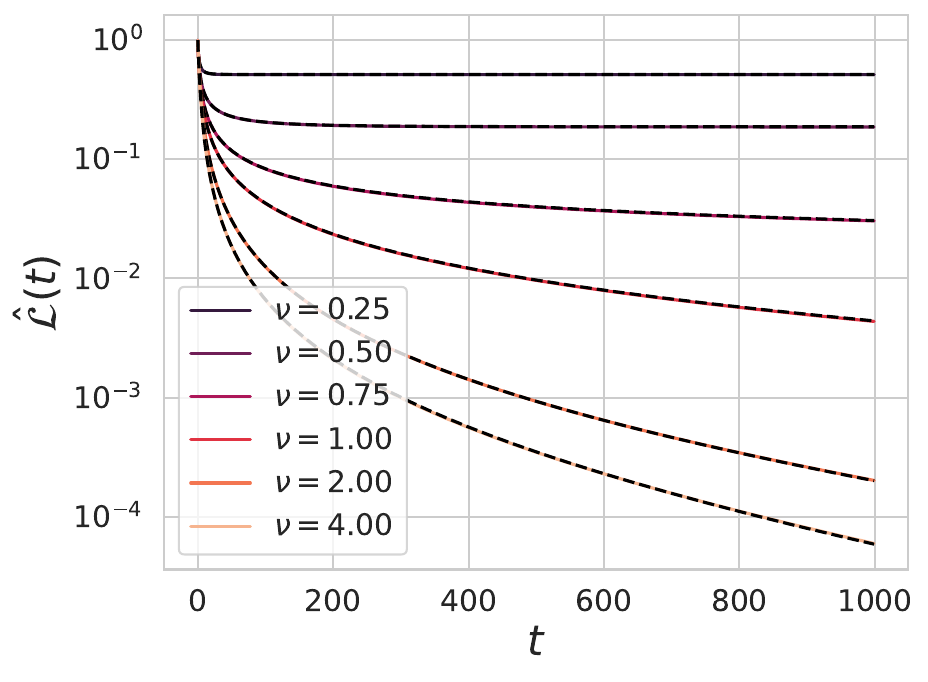}}
    \subfigure[Spectral Density $\alpha = 4$]{ \includegraphics[width=0.32\linewidth]{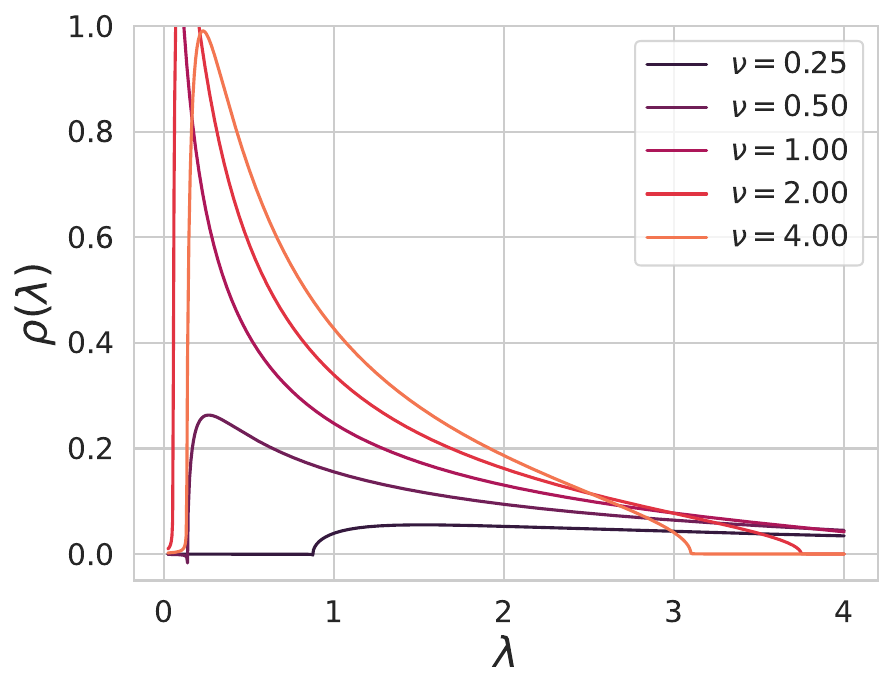} }
    \subfigure[Final Losses]{\includegraphics[width=0.32\linewidth]{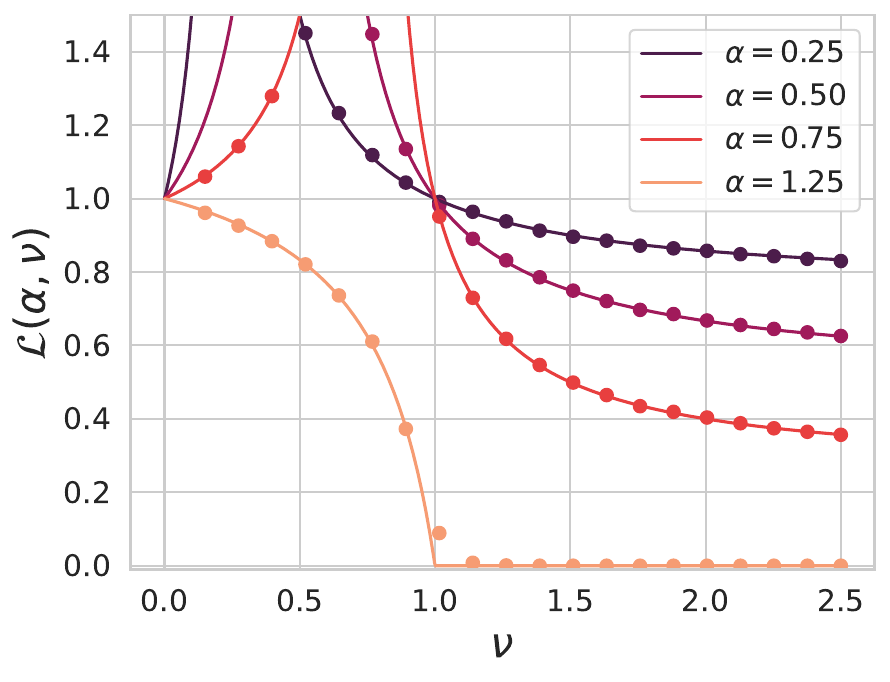}}
    \subfigure[Final Bias]{\includegraphics[width=0.32\linewidth]{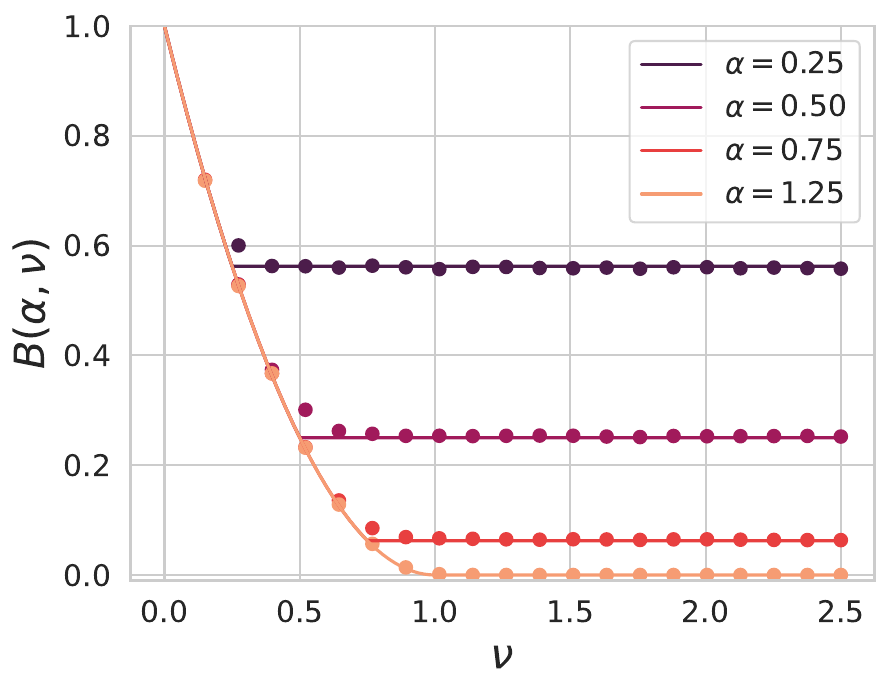}}
    \subfigure[Blowup at $\alpha=\nu$]{\includegraphics[width=0.32\linewidth]{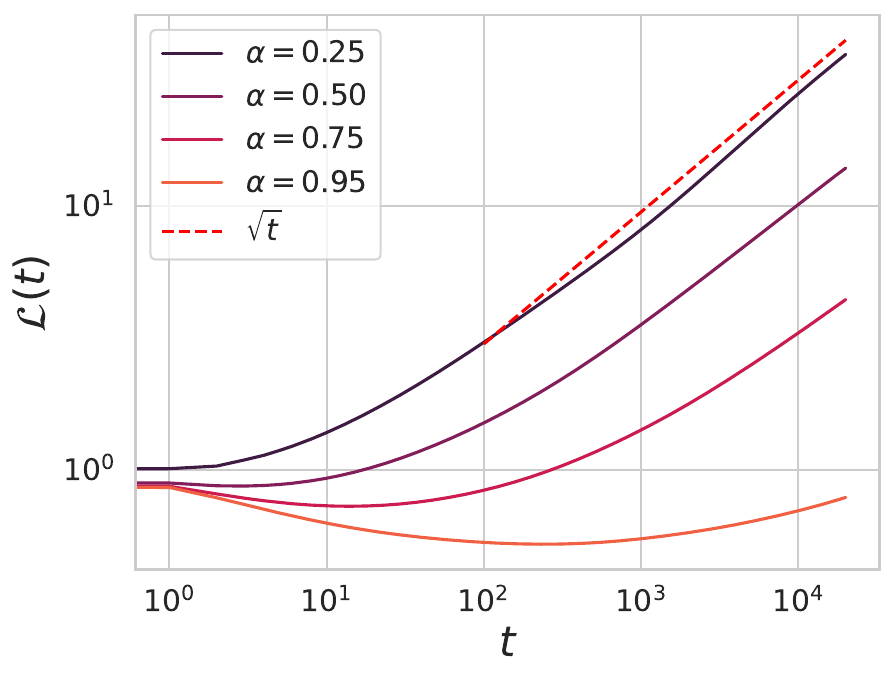}}
    \caption{Dynamics for isotropic random features fixed $\alpha$ and varying $\nu$ without any explicit label noise $\sigma^2=0$. (a) The test loss is non-monotonic for $\nu \leq 1$ due to misalignment between $\h_4(t)$ and $\h_0(t)$. (b) The train loss is monotonically decreasing over time and reaches zero for $\nu > \alpha$. (c) The spectrum is non-negative and can be obtained from $\mH(\omega)$. (d) Final loss over varying $\nu, \alpha$ and (e) final biases over $\nu,\alpha$. (f) Dynamics at the interpolation threshold $\nu=\alpha$ across varying $\alpha$. The loss exhibits a $\sqrt t$ blowup even in the absence of label noise due to the non-normal blowup. }
    \label{fig:isotropic_dynamics}
\end{figure}

\paragraph{Final Loss} The singular low frequency component satisfies $R_1(\omega) \mR_3(\omega) \sim i \omega r$ as $\omega \to 0$. Assuming that $\min\{ \alpha ,\nu \} < 1$, we have
\begin{align}
    &r = \frac{\min\{ \alpha, \nu \}}{1- \min\{ \alpha, \nu \}} 
    \ , \ \lim_{\omega \to 0} (i \omega) \mH(\omega) = \left[1- \min\{\alpha,\nu\} \right]_+
\end{align}
Similarly, we can access the final test and train losses
\begin{align}
    \lim_{t \to \infty} \mathcal L(t) = \lim_{t \to \infty} C_0(t,t) = \lim_{\omega,\omega' \to 0} (i\omega)(i\omega') \mC_0(\omega,\omega') = \begin{cases} \frac{1-\nu}{1-\nu/\alpha} & \nu < \alpha \ \& \  \min\{ \alpha , \nu\}  < 1
    \\
    \frac{1-\alpha}{1-\alpha/\nu} & \alpha < \nu  \ \& \  \min\{ \alpha , \nu\} < 1  
    \\
    0 & \min\{ \alpha , \nu\}  > 1 
    \end{cases} 
\end{align}
We see that this model displays \textit{double descent behavior} where the test loss diverges at $\alpha = \nu$ if $\min\{\alpha, \nu \} < 1$ as we illustrate in Figure \ref{fig:isotropic_dynamics} (d). Next, we investigate the behavior of the blowup at the interpolation threshold where $\alpha = \nu$.  

\subsubsection{Blowup Rate at the Interpolation Threshold}

When $\alpha = \nu$ there is an interesting asymptotic behavior as $t \to \infty$ which we can access by examining the low-frequency structure of $\mC(\omega,\omega')$. To access the long time behavior in this case, we have to expand $\mH$ to higher order in $i\omega$
\begin{align}
    i\omega \mH(\omega) \sim 1-\alpha + \sqrt{ \frac{\alpha^3}{1-\alpha}  (i\omega) } \ , \ i\omega \to 0
\end{align}
At low frequencies, the correlation function thus behaves like
\begin{align}
    \mC(\omega,\omega' ) 
    &\sim \sqrt{\frac{1-\alpha}{\alpha^3}} \frac{(1-\alpha)^2}{(\sqrt{i\omega} + \sqrt{i\omega'}) i\omega i\omega'} \ , \ \omega , \omega' \to \infty .
\end{align}
While the bias $\mathcal B(\omega,\omega') \equiv \frac{1}{N} \left< \h(\omega) \right> \cdot \left< \h(\omega') \right>= \mH(\omega) \mH(\omega') \propto \frac{1}{i\omega i\omega'}$ has simple poles at $\omega,\omega' = 0$, we see that the variance induces additional inverse square-root terms. Taking an inverse Fourier transform at large $t$ using a steepest descent approximation, we arrive at a $\sqrt{t}$ blowup for large $t$
\begin{align}
    C(t,t) \propto \int  \frac{d\omega d\omega'}{(2\pi)^2} \frac{\exp\left( i (\omega + \omega') t \right) }{(i\omega) (i\omega') \left[ \sqrt{i\omega} + \sqrt{i\omega'} \right] } \sim  \sqrt{t} 
\end{align}
This prediction matches experiments as we show in Figure \ref{fig:isotropic_dynamics} (f).


\subsubsection{Comparison of Ensembling and Bagging}

Similar to the analysis provided in section \ref{sec:bias_var_linear}, we can analyze the dynamics of the loss for the \textit{averaged predictor} over a model ensemble of size $E$ (meaning $E$ indepedent copies of $\bm A$) and a bagging over $B$ datasets ($B$ independently sampled data matrices $\bm\Psi$). We let $e \in \{1,...,E\}$ represent each ensemble member and $b \in \{1,...,B\}$ represent each dataset
\begin{align}
    \frac{d}{dt} \h_{e,b}(t) = - \left( \frac{1}{N_1} \bm A_e^\top \bm A_e \right) \left( \frac{1}{P} \bm\Psi_b^\top \bm \Psi_b   \right) \h_{e,b}(t)
\end{align}
and specifically are interested in the correlation and response functions for
\begin{align}
    \bar{\h}(t) = \frac{1}{E B} \sum_{e=1}^E \sum_{b=1}^B \h_{e,b}(t)
\end{align}
The exact DMFT equations can be averaged over different instances of $(e,b)$. The response functions are unchanged by this averaging operation but the correlation functions are altered. The key fact is that $u_{4,e,b}(t)$ are uncorrelated across separate ensemble members $(e,e')$ and $u_{2,e,b}(t)$ are uncorrelated across separate datasets $(b,b')$
\begin{align}
    \mC(\omega,\omega') = \frac{1}{1-\underbrace{\Gamma(\omega,\omega',E,B)}_{\text{Variance Reduction}}} \  \underbrace{\mH(\omega) \mH(\omega')}_{\text{Original Bias}} .
\end{align}
As $E,B \to \infty$, we have that $\lim_{E,B \to \infty} \Gamma(\omega,\omega', E, B) = 0$. Thus, the bias is indeed controlled only by $\mH(\omega)$ and is unaffected by bagging or ensembling, while the quantity $\Gamma$ which controls the variance is reduced by ensembling and bagging. 

\subsection{Online Stochastic Gradient Descent on Structured Random Features}

\begin{figure}
    \centering
    \subfigure[SGD varying Model size $N$]{\includegraphics[width=0.4\linewidth]{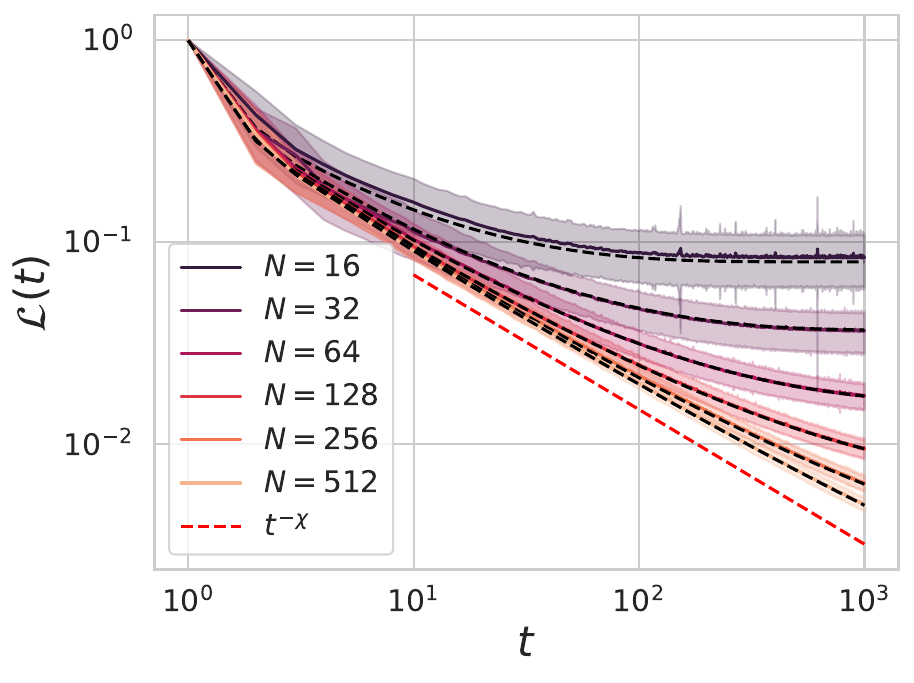}}
    \subfigure[SGD varying Batch size $B$, $N=64$]{\includegraphics[width=0.4\linewidth]{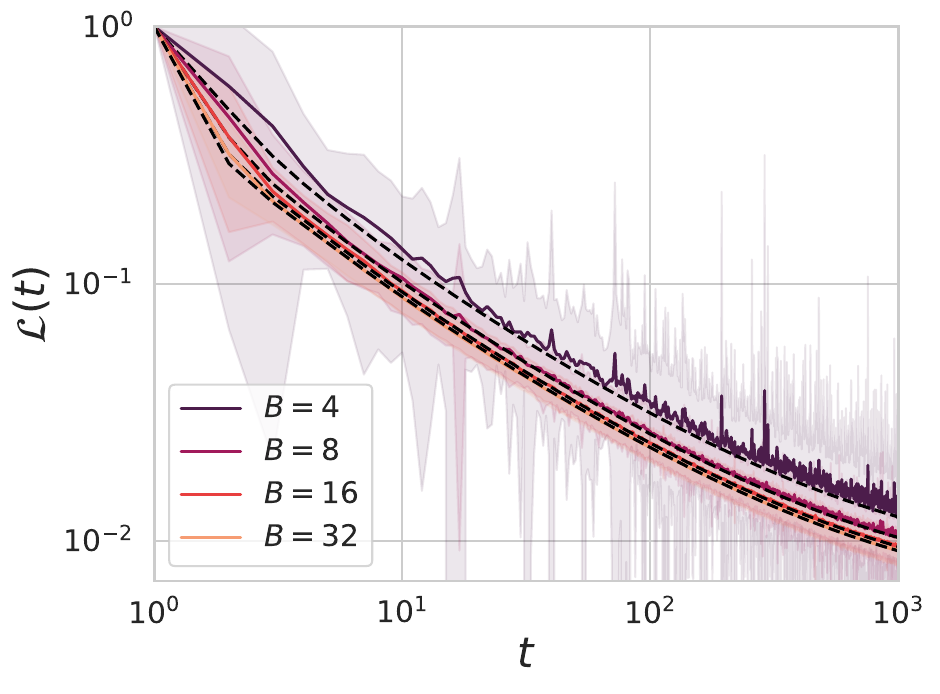}}
    \caption{Dynamics for online SGD under powerlaw random features. Dashed black lines are the theoretical predicitons, while colored errorbars represent standard deviations. (a) While the $N\to\infty$ limit of the loss obeys a powerlaw $\mathcal L \sim t^{-\chi}$, finite $N$ effects can cause the model to converge to a limiting loss value that scales as $N^{-\nu \chi}$. (b) Reducing the batch size $B$ leads to amplified variance in the correlation functions, while the bias dynamics (the dynamics of $\left< \h(t) \right>$) are independent of $B$ and depend on $N$ through the response function $R_3$.  }
    \label{fig:powerlaw_random_features_SGD_dynamics}
\end{figure}

The same type of technology can also be used to analyze stochastic gradient descent in discrete time, which evolves the error vector $\h(t)$ with the update rule
\begin{align}
    \bm h(t+1) = \bm h(t) - \eta \left( \frac{1}{N_1} \bm A^\top \bm A \right) \left( \frac{1}{B} \bm\Psi(t)^\top \bm \Psi(t) \right) \h(t) 
\end{align}
where the matrix $\bm A$ is fixed across iterations but the matrices $\bm\Psi(t) \in \mathbb{R}^{B \times N_0}$ are independently sampled data matrices at each step. The DMFT equations can be expressed as a simple set of decoupled stochastic linear equations
\begin{align}
    h_k(t) = w^\star_k - \eta \sum_{t'< t} \left[ u_k^4(t') -  \sum_{t''<t'} R_3(t',t'') u^2_k(t'') - \lambda_k \sum_{t''<t'} R_{3}(t',t'') h_k(t'')  \right] ,
\end{align}
where the noise processes $u^4_k(t)$ and $u^2_k(t)$ have the following correlation structure
\begin{align}
    \left< u_k^2(t) u_k^2(t') \right> = \underbrace{\delta_{t,t'} \ \frac{1}{B}  \lambda_k C_0(t,t) }_{\text{Uncorrelated SGD Noise}} \ , \  \left< u^4_k(t) u^4_k(t') \right> = \underbrace{\frac{1}{N_1} C_3(t,t')}_{\text{Correlated Noise from $\bm A$ }}  .
\end{align}
From the above result, we identify the key differences between online SGD and the finite dataset gradient flow regime.
\begin{enumerate}
    \item For online SGD, there is no response function generated from the random stream of data (ie no $R_1(t,t')$ arises in the dynamics). 
    \item For online SGD, the variance from limited batch size leads to a noise process $u^2_k(t)$ that is decorrelated across steps. For the case where the dataset is repeated across steps of training, there is a limiting loss set by $P$. However, taking either batchsize $B \to \infty$ in the online case or dataset $P \to \infty$ in the offline GD case recovers gradient descent on the population loss.
\end{enumerate}
We can introduce a matrix/vector notation for our sums over times $t$ up to some arbitrary cutoff time $T$ by letting $\bm H_k \in \mathbb{R}^{T \times T}$ and $\bm R_3 \in \mathbb{R}^{T \times T}$ and introduce the integrator matrix $\bm\Theta \in\mathbb{R}^{T \times T}$ with $\Theta_{t,t'} = \eta \Theta(t-t')$ where $\Theta(z)$ is the heaviside step function (indicator function for $t > t'$). Using this formalism, our response functions satisfy
\begin{align}
    &\bm R_3 = \bm I - \frac{1}{N_1} \sum_k \lambda_k \bm H_k \bm\Theta \bm R_3 = \left( \bm I + \frac{1}{N_1} \sum_k \lambda_k \bm H_k \bm\Theta  \right)^{-1}  \nonumber
    \\
    &\bm H_k = \bm I - \lambda_k \bm\Theta \bm R_3 \bm H_k = \left[  \bm I + \lambda_k \bm\Theta \left( \bm I + \frac{1}{N_1} \sum_\ell \lambda_\ell \bm H_\ell
    \bm\Theta \right)^{-1} \right]^{-1}
\end{align}
After solving for the response functions $\bm H_k$, we can introduce a matrix notation for the correlation matrix $\bm C_0 \in \mathbb{R}^{T \times T}$
\begin{align}
    \bm C_0 &= \sum_{k} \lambda_k \bm H_k \left[ (w^\star_k)^2 \bm 1 \bm 1^\top + \frac{1}{N_1} \bm\Theta \bm R_3 \bm C_2 \bm R_3^\top \bm\Theta^\top +  \frac{1}{B} \lambda_k  \bm\Theta \bm R_3  \ \text{diag}(\bm C_0) \  \bm R_3^\top \bm\Theta^\top  \right] \bm H_k^\top \nonumber
    \\
    \bm C_2 &= \sum_k \lambda_k^2  \ \bm H_k\left[ (w^\star_k)^2 \bm 1 \bm 1^\top + \frac{1}{N_1} \bm\Theta \bm R_3 \bm C_2 \bm R_3^\top \bm\Theta^\top \right] \bm H_k^\top \nonumber
    \\
    &+ \frac{1}{B}  \sum_k \lambda_k \left[ \bm I - \lambda_k \bm H_k \bm\Theta \bm R_3 \right] \text{diag}(\bm C_0) \left[ \bm I - \lambda_k \bm H_k \bm\Theta \bm R_3 \right]^\top
\end{align}
We plot this discrete time solution against SGD simulations in Figure \ref{fig:powerlaw_random_features_SGD_dynamics}. Additional simulations of this model can be found in \cite{bordelon2024dynamical}. A continuous time approximation of the above dynamics \cite{paquette2021sgd, paquette20244+, atanasov2025two, mignacco2022effective, mignacco2020dynamical}. 

\section{Symmetric and Asymmetric Free Products}

In this section, we move beyond Gaussian random matrices and consider general free products \cite{potters2020first, atanasov2024scaling, atanasov2025two, bordelon2025theory}. Consider two $N \times N$ matrices $\bm A$ and $\bm B$ and define the two types of free products
\begin{align}
    \bm M = \begin{cases}
        \bm O \bm B \bm O^\top \bm A & \text{Asymmetric Free Product}
        \\
        \bm A^{1/2} \bm O \bm B \bm O^\top \bm A^{1/2} & \text{Symmetrized Free Product}
    \end{cases} .
\end{align}
where $\bm O$ is a randomly sampled orthogonal matrix $N \times N$ matrix, drawn from the Haar measure. We would like to obtain correlation and response functions for the dynamical system 
\begin{align}
    \frac{d}{dt} \bm h(t) = - \bm M \bm h(t) + \delta(t) \h_0 ,
\end{align}
and compare the dynamics of the asymmetric and symmetrized free product. We are interested in the limiting dynamics in an appropriate $N \to \infty$ limit under assumed knowledge of the spectra of $\bm A, \bm B$. We will see that due to the fact that symmetric and asymmetric versions of the free-product $\bm M$ have identical spectra, the response function of the above dynamical system will be identical in either case. However, the correlation function dynamics will be distinct. To characterize this with DMFT, we write down the path integral for this system (defining $ i \hat{\h}(t) \equiv \bm\chi(t)$)
\begin{align}
    Z = \int \mathcal D \bm h \mathcal D \bm \chi \left< \exp\left( - \int dt \ \bm\chi(t) \cdot \left( \partial_t \bm h(t) - \bm M \h(t) \right) \right) \right>_{\bm O} = 1 .
\end{align}
As we outline in Appendix \ref{app:free_product}, after integrating out the dependence of the random orthogonal matrix $\bm O$ for large $N$, we can express $Z$ as an integral over a pair of $2 \times 2$ matrix valued functions $\bm\Sigma(\omega,\omega') \in \mathbb{R}^{2\times 2}$ and $\bm\Psi(\omega,\omega') \in \mathbb{R}^{2 \times 2}$
\begin{align}
    Z = \int \mathcal D \bm\Sigma \  \mathcal D \bm\Psi \  \exp\left(-\frac{N}{2} \mathcal S[\bm \Sigma,\bm\Psi] \right) , 
\end{align}
where the DMFT action $\mathcal S$ takes the form
\begin{align}
    \mathcal S[\bm \Sigma,\bm\Psi]  &= - \textbf{Tr} \ \bm\Psi \bm\Sigma - \frac{2}{N}\ln\mathcal Z_A(\bm\Psi) - \textbf{Tr} \ \hat{\bm\Sigma}_\star \bm\Sigma - \frac{1}{N} \textbf{Tr} \  \log\left( \hat{\bm\Sigma}_\star \otimes \bm I + \bm P \otimes \bm B \right) + \textbf{Tr}  \ \log\bm\Sigma , \nonumber
\end{align}
where we introduced the super-trace notation $\textbf{{Tr}} \ \bm\Sigma \bm\Psi = \int d\omega d\omega' \ \text{Tr} \bm\Sigma(\omega,\omega') \bm\Psi(\omega,\omega')$ represents both trace over the dimension of the matrix and integration over the frequency and the matrix $\bm P = \begin{bmatrix}
     0 & 1
     \\
     1 & 0
\end{bmatrix}$. The single site MGF $\mathcal Z_{A}(\bm\Psi)$ has the form
\begin{align}
    \mathcal Z_A(\bm\Psi) &= \int \exp\left(  -\frac{1}{2} \int d\omega d\omega' \left[\Psi_{\chi\chi}(\omega,\omega') \bm\chi(\omega) \cdot \bm\chi(\omega') + \Psi_{hh}(\omega,\omega') \h(\omega)^\top \bm A^2 \h(\omega')  \right] \right) \nonumber
    \\
    &\times \exp\left( - \int d\omega d\omega' \ \bm\chi(\omega) \cdot \left(i\omega \delta(\omega-\omega') \bm I +\Psi_{h\chi}(\omega,\omega') \bm A \right)  \h(\omega') \right)
\end{align} 
The order parameter $\hat{\bm\Sigma}_\star$ is an implicit function of $\bm\Sigma$ that satisfies the equation
\begin{align}
    \bm\Sigma = \frac{1}{N} \textbf{Tr} \left( \hat{\bm\Sigma}_\star \otimes \bm I + \bm P \otimes \bm B \right)^{-1}  .
\end{align}
As $N \to \infty$, the dominant contribution from the integral over $\{\bm\Sigma,\bm\Psi\}$ is the saddle point
\begin{align}
    \frac{\partial \mathcal S}{\partial \bm\Sigma} = 0 \ , \ \frac{\partial \mathcal S}{\partial \bm\Psi} = 0
\end{align}
At the saddle point (which dominates as $N \to \infty$), the structure of the $\bm\Psi$ and $\bm\Sigma$ 
\begin{align}
    \bm\Sigma(\omega,\omega') = 
    \begin{bmatrix}
        \Sigma_{hh}(\omega,\omega') & \Sigma_{h\chi}(\omega,\omega')
        \\
        \Sigma_{h\chi}(\omega,\omega') & 0 
    \end{bmatrix} \ , \ \bm\Psi(\omega,\omega') = 
    \begin{bmatrix}
        0 & \Psi_{h\chi}(\omega,\omega')
        \\
        \Psi_{\chi h}(\omega,\omega') & \Psi_{\chi\chi}(\omega,\omega')
    \end{bmatrix}
\end{align}
We will see that the diagonal and off-diagonal components of $\bm\Sigma(\omega,\omega')$ and $\bm\Psi(\omega,\omega')$ provide the correlation and response functions respectively.  

\subsection{Response Functions}

Under the saddle point equations, the off-diagonal entries of the order parameters decouple over frequencies and encode the response function
\begin{align}
    \Sigma_{h\chi}(\omega,\omega') = \Sigma_{\chi h}(\omega,\omega') = \delta(\omega-\omega')  \ \mathcal H(\omega) ,
\end{align}
where $\mH(\omega)$ is the usual single-frequency response function. Similarly, we have $\Psi_{h\chi}(\omega,\omega') = \Psi(\omega)\delta(\omega-\omega')$ and $\hat{\Sigma}_{h\chi}(\omega,\omega') = \delta(\omega-\omega') \hat\Sigma(\omega)$ where these single frequency functions satisfy the following 
\begin{align}
    &\Psi(\omega) = \mH(\omega)^{-1} - \hat{\Sigma}(\omega) \ , \ \mH(\omega) = \text{tr}\left( \hat\Sigma(\omega) + \bm B \right)^{-1}
    \\
    &\Psi(\omega) \mH(\omega) =  \text{tr}\bm A\left(  i\omega \Psi(\omega)^{-1} + \bm A \right)^{-1} = \text{tr} \bm B \left( \hat{\Sigma}(\omega)+ \bm B \right)^{-1} .
\end{align}
We see that it would be convenient to define the $\mathcal T_{\bm A}(\omega)$ transform of a matrix $\bm A$ and its inverse function $i\omega_{\bm A}(\mathcal T)$
\begin{align}
    \mathcal T_{\bm A}(i \omega) \equiv \text{tr} \bm A \left( i\omega  + \bm A \right)^{-1} \ , \ i\omega_{\bm A}(\mathcal T) = \mathcal T^{-1}_{\bm A}(\mathcal T)
\end{align}
Using this definition, we find the simple relationship between the matrices $\bm A, \bm B, \bm M$
\begin{align}
    &\mathcal T_{\bm M}(i\omega) = \text{tr}\bm A\left( i\omega/\Psi(\omega) + \bm A \right)^{-1} = \mathcal{T}_{\bm A}( i\omega_{\bm A}) = \mathcal T_{\bm B}( i \omega_{\bm B}) 
    \\
    &i\omega_{\bm A} = i\omega / \Psi(\omega) \ , \ i \omega_{\bm B} = \hat{\Sigma}(\omega)
\end{align}
We can reformulate the subordination relation as an equivalence in the $\mathcal T$ transforms computed across these matrices
\begin{align}
    \mathcal T_{\bm A}(i \omega_{\bm A}) =  \mathcal T_{\bm B}(i \omega_{\bm B})  = \mathcal T_{\bm M}(i\omega) = \mathcal T
\end{align}
where from the saddle point equations, the variables $\{i\omega_{\bm A}, i\omega_{\bm B}, i\omega, \mathcal T\}$ satisfy
\begin{align}
    i\omega_{A }(\mathcal T) \  i\omega_B(\mathcal T) =  \frac{1 - \mathcal T}{\mathcal T} \  i \omega . 
\end{align}
This determines $\mathcal T(i\omega)$ from which we can infer the response function function $\mathcal H(\omega) = \frac{1}{i\omega} \left[ 1 - \mathcal T(i\omega) \right]$. The eigenvalue density can, as before, be obtained from $\rho(\lambda) = \frac{1}{\pi} \lim_{\epsilon \to 0}  \Im \ \mH(i\lambda - \epsilon)$. 

\subsection{Correlation Functions}
While the response functions were identical across asymmetric and symmetrized cases, we now turn to the correlation functions $\Sigma_{hh}(\omega,\omega')$ which are distinct in these two cases. We first describe the asymmetric case before 

\subsubsection{Asymmetric Case} 

In the asymmetric case, where $\bm M = \bm O \bm B \bm O^\top \bm A$, we can state the main result for the correlation function in terms of a deterministic equivalent\footnote{The symbol $\bm V \simeq \bm V'$ for deterministic equivalent indicates asymptotic equivalence of traces against test matrices $\lim_{N \to \infty} \frac{\text{tr} \bm V \bm D}{\text{tr} \bm V' \bm D} = 1$ where $\bm D$ is an arbitrary test matrix.} 
\begin{align}
    \bm h(\omega) \bm h(\omega')^\top \simeq \left( i\omega + \Psi(\omega) \bm A \right)^{-1} \left[ \h_0 \h_0^\top - \Psi_{\chi\chi}(\omega,\omega') \bm I \right] \left( i\omega' + \Psi(\omega') \bm A \right)^{-1}
\end{align}
The function $\Psi_{\chi\chi}(\omega,\omega')$ is determined by its saddle point equation 
\begin{align}
    \Psi_{\chi\chi}(\omega,\omega') = - \Sigma_{hh}(\omega,\omega') \left[ \Sigma(\omega)^{-1} \Sigma(\omega')^{-1} - \left( \text{tr}(\hat\Sigma(\omega) + \bm B )^{-1} (\hat\Sigma(\omega') + \bm B)^{-1} \right)^{-1} \right] ,
\end{align}
and $\Sigma_{hh} \equiv \frac{1}{N} \left< \h(\omega)^\top \bm A^2 \h(\omega') \right> = \text{tr} \bm A^2 \left<\h(\omega') \h(\omega)^\top \right>$ satisfies
\begin{align}
    \Sigma_{hh}(\omega,\omega') =& \frac{1}{1-\Gamma(\omega,\omega')} \  \text{tr} \bm A^2 \left( i\omega + \Psi(\omega) \bm A \right)^{-1} \h_0 \h_0^\top \left( i\omega' + \Psi(\omega') \bm A \right)^{-1} \nonumber
    \\
    \Gamma(\omega,\omega') =& \left[ \Sigma(\omega)^{-1} \Sigma(\omega')^{-1} - \left( \text{tr}(\hat\Sigma(\omega) + \bm B )^{-1} (\hat\Sigma(\omega') + \bm B)^{-1} \right)^{-1} \right]  \nonumber
    \\
    &\times \text{tr} \bm A^2 \left( i\omega + \Psi(\omega) \bm A \right)^{-1} \left( i\omega' + \Psi(\omega') \bm A \right)^{-1} .
\end{align}
From the deterministic equivalent expression, one can compute traces $\text{tr} \bm D \h(\omega) \h(\omega')^\top$ against arbitrary test matrices $\bm D$. 

\subsubsection{Symmetrized Case} 

In the symmetrized case where $\bm M = \bm A^{1/2} \bm O \bm B \bm O^\top \bm A^{1/2}$, the outer product $\h(\omega)\h(\omega')^\top$ has the following deterministic equivalent
\begin{align}
    \h(\omega)\h(\omega')^\top \simeq \left(i \omega + \Psi_{h\chi}(\omega) \bm A \right)^{-1} \left[ \h_0 \h_0^\top - \Psi_{\chi\chi}(\omega,\omega') \bm A \right] \left(i \omega' + \Psi_{h\chi}(\omega') \bm A \right)^{-1} , 
\end{align}
which contains an additional factor of $\bm A$ in the variance term. As before, the function $\bm\Psi_{\chi\chi}(\omega,\omega')$ is determined by its saddle point equation
\begin{align}
    \Psi_{\chi\chi}(\omega,\omega') &= - \Sigma_{hh}(\omega,\omega') \times  \left[ \Sigma(\omega)^{-1} \Sigma(\omega')^{-1} - \left( \text{tr}(\hat\Sigma(\omega) + \bm B )^{-1} (\hat\Sigma(\omega') + \bm B)^{-1} \right)^{-1} \right] .
\end{align}
However, for this symmetrized ensemble, the definition for $\Sigma_{hh}(\omega,\omega') \equiv \frac{1}{N} \left< \h(\omega)^\top  \bm A \h(\omega') \right>$ is different than the asymmetric case (by a factor of $\bm A$) and has the form 
\begin{align}
    \Sigma_{hh}(\omega,\omega') &= \frac{1}{1-\Gamma(\omega,\omega')} \text{tr} \bm A \left(i \omega + \Psi_{h\chi}(\omega) \bm A \right)^{-1} \h_0 \h_0^\top \left(i \omega' + \Psi_{h\chi}(\omega') \bm A \right)^{-1} 
    \\
    \Gamma(\omega,\omega') =& \left[ \Sigma(\omega)^{-1} \Sigma(\omega')^{-1} - \left( \text{tr}(\hat\Sigma(\omega) + \bm B )^{-1} (\hat\Sigma(\omega') + \bm B)^{-1} \right)^{-1} \right]  \nonumber
    \\
    &\times \text{tr} \bm A^2 \left( i\omega + \Psi(\omega) \bm A \right)^{-1} \left( i\omega' + \Psi(\omega') \bm A \right)^{-1}
\end{align}
We thus see that the symmetrized case has a different formula for $\Sigma_{hh}(\omega,\omega')$ and $\Psi_{\chi\chi}(\omega,\omega')$, resulting in different variance in the dynamics.

\subsection{Free Product of Projections} 
\begin{figure}
    \centering
    \subfigure[Varying Projection Ratio $\alpha$ with $\beta=0.5$]{\includegraphics[width=0.45\linewidth]{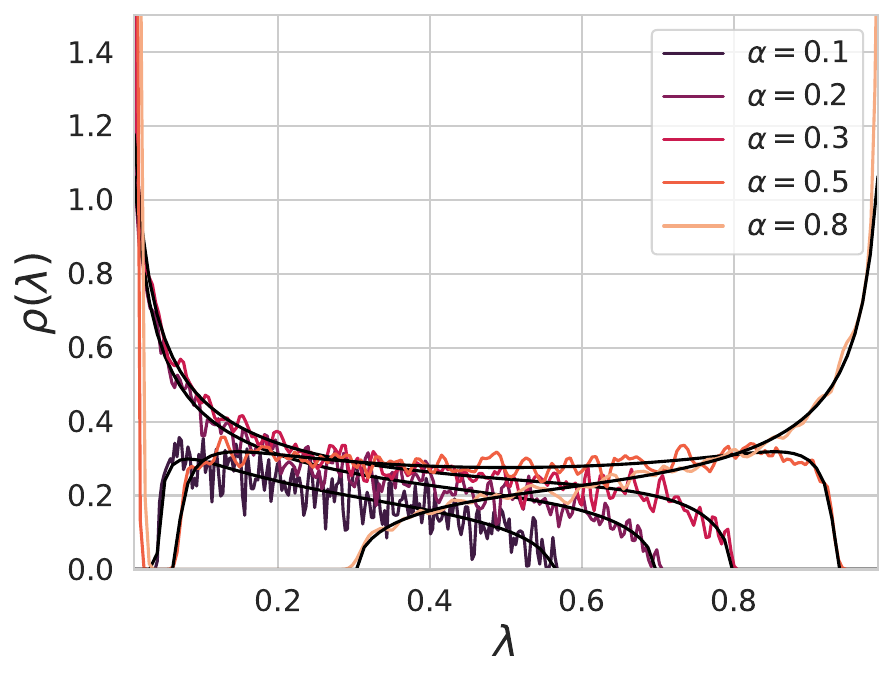}}
    \subfigure[Response Functions $\beta=0.8$]{\includegraphics[width=0.45\linewidth]{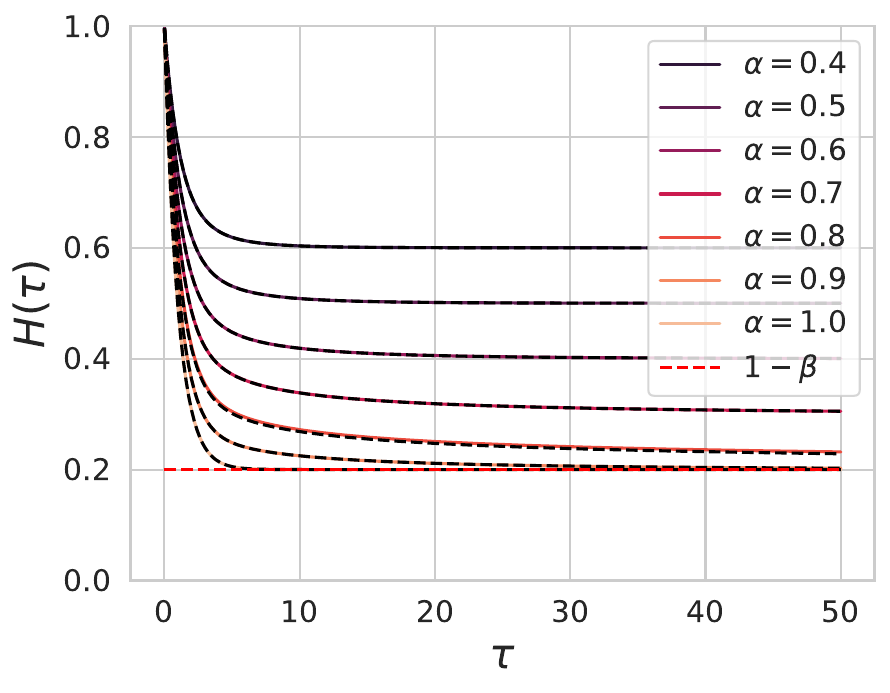}}
    \caption{One-point functions and spectral densities for a free product of orthogonal projections. (a) The eigenvalue density $\rho(\lambda)$ for the free product $\bm M$ for varying $\alpha$ at fixed $\beta$. (b) The response function $H(\tau)$ visualized across varying $\alpha$ relaxes monotonically to $\lim_{\tau \to \infty} H(\tau) = 1-\min(\alpha,\beta)$, which measures the dimension of the nullspace of $\bm M$.}
    \label{fig:one_point_prod_projections}
\end{figure}

In this section, we consider the case where $\bm A$ and $\bm B$ are $N\times N$ matrices with rank $\alpha N$ and $\beta N$ respectively. Thus the spectral measures for $\bm A$ and $\bm B$ are 
\begin{align}
    &\rho_A(\lambda) = \alpha \delta(\lambda-1) + (1-\alpha) \delta(\lambda) 
    \\
    &\rho_B(\lambda) = \beta \delta(\lambda-1) + (1-\beta) \delta(\lambda).
\end{align}
We now can express the $\mathcal T$-transforms for these matrices 
\begin{align}
    \mathcal T = \text{tr} \bm A\left( i \omega_{\bm A} + \bm A \right)^{-1} = \frac{\alpha}{i\omega_{\bm A} + 1} = \text{tr} \bm B \left( i \omega_{\bm B} + \bm B \right)^{-1} = \frac{\beta}{i\omega_{\bm B} + 1}  .  
\end{align}
Rearranging this relationship, our defining equation between $\mathcal T$ and the original frequency $i\omega$ is thus 
\begin{align}
    \left( - 1 + \frac{\alpha}{\mathcal T} \right)\left( - 1 + \frac{\beta}{\mathcal T} \right) = \frac{1 - \mathcal T}{\mathcal T} \  i \omega 
\end{align}
Using the fact that $\mathcal T = 1 - i\omega  \ \mH(\omega)$, we can also express the response function
\begin{align}
    \mH(\omega) = \frac{1}{2(i\omega) (1 + i\omega) } \left[ (2+i\omega-\alpha-\beta) - \sqrt{(2 + i\omega - \alpha-\beta)^2 - 4 (1+i\omega)(1-\alpha)(1-\beta) } \right]
\end{align}
From this response function $\mathcal H(\omega)$, we can deduce the eigenvalue density
\begin{align}
    \rho(\lambda) =& \ \ \delta(\lambda) [1 - \min(\alpha,\beta) ]_+ +  \delta(\lambda-1) \left[ \alpha + \beta - 1 \right]_+  \nonumber
    \\
    &+ \frac{1}{2\pi \lambda (1-\lambda)} \sqrt{ \left[ 4(1-\lambda)(1-\alpha)(1-\beta) - (2 - \lambda -\alpha-\beta)^2 \right]_+   } ,
\end{align}
which has two Dirac masses at $\lambda = 0$ and $\lambda = 1$ and a bulk density with support $\lambda \in [\lambda_+ , \lambda_{-}]$ where $\lambda_{\pm} = (\alpha + \beta - 2\alpha\beta) \pm \sqrt{\alpha\beta (1-\alpha)(1-\beta)}$. 
We plot the response function and associated eigenvalue density in Figure \ref{fig:one_point_prod_projections}.

\begin{figure}
    \centering
    \subfigure[Asymmetric Dynamics $\beta = 0.8$]{\includegraphics[width=0.4\linewidth]{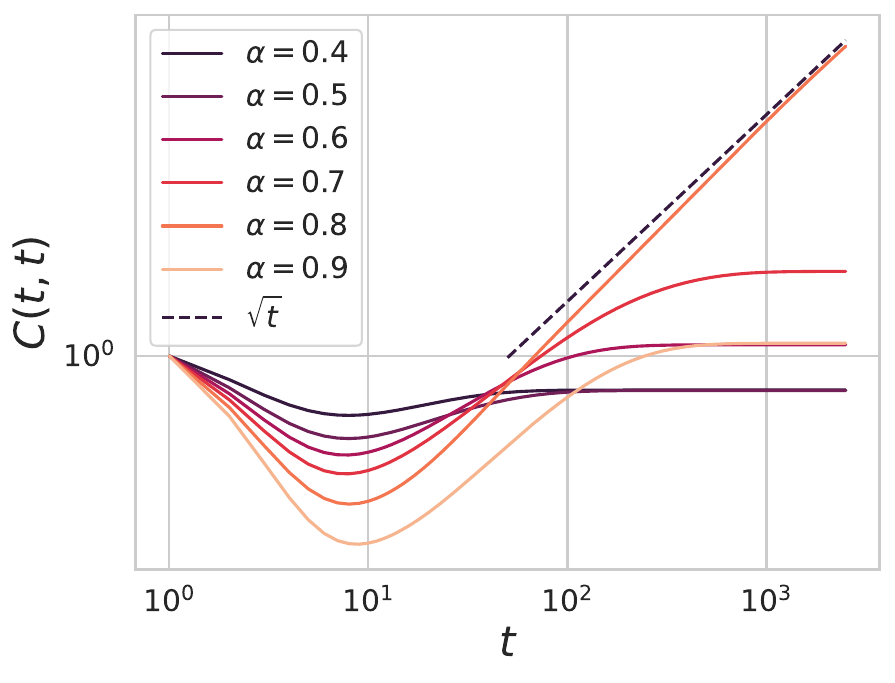}}
    \subfigure[Asymmetric Final Correlation]{\includegraphics[width=0.4\linewidth]{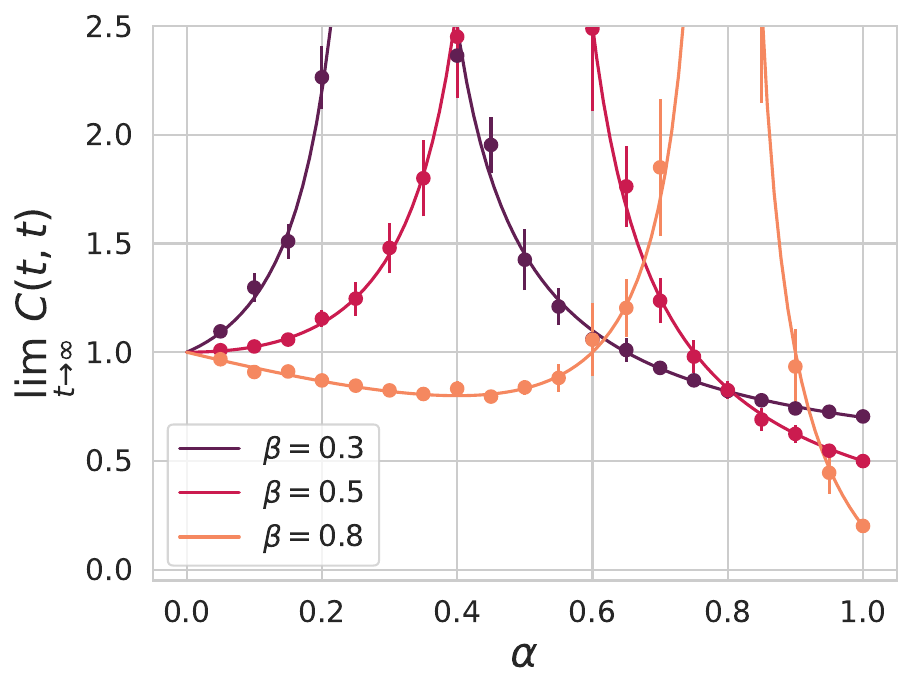}}
    \subfigure[Symmetrized Dynamics $\beta = 0.8$]{\includegraphics[width=0.4\linewidth]{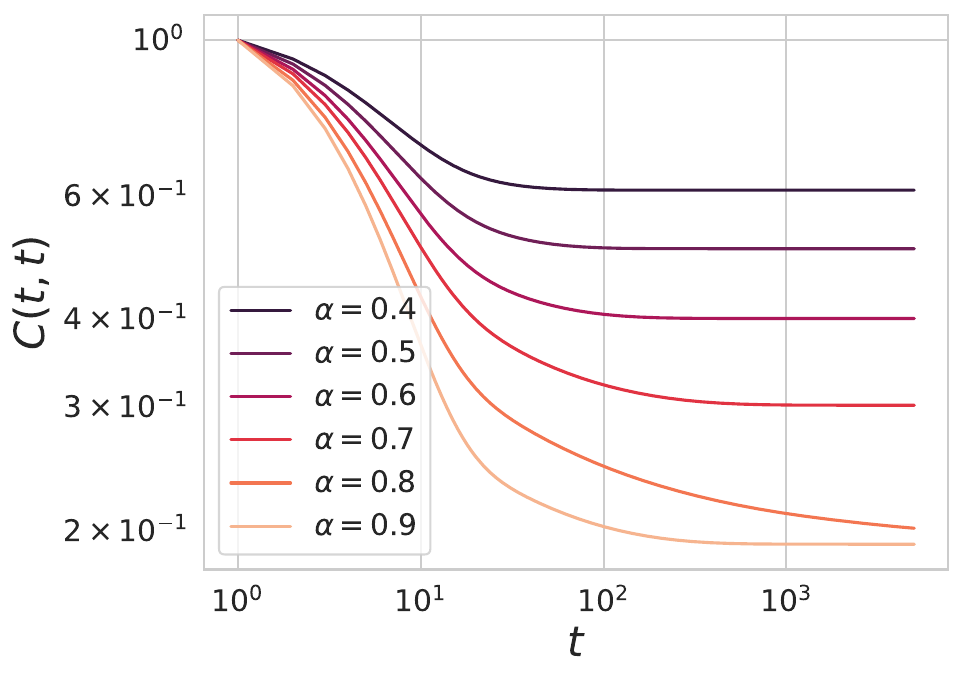}}
    \subfigure[Symmetrized Final Correlation]{\includegraphics[width=0.4\linewidth]{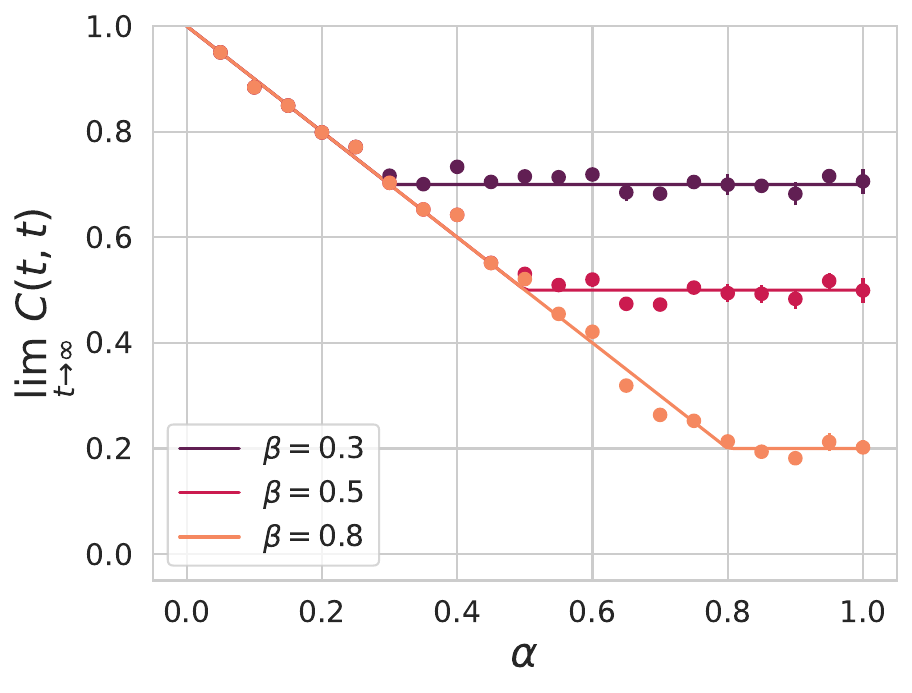}}
    \caption{Despite possessing identical spectra, the asymmetric free product and the symmetric free product exhibit distinct dynamics for their two-point correlation functions. The asymmetric model can exhibit divergence in their correlation $\lim_{t\to\infty} C(t,t)$ at equal aspect ratio $\alpha = \beta$ due to the non-normality of the dynamics, however the symmetrized model's final correlation decreases monotonically in both $\alpha,\beta$.  }
    \label{fig:two_point_correlation_dynamics}
\end{figure}

\paragraph{Deterministic Equivalent Comparison} For the free product of projection matrices, the deterministic equivalence for the asymmetric and symmetric case have the form 
\begin{align}
    \h(\omega) \h(\omega')^\top &\simeq \left( i\omega + \Psi(\omega) \bm A \right)^{-1} \h_0 \h_0^\top \left( i\omega' + \Psi(\omega') \bm A \right)^{-1}  \nonumber
    \\
    &+ \begin{cases}
        \frac{\Gamma(\omega,\omega')}{1-\Gamma(\omega,\omega')}  \left( i\omega + \Psi(\omega) \bm A \right)^{-1} \left( i\omega' + \Psi(\omega') \bm A \right)^{-1} & \text{Asymmetric}
        \\
        \frac{\Gamma(\omega,\omega')}{1-\Gamma(\omega,\omega')}  \  \bm A \  \left( i\omega + \Psi(\omega) \bm A \right)^{-1} \left( i\omega' + \Psi(\omega') \bm A \right)^{-1} & \text{Symmetrized}
    \end{cases} .
\end{align}
We see that the bias term (the first term) is identical across both cases, but that the variance term differs by a factor of $\bm A$ in the symmetrized and asymmetric cases. The function $\Gamma(\omega,\omega')$ which controls the variance has the following form
\begin{align}
    \Gamma(\omega,\omega') &= \frac{1}{\alpha} - \frac{\alpha}{(i\omega + \Psi(\omega))(i\omega'+\Psi(\omega'))} \times \left[ \frac{\beta}{(i\omega_{\bm B}+1)(i\omega_{\bm B}' + 1)} + \frac{1-\beta}{i\omega_{\bm B} i\omega_{\bm B}'} \right]^{-1} .
\end{align}
For isotropic and random initial conditions $\left< \h_0 \h_0^\top \right> = \bm I$, the correlation functions have the form
\begin{align}
    \mC(\omega,\omega') = 
    \begin{cases}
        \frac{1}{1-\Gamma(\omega,\omega')} \left[ \frac{\alpha}{(i\omega + \Psi(\omega) )(i\omega' + \Psi(\omega'))} + \frac{1-\alpha}{i\omega i\omega'} \right] & \text{Asymmetric}
        \\
        \frac{1}{1-\Gamma(\omega,\omega')} \left[ \frac{\alpha}{(i\omega + \Psi(\omega) )(i\omega' + \Psi(\omega'))} \right] +  \frac{1-\alpha}{i\omega i\omega'}  & \text{Symmetrized}
    \end{cases}
\end{align}
In the symmetrized case, the null-space of the matrix $\bm A$ (which has dimension $(1-\alpha)N$) does not interact with the variance term that appears in $\mC(\omega,\omega')$, while in the asymmetric case, it is amplified by a factor $\frac{1}{1-\Gamma(\omega,\omega')}$, indicating interaction with the variance of the dynamics. In the next section we will show that this implies the final value of $\lim_{t \to \infty} C(t,t)$ cannot diverge for the symmetrized matrix but the final value can diverge when $\alpha = \beta$ for the asymmetric case, similar to the random feature model at $\nu = \alpha$.

\paragraph{Final Correlation Values: Asymmetric vs Symmetrized}

To illustrate the impact of the the non-normality on the final value of the correlation function, we can now evaluate the $\omega \to 0$ limit. The $\Psi$ function has the following behavior
\begin{align}
    \Psi(\omega) = \frac{i\omega}{i\omega_{A}} \sim 
    \begin{cases}
        \frac{i\omega \beta}{\alpha - \beta} & \alpha > \beta 
        \\
        \frac{\beta-\alpha}{1-\alpha} & \alpha < \beta
    \end{cases} \ , \ i \omega \to 0 .
\end{align}
The final values of the correlation function have the following form
\begin{align}
    &\lim_{t\to \infty} C(t,t) = 
    \begin{cases}
        |\alpha - \beta| + \frac{\max(\alpha,\beta)(1-\max(\alpha,\beta))}{|\alpha - \beta|} & \text{Asymmetric}
        \\
         [1- \min(\alpha,\beta)]_+ & \text{Symmetrized}
    \end{cases}
\end{align}
We note that this loss curve exhibits a divergence at $\alpha = \beta$, reminiscent of the isotropic random feature model where the matrices had independent entries where a divergence occurred at $\alpha = \nu$. While the structure of the loss curve is distinct from the random feature model due to distinct spectra, the $\sqrt{t}$ blowup at the interpolation threshold is universal.



\section{Non-Hermitian Systems with Complex Spectra}\label{sec:complex_spectra}

Unfortunately, the response function for a DMFT does not always capture the full spectrum of a random matrix. In fact this can lead to problems for matrices whose spectral densities are not confined to the real line in the complex plane. 

\paragraph{Standard DMFT Response and Spectra in Complex Plane} To see an example of the failure of the standard response function to capture a spectrum that is extended in the complex plane, consider the case of an asymmetric random Gaussian matrix $M_{ij} \sim \mathcal{N}(0, 1/N)$. The behavior of this system as $N \to \infty$ is governed by the following DMFT equation
\begin{align}
   &\text{DMFT Equation:} \quad  \frac{d}{dt} h(t) = u(t) \ , \ u(t) \sim \mathcal{GP}(0, C(t,t')) 
   \\
   &\implies \text{Trivial Response Function:} \quad  \mR(\omega)  = \frac{1}{i\omega}  ,
\end{align}
which is indistinguishable from the response function for $\bm M = \bm 0$. Thus the spectral density of $\bm M$ cannot be obtained by studying the simple dynamical system driven by the matrix $\bm M$ \footnote{This failure is closely connected to the failure of moment methods and standard expansions of the resolvent that use $\text{tr} \left< \bm M^k \right>$ to compute resolvents for generic asymmetric matrices.}.

\paragraph{Hermitianization} To counteract this problem, we can consider a dynamical system driven by a \textit{hermitian matrix} which relaxes to a fixed point that depends on the resolvent of the matrix \cite{girko1985circular, feinberg9703118non, baron2022eigenvalues, cui2024elementary}. We consider the following flows 
\begin{align}
    \frac{d}{dt} \h(t) = - \left( \bm M - z \right)^{\dagger} \left[ \left(\bm M - z \right) \h(t,z) + \bm b \right]  , 
\end{align}
where $z \in \mathbb{C}$ is an arbitrary complex argument. Since the matrix $\left( \bm M - z \right)^\dagger \left( \bm M - z \right)$ is Hermitian and positive semidefinite, the dynamics will relax to the following long time limit
\begin{align}
    \lim_{t \to \infty} \bm h(t) = \left( z - \bm M \right)^{-1} \bm b
\end{align}
The resolvent of the matrix for complex $z = x + i y \in \mathbb{C}$ and the density $\rho(z)$ can be obtained from the long time limit 
\begin{align}
    \mathcal G(z) = \lim_{t \to \infty} \text{tr} \left( \frac{\partial \h(t)}{\partial \bm b^\top} \right) = \text{tr} \left( z - \bm M \right)^{-1} \quad \rho(z) = \frac{1}{\pi} \partial_{z^\star} \mathcal G(z)   
\end{align}
from which the spectral density in the complex plane at point $z \in \mathbb{C}$ can be obtained\footnote{We use the definition $\partial_{z^\star} = \frac{1}{2}(\partial_x + i \partial_y)$ for $z=x+iy$ so that $\partial_{z^\star} z^\star = 1$ and $\partial_{z^\star}  \ \frac{1}{z}  = \delta(z)$.}. Fortunately, this \textit{Hermitianized} dynamical system does generate response functions in its associated DMFT equations that enable computation of the complex spectrum from the long time limit \footnote{For readers familiar with Hermitianization, often a regulator is introduced when computing the resolvent of the Hermitianized matrix \cite{sommers1988spectrum}. In our setup, the finite time plays the role of the regulator, with the long-time limit $t\to\infty$ ($i\omega \to 0$) giving the final result.}.

\subsection{Ginibre Matrices and the Circular Law} 
The simplest such example is a completely asymmetric random matrix $\bm M = \frac{1}{\sqrt N} \bm A$ where $A_{ij} \sim \mathcal{N}(0, 1)$ have real entries with no symmetry requirement. As before, we break up the dynamics into two separate processes whose defining equations are each linear in $\bm M$
\begin{align}
    \frac{d}{dt} \h_0(t) = - \left( \bm M -  z \right)^\dagger \h_1(t) \ , \ \h_1(t) = \left( \bm M - z \right) \h_0(t) + \bm b  .
\end{align}
The long time limit of $\h_0$ is a function of the matrix resolvent 
\begin{align}
    \lim_{t \to \infty} \h_0(t) = \left( z - \bm M \right)^{-1} \bm b
\end{align}
We can therefore investigate the limit of the dynamical system we started with. These dynamics in the $N \to \infty$ limit, can be described by the following DMFT equations 
\begin{align}
    &\frac{d}{dt} h_0(t) = - h^2(t) \ , \ h^2(t) = \xi_2(t) + \int  dt' \ R_1(t,t') h_0(t') - z^\star h_1(t) 
    \\ 
    &h_1(t) = \xi_1(t) + \int dt' \ R_{0,2}(t,t') h_1(t') - z h_0(t) + b
\end{align}
where $R_1(t,t') = \frac{\partial h_1(t)}{\partial \xi_1(t')}$ and $R_{0,2}(t,t') = \frac{\partial h_0(t)}{\partial \xi_2(t')}$ are the response functions that arise from the DMFT equations. Defining $\mH(t,t') = - R_{0,2}(t,t')$ and taking a Fourier transform, we find
\begin{align}
    i\omega = \frac{1}{\mH(\omega) (1+\mH(\omega))^2} \left[ 1 - (|z|^2 - 1) \mH(\omega)  \right] .
\end{align}
We are interested in the solutions to this equation as $i \omega \to 0$. One solution is $\mH(\omega)= \frac{1}{|z|^2 - 1}$, but there are two other solutions $\mH(\omega) \sim \pm (i\omega)^{-1/2} \left[1-|z|^2 \right]^{1/2}$. For a given value of $|z|^2$, we select the branch which gives an analytic function of $z$ and decays like $\mG(z) \sim \frac{1}{z}$ for large $|z|$. Thus for $|z|>1$, we choose $\mH(\omega) = \left[|z|^2-1\right]^{-1}$, while for $|z|<1$ we choose the diverging solutions. This results in the following expression for the resolvent and eigenvalue density,
\begin{align}
    \mG(z) = \lim_{\omega \to 0} \frac{   z^\star}{1 + \mH(\omega)^{-1}} = \begin{cases}
        z^\star & |z|^2 \leq 1
        \\
        \frac{1}{z} & |z|^2 > 1 
    \end{cases} \implies \rho(z)  = \begin{cases}
        \frac{1}{\pi} &  |z|^2 < 1
        \\
        0 & |z|^2 > 1
    \end{cases}.
\end{align}
This recovers the circular law, where the eigenvalue density is uniform in a unit disk in the complex plane \cite{ginibre1965statistical, sommers1988spectrum}. 

\subsection{Diagonally Modulated Gaussian}

\begin{figure}
    \centering
    \subfigure[Bernoulli Modulated Gaussian]{\includegraphics[width=0.35\linewidth]{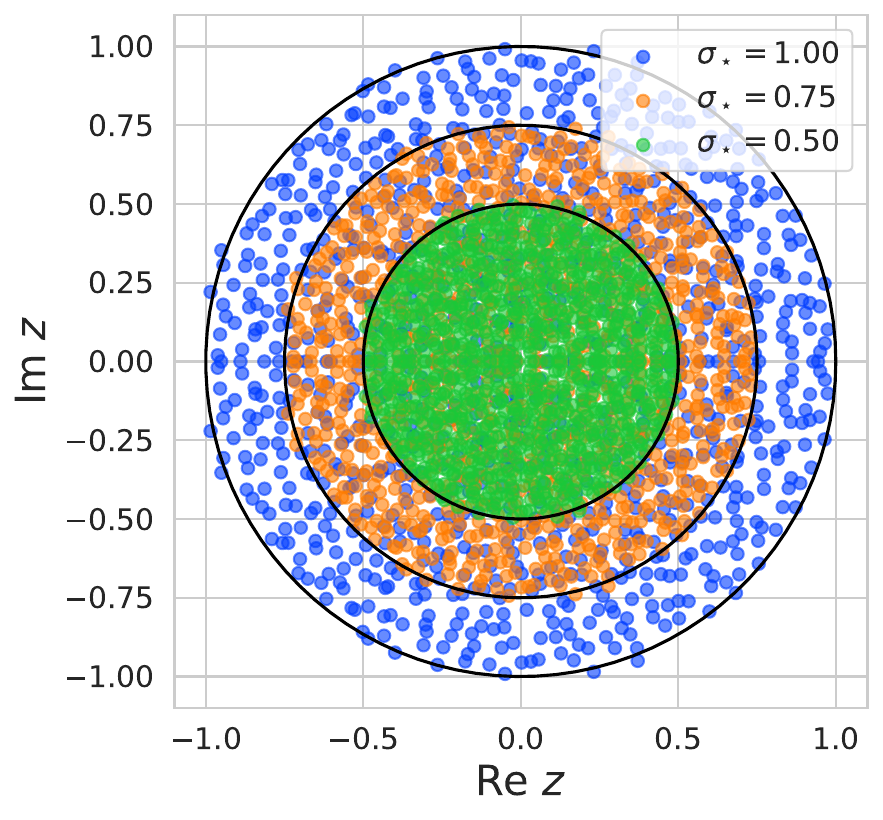}}
    \subfigure[Varying Diagonal Measure $\mu(\sigma)$]{\includegraphics[width=0.425\linewidth]{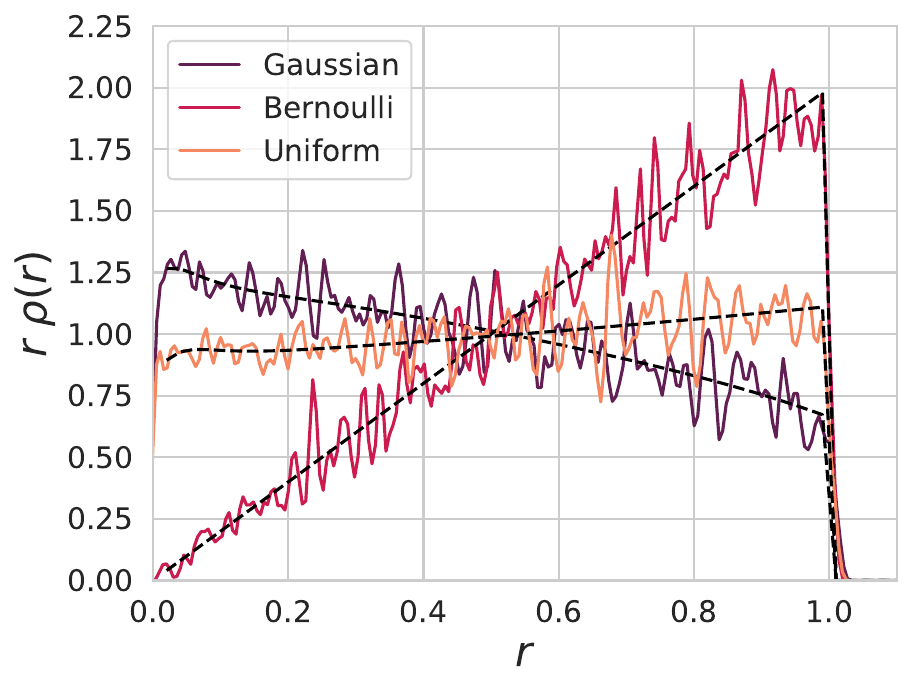}}
    \caption{Eigenvalue densities in the complex plane for generic asymmetric matrices can be computed with the Hermitianized DMFT. (a) The eigenvalue densities obey a rescaled circular law when there is diagonal matrix whose entries are $\pm \sigma_\star$ with equal probability. (b) The eigenvalue densities in the bulk depend on probability density of $\mu(\sigma)$. We compare Gaussian $\mu(\sigma) = \mathcal{N}(0,1)$, Bernoulli $\mu(\sigma) = \frac{1}{2} \delta(\sigma-\sigma_\star) + \frac{1}{2} \delta(\sigma+\sigma_\star)$ and uniform $\mu(\sigma) = \mathcal{U}\left[0, \sqrt{3} \right]$. While all three eigenvalue densities have the same support $|z| \leq 1$, the densities have different shapes in the bulk. Dashed black lines are the solution to the self-consistent equation for $\mG(z)$ while the colored lines are kernel density plots for a single $N=5000$ realization.  }
    \label{fig:complex_plane_densities}
\end{figure}

To see how this method can be used for slightly more interesting ensembles, consider a Gaussian asymmetric matrix $\bm A$ which is multiplied by a diagonal matrix with entries $\sigma_i$, each of which are iid draws from some distribution $\mu(\sigma)$
\begin{align}
    \bm M = \frac{1}{\sqrt N} \  \bm A 
    \ \text{diag}(\bm\sigma) \quad , \quad  \sigma_i \sim \mu(\sigma) .
\end{align}
This ensemble is of interest in theoretical neuroscience since it describes the eigenvalues of the Jacobian for a randomly connected recurrent neural network \cite{sompolinsky1988chaos, helias2020statistical}. The DMFT equations for the Hermitianized dynamical system take the form
\begin{align}
    &\frac{d}{dt} h_0(t) = - h_2(t) \ , \ h_2(t)= \xi_2(t) +  \sigma^2 \int dt' R_1(t,t') h_0(t') - z^\star h_1(t) \nonumber
    \\
    &h_1(t) = \xi_1(t) + \int dt' R_{0,2}(t,t') h_1(t') - z h_0(t') + b.    
\end{align}
In the above expressions, the dynamics for $h_0$ are modulated by the random variable $\sigma \sim \mu(\sigma)$. The response functions $R_1(t,t')$ and $R_{0,2}(t,t')$ are computed as
\begin{align}
    R_1(t,t') = \left< \frac{\partial h_1(t)}{\partial \xi_1(t')} \right>_{\sigma} \ , \ R_{0,2}(t,t') =  \left< \sigma^2 \frac{\partial h_0(t)}{\partial \xi_2(t')} \right>_{\sigma}
\end{align}
where the average $\left< \cdot \right>$ is over the measure $\mu(\sigma)$. To close the equations for the response functions, it will be advantagous to introduce a function $\mH_\sigma(\omega) \equiv -\int d\tau e^{-i\omega \tau} \frac{\partial h_0(t+\tau)}{\partial \xi_2(t)}$ which is the response function \textit{conditional on a particular value of $\sigma$} . 
\begin{align}
    \mH_\sigma(\omega) \equiv \frac{1}{i\omega + \sigma^2 \mR_1(\omega) + \frac{|z|^2}{ 1 - \mR_{0,2}(\omega) }} 
\end{align}
From this function, we can compute the averaged Fourier-transformed response functions $\mR_1(\omega)$ and $\mR_{0,2}(\omega)$ as
\begin{align}
    \mR_{0,2}(\omega) = - \left< \sigma^2 \mH_\sigma(\omega) \right> \ , \ \mR_{1}(\omega) = \frac{1}{1 + \left< \sigma^2 \mH_\sigma(\omega) \right> }  \left[ 1 - \frac{|z|^2 \left< \mH_\sigma(\omega) \right>}{1 + \left< \sigma^2 \mH_\sigma(\omega) \right>}  \right]
\end{align} 
where $\left< \cdot \right>$ represents an average over $\rho(\sigma)$.  Lastly we can recover the limiting resolvent and spectral density from
\begin{align}
    \mG(z) = \lim_{\omega \to 0} \frac{\left< \mH_\sigma(\omega) \right> z^\star}{1 + \left< \sigma^2 \mH_\sigma(\omega) \right> }  \ , \ \rho(z) = \frac{1}{\pi} \partial_{z^\star} \mG(z) 
\end{align}
The defining equation for $\mH_\sigma(\omega)$ depends on the random variable $\sigma$ two moments $\left< \mH(\omega) \right>$ and $\left< \mH(\omega) \sigma^2 \right>$ 
\begin{align}
    \mH_\sigma(\omega) = \frac{( 1+\left<\sigma^2 \mH_\sigma(\omega) \right>)^2 }{ i\omega (1+\left<\sigma^2 \mH_\sigma(\omega) \right>)^2 + \sigma^2 \left( 1 + \left<\sigma^2 \mH_\sigma(\omega)\right> - |z|^2 \left<\mH_\sigma(\omega) \right> \right) + |z|^2 (1+\left<\sigma^2 \mH_\sigma(\omega) \right>) }  
\end{align}
For the first solution, we will investigate when $\left< \sigma^2 \mH_\sigma(\omega) \right> = \mathcal{O}(1)$ as $i\omega \to 0$. In this case, 
\begin{align}
    \mG(z)= \lim_{\omega \to 0} 
    \frac{z^\star \left<\mH_\sigma(\omega) \right>}{1 + \left<\sigma^2 \mH_\sigma(\omega) \right>} = \frac{1}{z} 
\end{align}
In the other case, we investigate $\left< \sigma^2 \mH(\omega) \right>$ which diverges as $i \omega \to 0$. This leads to the condition
\begin{align}
    1 = \left< \frac{\sigma^2  }{ |z|^2 + \sigma^2 \left( 1 - z \mG(z)  \right) } \right> .
\end{align}
The boundary separating these two solutions occurs when this solution approaches $\mG(z) = \frac{1}{z}$, which occurs for $|z| = \sqrt{ \left< \sigma^2 \right>}$. The eigenvalue density in the bulk region can be obtained by differentiating the above equation, resulting in the density
\begin{align}
    \rho(z) = \frac{1}{\pi} \left< \frac{\sigma^2}{\left[|z|^2 + \sigma^2 \left( 1 - z \mG(z)  \right) \right]^2} \right> \left< \frac{\sigma^4 }{\left[|z|^2 + \sigma^2 \left( 1 - z \mG(z)  \right) \right]^2} \right>^{-1}  \Theta\left( \left<\sigma^2\right> - |z|^2 \right) 
\end{align}
If the density of diagonal values is Bernoulli over $\{\pm \sigma_\star \}$ with probability $1/2$, (i.e. $\mu(\sigma) = \frac{1}{2} \delta(\sigma-\sigma_\star) + \frac{1}{2} \delta(\sigma + \sigma_\star)$) then we recover the previous result (the circular law) where $\rho(z) = \frac{1}{\pi \sigma^2_\star} \Theta(\sigma^2_\star-|z|^2)$ as we show in Figure \ref{fig:complex_plane_densities} (a). However, for other distributions $\mu(\sigma)$ variables, the density in the bulk is generally non-uniform as we show in Figure \ref{fig:complex_plane_densities} (b).

\section{Beyond Linear TTI Structure, Evolving Matrices}

In the previous examples, the random matrices appearing in the dynamics were frozen. Since the dynamics for $\h$ were linear, the response functions were time-translation invariant (TTI) and could be obtained directly from a Fourier transform. Subsequently, the correlation functions could be computed in closed form as a two-variable Fourier transform. However, DMFT is arguably most useful in settings beyond linear dynamics. In this section, we illustrate a few simple examples where the resulting DMFT equations remain Gaussian, which enables exact analytical computations of the correlations. A number of interesting high dimensional systems with this Gaussian property have been recently shown to capture similar phenomena such as transitions to chaos, aging dynamics, and potential separations from mean field statics \cite{fournier2023statistical, fournier2025non}.

\subsection{Toy Example: Simple (Anti)-Hebbian Linear Dynamics}

\begin{figure}
    \centering
    \subfigure[Equal Time Correlation]{\includegraphics[width=0.45\linewidth]{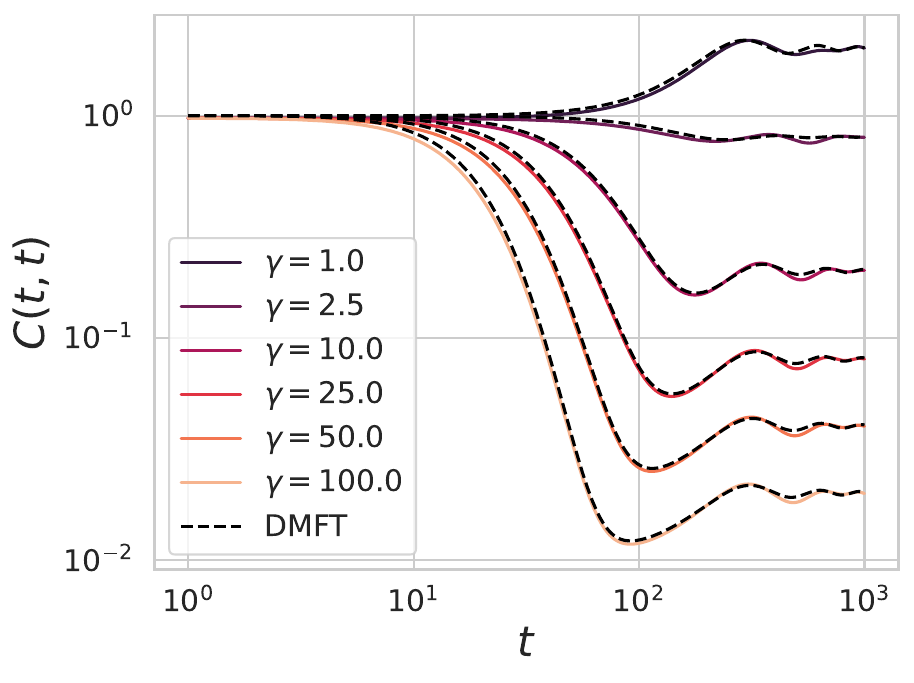}}
    \subfigure[Correlation with Initial Condition]{\includegraphics[width=0.45\linewidth]{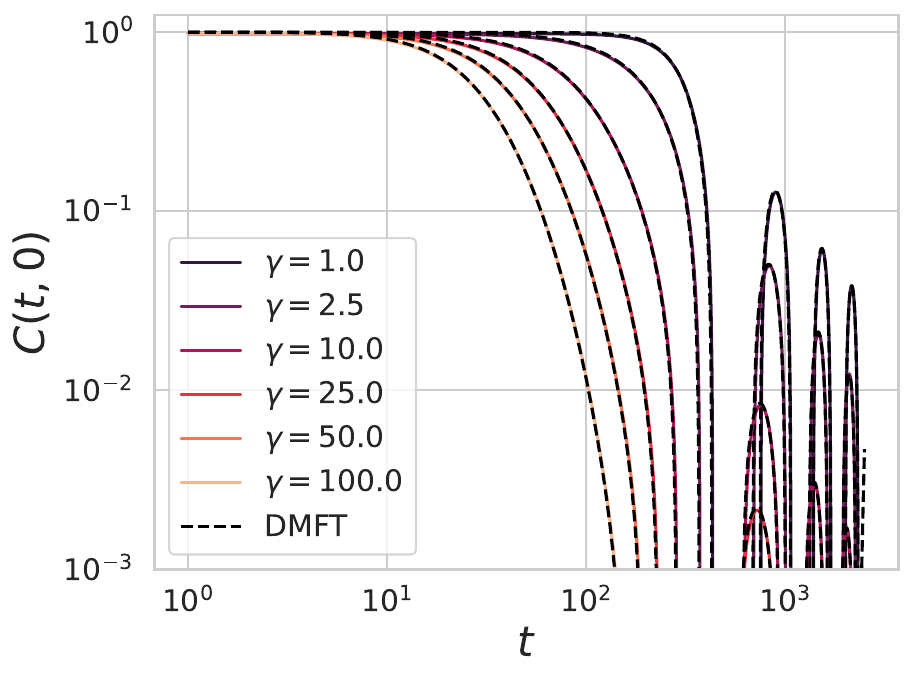}}
    \caption{A random linear RNN with anti-Hebbian dynamics are well predicted by the non-stationary DMFT. (a) The equal time correlation function undergoes transient dynamics before relaxing to an approximate steady state which depends on $\gamma$. (b) The correlation between the state $\h(t)$ and the initial condition $\h(0)$ undergoes nontrivial oscillations, the scale of which depend on $\gamma$.  }
    \label{fig:rnn_hebb_model}
\end{figure}

Consider a linear dynamical system with dynamical connectivity $\bm M(t)$. While this model was studied in a nonlinear RNN with asymmetric initial connectivity \cite{clark2024theory}, we will take $\bm M(0)$ to be a GOE matrix and study a linear RNN with adaptive weights
\begin{align}
    \frac{d}{dt} \h(t) = - \bm M(t) \h(t) \ , \ \frac{d}{dt} \bm M(t) = \frac{\gamma}{N} \h(t) \h(t)^\top . 
\end{align}
In the above system, the matrix $\bm M$ has two random components, a random initial matrix $\bm M(0)$ and a dynamical spike $\gamma N^{-1} \int_0^t dt' \  \h(t') \h(t')^\top$ which depends on the random variables $\h(t)$ that themselves depend on $\bm M(0)$. Fortunately, one can still easily compute the DMFT system to characterize the $N \to \infty$ limit. As before, the single-site stochastic process for $h(t)$ depends only on the correlation and response  
\begin{align}
    &\frac{d}{dt} h(t) =  u(t) + \int_0^t dt' R(t,t') h(t')  - \gamma \int_0^t dt' \  C(t,t') h(t') \ , \ u(t) \sim \mathcal{GP}(0, C(t,t')) .
\end{align}
In the above, the only two order parameters that arise are the correlation $C(t,t')$ and response $R(t,t')$. We note that this system, unlike the linear dynamics ($\gamma= 0$), is non-TTI and the response functions cannot be solved for directly through Fourier transform. Rather, we are left with a coupled set of integro-differential equations for our two-time order parameters
\begin{align}
    &\frac{\partial}{\partial t} C(t,t') = \int_0^t dt'' \left[R(t',t'') C(t'',t) + R(t,t'') C(t'',t') - \gamma C(t,t'') C(t'',t') \right] \nonumber
    \\
    &\frac{\partial}{\partial t} R(t,t') = \delta(t-t') + \int_0^t dt'' \left[ R(t,t'') R(t'',t') -\gamma C(t,t'') R(t'',t') \right] .
\end{align}
These integro-differential equations capture the interesting transient and late time dynamics of this model as we show in Figure \ref{fig:rnn_hebb_model}. Despite no longer being exactly solveable in frequency space, the DMFT equations can be integrated in real time to capture the asymptotic $N \to \infty$ limit of the dynamics.

\subsection{Deep Linear ResNets under Random Initialization}

\begin{figure}
    \centering
    \subfigure[Varying Richness $\gamma_0$]{\includegraphics[width=0.32\linewidth]{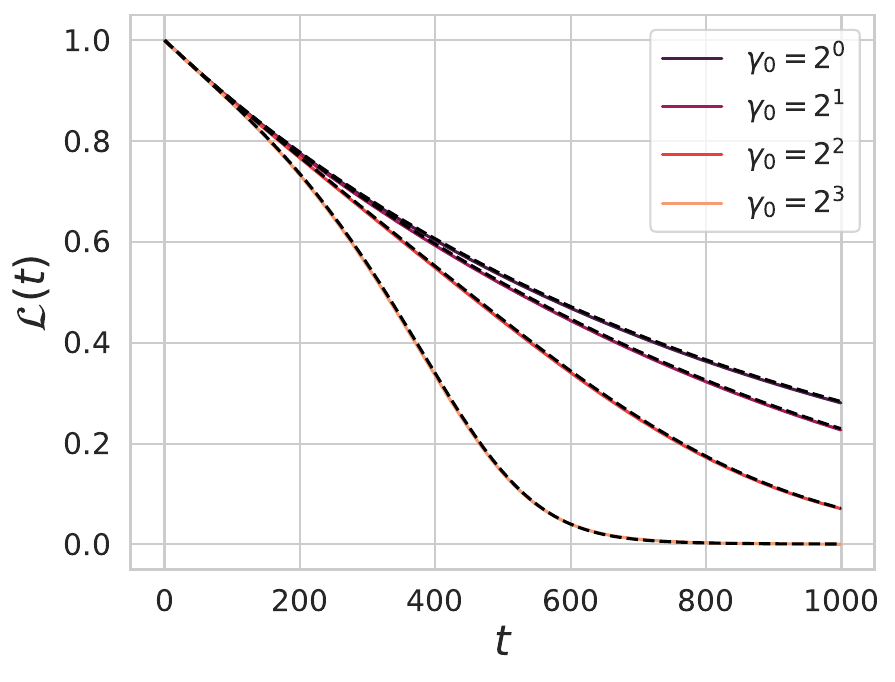}}
    \subfigure[Varying Depth $L$ with $\beta = L^{-1/2}$]{\includegraphics[width=0.32\linewidth]{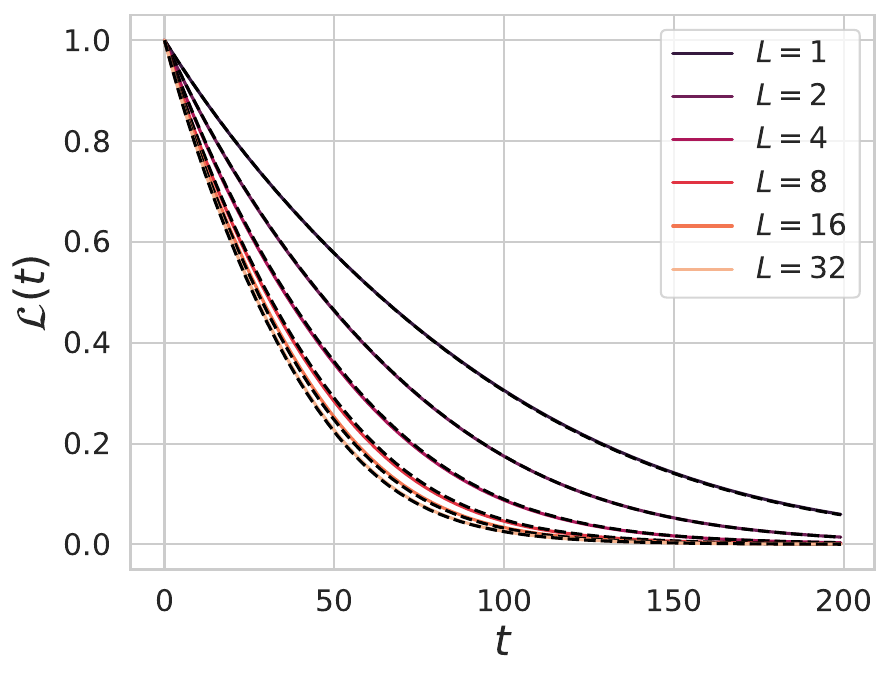}}
    \subfigure[Varying Width $\nu$]{\includegraphics[width=0.32\linewidth]{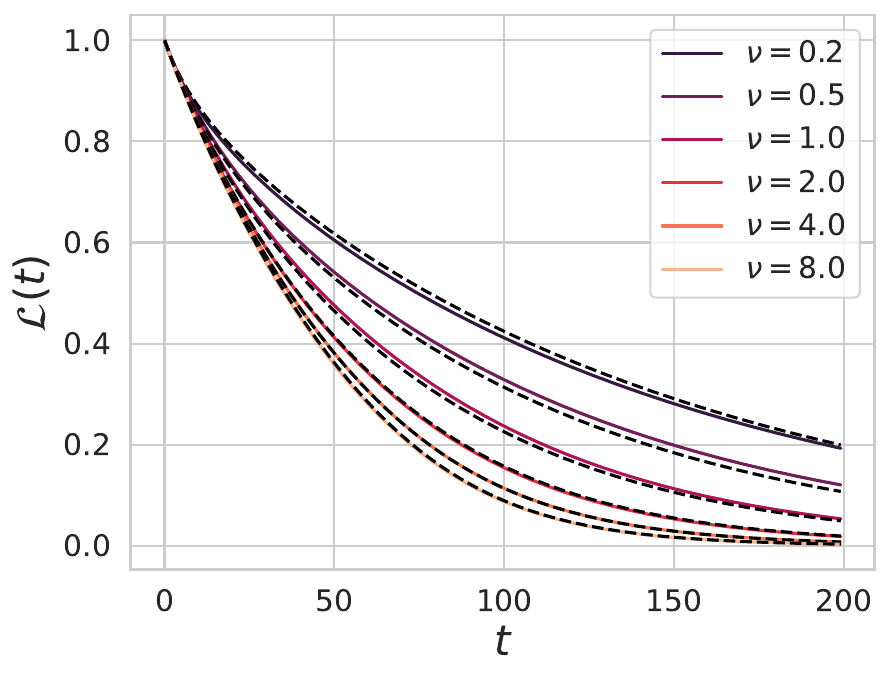}}
    \caption{The dynamics of a deep linear residual network trained with gradient descent. (a) Increasing the parameter $\gamma_0$ induces more significant changes in the hidden weights of the network, leading to nonlinear dynamics (the $\gamma_0 \to 0$ limit results in linear, termed \textit{lazy} dynamics \cite{chizat2019lazy,bordelon2022self}). (b) Varying the depth of the residual network converges to a stable predictor if $\beta = \beta_0 / \sqrt{L}$. (c) Increasing the width to input-dimension ratio $\nu = N_1/ N_0$ reduces the test loss. }
    \label{fig:deep_resnet_dynamics}
\end{figure}

Another setting where DMFT equations still provide an exact description of the asymptotic dynamics, without having a TTI description is the training dynamics of randomly initialized infinite width neural networks \cite{bordelon2022self, bordelon2023depthwise}. In this model, all weights of the network start off as random matrices, but experience structured updates from training. In this setting, we will briefly describe a special case of deep linear networks trained on random data in a proportional scaling regime for input dimension $N_0$, hidden width $N_1$ and dataset size $P$ \cite{bordelon_deep_linear}   
\begin{align}
    N_0, N_1, P  \to \infty \ , \  \frac{P}{N_0} = \alpha  \ , \  \frac{N_1}{N_0} = \nu .
\end{align}
We consider a $L$ hidden layer linear network $f(\x)$ and a noisy linear target function $y(\x)$ 
\begin{align}
    f(\x) = \frac{\sqrt{N_0}}{\gamma_0 N_1 } \  \bm w^L \cdot   \prod_{\ell=1}^{L-1} \left( \bm I  +  \frac{\beta}{\sqrt{N_1}} \bm W^\ell \right) \left( \frac{1}{\sqrt{N_0}} \bm W^0 \right) \x  \quad , \quad  y(\x) = \frac{1}{\sqrt{N_0}} \bm \beta_\star \cdot \bm x + \sigma \epsilon 
\end{align}
We train all parameters $\bm\theta \in \{ \w^L , \W^{L-1},..., \W^0 \}$ with gradient flow with learning rate $\eta =  N_1 \gamma_0^2$ on a training loss $\mathcal{\hat L} = \frac{1}{P} \sum_{\mu=1}^P \left( f(\x_\mu ,t) - y_\mu \right)^2$ defined over dataset $\mathcal D = \{ (\x_\mu, y_\mu) \}_{\mu=1}^P$ 
\begin{align}
    \frac{d}{dt} \bm\theta(t) = - \eta \frac{\partial}{\partial \bm\theta} \  \hat{\mathcal L}(t)  . 
\end{align}
We aim to characterize not only the training loss but also the test loss, which is an average over the population distribution for $\x$ and $\epsilon$.
\begin{align}
    \mathcal L(t) = \left< \left( f(\x,t) - y(\x) \right)^2  \right>_{\x, \epsilon} 
\end{align}
The vector $\bm v(t)$ which controls the test loss $\mathcal L(t) = \frac{1}{N_0} \v(t) \cdot \v(t) + \sigma^2$ is defined as
\begin{align}
    \bm v(t) \equiv \bm\beta_\star - \frac{\sqrt{N_0}}{\gamma_0 N_1} \bm W^0(t)^\top \left[ \prod_{\ell=1}^{L-1} \left( \bm I  +  \frac{\beta}{\sqrt{N_1}} \bm W^\ell(t) \right) \right]^\top  \w^L(t) .
\end{align}
We note that the gradient flow dynamics on the weights $\W^\ell(t)$ can be expressed in terms of its initial condition $\W^\ell(t)$ and a low rank update expressed in terms of vectors $\h^\ell(t)$ and $\g^\ell(t)$
\begin{align}
    &\w^L(t) = \w^L(0) + \gamma_0 \int_0^t dt' \ \h^L(t')  \nonumber
    \\
    &\W^\ell(t) = \W^\ell(0) + \frac{ \gamma_0 \beta}{\sqrt{N_1}} \int_0^t dt' \  \g^{\ell+1}(t') \h^\ell(t')^\top   \nonumber
    \\
    &\W^0(t) = \W^0(0) + \frac{ \gamma_0}{\sqrt{N_0}} \int_0^t dt' \ \g^1(t') \h^0(t')^\top 
\end{align}
where the vectors $\h^\ell(t)$ and $\g^\ell(t)$ are defined as forward pass and backward pass variables 
\begin{align}
    &\h^0(t) = \frac{\sqrt{N_0}}{P} \bm X^\top \bm\Delta(t) \ , \ \bm\Delta(t) = \frac{1}{\sqrt{N_0}} \bm X \v(t) + \sigma \epsilon  \nonumber
    \\
    &\h^{1}(t) = \frac{1}{\sqrt{N_0}} \bm W^0(t) \h^0(t) \ , \ \h^{\ell+1}(t) = \h^\ell(t) +  \frac{\beta}{\sqrt{N_1}} \W^\ell(t) \h^\ell(t) \ , \ \ell \in \{ 1 ,..., L-1 \}  \nonumber
    \\
    &\g^L(t) = \w^L(t)  \quad , \quad \g^\ell(t) = \g^{\ell+1}(t) +  \frac{\beta}{\sqrt{ N_1 }} \bm W^\ell(t)^\top \g^{\ell+1}(t)  \ , \ \ell \in \{ 1, ... , L-1 \} .
\end{align}
To isolate the dependence of these variables on the initial random matrices, we introduce variables $\bm \chi^\ell(t)$ and $\bm \xi^\ell(t)$ which depend explicitly on the initial conditions for the weights $\W^\ell(0)$ 
\begin{align}
    &\bm\chi^1(t) = \frac{1}{\sqrt{N_0}} \bm W^0(0) \h^0(t) \ , \ 
    \bm\chi^{\ell+1}(t) = \frac{1}{\sqrt{N_1}} \bm W^\ell(0) \h^\ell(t) \nonumber 
    \\ 
    &\bm\xi^0(t) = \frac{\sqrt{N_0}}{N_1} \bm W^0(0)^\top \g^1(t) \ , \ 
    \bm\xi^\ell(t) = \frac{1}{\sqrt{N_1}} \bm W^\ell(0)^\top \g^{\ell+1}(t) 
\end{align}
The full DMFT equations in the proportional scaling limit $P/N_0 = \alpha$ and $N_1/N_0 = \nu$ are thus 
\begin{align}
    &v(t) = \beta_\star - \frac{1}{\gamma_0} r^0(t) -  \int_0^t dt' \left[ \gamma_0^{-1} R_{gu}^1(t,t')  + C_g^1(t,t') \right] h^0(t')  \ , \ r^0(t) \sim \mathcal{N}\left(0, \frac{1}{\nu} C_g^1(t,t') \right)  \nonumber 
    \\
    &h^0(t) = u^0(t) + \int_0^t dt' R_\Delta(t,t') v(t') \ , \ u^0(t) \sim \mathcal{GP}\left(0, \frac{1}{\alpha} C_\Delta(t,t') \right) \nonumber 
    \\
    &\Delta(t) = u_\Delta(t) + \frac{1}{\alpha} \int_0^t R_{vu}^0(t,t') \Delta(t')  + \sigma \epsilon  \ , \ u_\Delta(t) \sim \mathcal{GP}\left(0, C_v(t,t')   \right) \ , \ \epsilon \sim \mathcal{N}(0,1 ) \nonumber 
    \\
    &h^1(t) = u^1(t) +  \int_0^t dt' \left[ \frac{1}{\nu} R_{hr}^0(t,t')  + \gamma_0 C_h^0(t,t') \right] g^1(t') \ , \ u^1(t) \sim \mathcal{GP}\left( 0 , C_h^0(t,t') \right) \nonumber 
    \\
    &h^{\ell+1}(t) = h^\ell(t) + \beta u^{\ell+1}(t) + \beta \int_0^t dt' \left[ R_{hr}^{\ell}(t,t') + \gamma_0 \beta C_h^{\ell}(t,t') \right] g^{\ell+1}(t') \ , \ u^{\ell+1}(t) \sim \mathcal{GP}\left( 0 , C_h^\ell(t,t') \right) \nonumber 
    \\
    &g^L(t) = r^L + \gamma_0 \int_0^t dt' h^L(t') \ , \ r^L \sim \mathcal{N}(0,1) \nonumber
    \\
    &g^\ell(t) = g^{\ell+1}(t) + \beta r^\ell(t) + \beta \int_{0}^t dt' \left[ R_{gu}^{\ell+1}(t,t') + \gamma_0 \beta C_g^{\ell+1}(t,t') \right] h^\ell(t') \ , \ r^\ell(t) \sim \mathcal{GP}\left(0, C_g^{\ell+1}(t,t') \right) .
\end{align}
Despite the original loss function being non-convex in the original trainable parameters $\bm\theta$, this system retains Gaussianity of all hidden fields in the proportional limit. The correlation functions $C$ in the above dynamics are defined as
\begin{align}
    &C_\Delta(t,t') = \left<\Delta(t) \Delta(t') \right> \ , \ C_v(t,t') = \left< v(t) v(t') \right> \nonumber
    \\
    &C_h^\ell(t,t') = \left< h^\ell(t) h^\ell(t') \right> \ , \ C_g^\ell(t,t') = \left< g^\ell(t) g^\ell(t') \right>
\end{align}
We note that the test loss and train loss can be computed from htese correlations
\begin{align}
    \mathcal L(t) = C_v(t,t) + \sigma^2 \ , \ \hat{\mathcal L}(t) = C_\Delta(t,t) .
\end{align}
The response functions, likewise can be computed as
\begin{align}
    R_{vu}^0(t,t') = \frac{\partial v(t)}{\partial u^0(t')} \ , \ R_\Delta(t,t') = \frac{\partial \Delta(t)}{\partial u_\Delta(t')} \nonumber
    \\
    R_{hr}^\ell(t,t') = \frac{\partial h^\ell(t)}{\partial r^\ell(t')} \ , \ R_{gu}^\ell(t,t') = \frac{\partial g^\ell(t)}{\partial u^\ell(t')} 
\end{align}
These equations close directly at the level of correlation and response, however the systems dynamics are no longer time-translation invariant so both correlation and response must be solved for simultaneously (unlike the previous examples where response functions could be solved for independently of the correlations). We plot examples of these equations compared to wide but finite networks in Figure \ref{fig:deep_resnet_dynamics}. Finite $\nu$ and finite $\alpha$ generate corrections to both the bias and variance dynamics of the model. Further if the branch scale parameter $\beta$ is scaled as $\beta = \frac{\beta_0}{\sqrt L}$, the above dynamics converge to a well defined infinite depth limit as $L \to \infty$. A discrete time version of these equations was explored in \cite{bordelon_deep_linear}, which enabled comparisons of how optimal learning rate depends on $\nu$ and $L$ in different parameterizations, providing a simple tractable example where the hyperparameter transfer effect can be characterized theoretically \cite{yang2021tuning}. 

\section{Discussion} 
In this work, we provided a pedagogical overview of DMFT ideas to analyze the evolution of dynamical systems in high dimensions. Our focus was on dynamical systems that admit single site Gaussian processes in the high dimensional limit. However, the potential application areas of DMFT are far more diverse. While we examined the use of DMFT methods for simple models that yield Gaussian processes, many problems generate asymptotic descriptions that are non-Gaussian including descriptions of training dynamics of nonlinear models with random data \cite{gerbelot2022rigorous, mignacco2020dynamical, mignacco2022effective}, training deep nonlinear networks from random initialization \cite{bordelon2022self, bordelon2023depthwise}, and Hebbian learning in nonlinear RNNs \cite{clark2024theory}. In such cases, the single site equations can still be solved with Monte-Carlo sampling to estimate the non-Gaussian single site equations \cite{roy2019numerical}. Approximations which make the limiting equations as tractable as the examples provided in this work are an active area of research \cite{montanari2025dynamical,fournier2023statistical}. In general, DMFT is a powerful tool which enables insights into a large variety of complex dynamical systems, including those arising in machine learning theory. We hope that this note inspires future research in this direction.

\section*{Acknowledgements}
We thank Clarissa Lauditi and Haiping Huang for comments on this draft and thank David Clark, Jacob Zavatone-Veth, Alex Atanasov, Courtney Paquette, Elliot Paquette, Francesco Mori, Jingfeng Wu, Yue Lu for insightful discussions. B.B. acknowledges support from the Center of Mathematical Sciences and Applications (CMSA) of Harvard University. C.P. is supported by an NSF CAREER Award (IIS-2239780), DARPA grants DIAL-FP-038 and AIQ-HR00112520041, the Simons Collaboration on the Physics of Learning and Neural Computation, and the William F. Milton Fund from Harvard University. This work has been made possible in part by a gift from the Chan Zuckerberg Initiative Foundation to establish the Kempner Institute for the Study of Natural and Artificial Intelligence.

\bibliographystyle{unsrt}
\bibliography{refs}

\appendix

\section*{Appendix}

\section{Path Integral Approach to DMFT}\label{app:path_integral}

In this section, we describe the basic machinery of the path integral. As a starting point, we take a discretization of the equations of motion with timestep $\delta t$ 
\begin{align}
    \h_{n+1}  = \h_{n} - (\delta t) \ \bm M \h_n  + (\delta t) \ \bm j_n 
\end{align}
where $\h_{n} \equiv \h(t)|_{t=n (\delta t)}$ as $\delta t \to 0$. We will first express the path integral in discrete time for finitely many steps before describing its continuous limit \footnote{While we will express a formal path integral in continuous time, one could also properly take the saddle point in discrete time over finitely many steps (compared to the number of dimensions $N$) and then subsequently take a continuum limit of the resulting DMFT equations. This procedure generates the same final result. }
\begin{align}
    Z[\bm \zeta] = \int \left(\prod_{n=1}^\infty d\h_{n} \right) \ \left< \prod_{n=0}^\infty \delta\left(\h_{n+1} - \h_n + (\delta t) \bm M \h_n - \bm j_n \right) \right>_{\bm M} \exp\left(  (\delta t) \sum_{n=0}^\infty \bm \zeta_n \cdot \h_n \right) 
\end{align}
where $\delta(\cdot)$ is a Dirac delta function and the average $\left< \right>$ is computed over the random matrix $\bm M$. From this convention (known as the Ito convention \cite{crisanti2018path}), the moment generating function trivially satisfies $Z[\bm 0] = 1$. A key identity which we will utilize repeatedly is the Fourier integral representation of the dirac delta function
\begin{align}
    \delta(z) = \int_{-\infty}^\infty \frac{d\hat z}{2\pi} \exp( i \hat z z)  .
\end{align}
where the $\hat z$ integral runs over the real axis. Applying this to each timestep for $\h_{n}$ we have 
\begin{align}
    Z =  \int \left(\prod_{n=1}^\infty \frac{d\h_n d\hat{\h}_n}{(2\pi)^N}  \right) &\left< \exp\left(i  (\delta t)  \sum_{n=0}^\infty \hat{\h}_{n+1} \cdot \left[ (\h_{n+1} - \h_{n})/(\delta t) + \bm M \h_n - \bm j_n  \right]   \right) \right>_{\bm M} \nonumber
    \\
    &\exp\left(  (\delta t) \sum_{n=0}^\infty \bm \zeta_n \cdot \h_n  \right)
\end{align}
Now, taking the $dt \to 0$ limit, we define the following formal measure over the functions $\h(t), \hat{\h}(t)$
\begin{align}
    \mathcal D \h \mathcal D\hat{\h} \equiv \lim_{\delta t \to 0} \left( \prod_{n=0}^{\infty} \frac{d\h_n d\hat{\h}_n}{(2\pi)^N} \right)  .
\end{align}
Using this notation, we arrive at our formal path integral
\begin{align}
    Z = \int \mathcal D\h \mathcal D \hat{\h} \left< \exp\left( i \int dt \  \hat{\h}(t)  \cdot \left[ \partial_t \h(t) + \bm M \h(t) - \bm j(t) \right]  + \int dt \  \bm \zeta(t) \cdot \h(t) \right) \right>_{\bm M} .
\end{align}
This is the starting point for the various path integral computations we utilize in this work. Depending on the matrix $\bm M$, one arrives at different effective descriptions of the limiting dynamics in terms of a low dimensional set of dynamical order parameters $\bm Q$ (which are usually a collection of correlation and response functions, see examples in coming sections). The general form for $\bm Z$ takes the form
\begin{align}
    Z = \int \mathcal D \bm Q \exp\left( - N \mathcal S[\bm Q] \right)
\end{align}
where $\mathcal S$ is a $\mathcal O(1)$ action. The $N \to \infty$ limit is dominated by the saddle point
\begin{align}
    \frac{\partial \mathcal S[\bm Q]}{\partial \bm Q} = 0 .
\end{align}
These equations will provide us the DMFT equations for correlation and response functions in each of our examples as we outline in the coming sections for the main examples in the main text.
In this setting, one can characterize higher order moments of $\bm Q$ over the distribution induced by the DMFT action $\mathcal S$. Schematically, leading order corrections to the mean and variance of $\bm Q$ can be obtained from higher order derivatives of the mean field action (disregarding matrix indices)
\begin{align}
    &\left< \bm Q \right> \sim \bm Q_\star - \frac{1}{2 N}  \left( \frac{\partial^3 \mathcal S}{\partial \bm Q^3} \right) \cdot \left( \frac{\partial^2 \mathcal S}{\partial \bm Q^2} \right)^{-2} + \mathcal{O}(N^{-2}) \nonumber
    \\
    &\left< (\bm Q-\bm Q_\star)^2 \right> \sim \frac{1}{N} \left( \frac{\partial^2 \mathcal S}{\partial \bm Q^2} \right)^{-1} + \mathcal{O}(N^{-2})
\end{align}
Thus, at large but finite $N$ the order parameters $\bm Q$ deviate from the saddle point $\bm Q_\star$ through a $\mathcal{O}(N^{-1/2})$ zero mean fluctuation and a $\mathcal{O}(N^{-1})$ mean shift. 

\section{GOE Path Integral Derivation}\label{app:goe_path_integral}

For the GOE matrix, the average over the matrix $\bm M$ gives the following
\begin{align}
    \ln \left< \exp\left(  i \int dt \hat{\h}(t)^\top \bm M \h(t) \right)\right> =  - \frac{1}{2} \int dt dt' \  \hat{\h}(t) \cdot \hat{\h}(t') \  \underbrace{\frac{1}{N} \h(t) \cdot \h(t') }_{C(t,t')}   \nonumber
    \\
    - \frac{1}{2} \int dt dt' \   \hat{\h}(t) \cdot \h(t') \ \underbrace{\frac{1}{N} \h(t) \cdot \hat{\h}(t')}_{i R(t,t')} 
\end{align}
We introduced the following correlation and response function
\begin{align}
    C(t,t') = \frac{1}{N} \h(t) \cdot \h(t') \ , \ R(t,t') = - \frac{i}{N} \h(t) \cdot \hat{\h}(t') . 
\end{align}
To enforce the definition of these new variables, we need introduce the following resolution of the identity. We start in discrete time, with $C(t,t') = C_{n,n'}$ where $n (\delta t) = t$ and $n' (\delta t) = t'$
\begin{align}
    1 &= N \int dC_{n,n'} \delta( N C_{n,n'} - \h_n \cdot \h_{n'} )  \nonumber
    \\
    &=  \int \frac{d C_{n,n'} d\hat{C}_{n,n'}}{4\pi i N^{-1} (\delta t)^{-2} } \exp\left( \frac{1}{2} (\delta t)^2 \   \hat{C}_{n,n'} \left[ N C_{n,n'} - \h_n \cdot \h_{n'} \right]  \right)  \nonumber
    \\
    1 &= N \int dR_{n,n'} \delta( N R_{n,n'} + i \h_n \cdot \hat{\h}_{n'} )  \nonumber \nonumber
    \\
    &= \int \frac{dR_{n,n'} d\hat{R}_{n,n'}}{4 \pi i N^{-1} (\delta t)^{-2}} \exp\left( - \frac{1}{2} (\delta t)^2 \hat{R}_{n',n} \left[ N R_{n,n'} + i \h_{n} \cdot \hat{\h}_{n'} \right]  \right)
\end{align}
We multiply the path integral $Z$ by these integral expression for each pair of $(n,n')$, corresponding to pairs of time. We define the limiting measure over the functions $C(t,t'),\hat{C}(t,t'), R(t,t'), \hat{R}(t,t')$ has the expression 
\begin{align}
    \mathcal D C \mathcal D \hat{C} = \lim_{(\delta t) \to 0} \left( \prod_{n=0}^\infty \prod_{n'=0}^\infty \frac{d C_{n,n'} d\hat{C}_{n,n'}}{4\pi i N^{-1} (\delta t)^{-2} }  \right)
    \\
    \mathcal D R \mathcal D \hat{R} = \lim_{(\delta t) \to 0} \left( \prod_{n=0}^\infty \prod_{n'=0}^\infty \frac{d R_{n,n'} d\hat{R}_{n,n'}}{4\pi i N^{-1} (\delta t)^{-2} }  \right)
\end{align}
Using this notation, we can now notice that all expressions involving $\h(t)$ or $\hat{\h}(t)$ decouple as a sum over each of the $N$ sites. This allows us to 
\begin{align}
    Z &= \int \mathcal DC \mathcal D\hat{C} \mathcal D R \mathcal D \hat{R} \exp\left( - N \mathcal S[C,\hat C, R,\hat R] \right)
    \\
    \mathcal S &= - \frac{1}{2} \int dt dt' \left[ \hat{C}(t,t') C(t,t') -  \hat R(t,t') R(t',t)   \right] - \ln \mathcal Z 
\end{align}
where $\mathcal Z$ is a single-site moment generating function
\begin{align}
    \mathcal Z = \int \mathcal Dh \mathcal D\hat{h} &\exp\left( -\frac{1}{2} \int dt dt' \left[ \hat{C}(t,t') h(t) h(t')  + C(t,t') \hat{h}(t) \hat{h}(t') \right] \right) \nonumber
    \\
    &\exp\left( i \int dt \  \hat{h}(t) \left[ \partial_t h(t) - \frac{1}{2} \left(\hat{R}(t,t') + R(t,t') \right) h(t') \right] \right)
\end{align}
The $N \to \infty$ limit is governed by the saddle point equations
\begin{align}
    \frac{\partial \mathcal S}{\partial C(t,t')} &= - \frac{1}{2} \hat{C}(t,t') + \frac{1}{2} \left< \hat h(t) \hat h(t') \right> = 0
    \\
    \frac{\partial \mathcal S}{\partial \hat C(t,t')} &= - \frac{1}{2} \hat{C}(t,t') + \frac{1}{2} \left< \hat h(t) \hat h(t') \right> = 0
    \\
    \frac{\partial \mathcal S}{\partial R(t',t)} &=  \frac{1}{2} \hat{R}(t,t') + \frac{i}{2} \left< h(t) \hat{h}(t') \right> = 0
    \\
    \frac{\partial \mathcal S}{\partial \hat R(t',t)} &= \frac{1}{2} R(t,t') + \frac{i}{2} \left< h(t) \hat{h}(t') \right> = 0
\end{align}
where by $\left< \right>$ we mean an average over the single-site distribution. Let $G\left[h, \hat h\right]$ be an arbitrary functional of $h(t)$ and $\hat h(t)$, then the single site average $\left< G\left[h, \hat h\right] \right>$ is
\begin{align}
   \left< G\left[h, \hat h\right] \right> = \frac{1}{\mathcal Z} \int \mathcal Dh \mathcal D\hat{h} &\exp\left( -\frac{1}{2} \int dt dt' \left[ \hat{C}(t,t') h(t) h(t')  + C(t,t') \hat{h}(t) \hat{h}(t') \right] \right) \nonumber
    \\
    &\exp\left( i \int dt \  \hat{h}(t) \left[ \partial_t h(t) - \frac{1}{2} \left(\hat{R}(t,t') + R(t,t') \right) h(t') \right] \right) \times  G\left[h, \hat h\right]
\end{align}
These expressions imply that $\{C, \hat C, R, \hat R\}$ all take on deterministic values in the $N \to \infty$ limit. Further, we learned that $R(t,t') = \hat R(t,t')$ at the saddle point. To simplify the expressions, we will linearize the quadratic term in $\hat h(t)$ at the expense of introducing a new Gaussian field $u(t)$
\begin{align}
    \exp\left( -\frac{1}{2} \int dt dt' C(t,t') \hat h(t) \hat h(t') \right) =  \left< \exp\left( - i \int dt \ u(t) \ \hat h(t)  \right) \right>_{u \sim \mathcal{GP}(0, C)}
\end{align}
After introducing this new variable, we note that the response functions can be expressed as derivatives with respect to $u(t)$
\begin{align}
    R(t,t') &= \left<  h(t) i \hat h(t) \right> \nonumber
    \\
    &= - \frac{1}{\mathcal Z} \int \mathcal Dh \mathcal D\hat h  \ h(t) \left<  \frac{\partial}{\partial u(t')} \exp\left( i \int dt \hat{h}(t) \left[ \partial_t h(t) - u(t) - \int dt' R(t,t') h(t')  \right]  \right) \right> \nonumber
    \\
    &= \frac{1}{\mathcal Z} \int \mathcal Dh  \frac{\partial }{\partial u(t')} h( \{ u(\cdot ) \}, t)
\end{align}
where we integrated by parts after utilizing the fact that the integral over $\hat h(t)$ collapses to a Dirac mass after introduction of the $u(t)$ variable
\begin{align}
    &\int \mathcal D \hat h(t) \exp\left( i \int dt  \hat{h}(t) \left[ \partial_t h(t) - u(t) - \int dt' R(t,t') h(t')  \right] \right)  \nonumber
    \\
    &\propto \prod_t \delta\left(  \partial_t h(t) - u(t) -  \int dt' R(t,t') h(t')  \right) .
\end{align}
In the above expression, we let $h( \{ u(\cdot ) \}, t)$ represent $h$ at time $t$ as a functional of $u(t)$ that is the solution to the ODE
\begin{align}
    \partial_t h(t) = u(t) + \int dt' R(t,t') h(t')   \ , \ u(t) \sim \mathcal{GP}(0, C(t,t'))
\end{align}
This expression coupled with the formulas $C(t,t') = \left< h(t) h(t') \right>$ and $R(t,t') =  \left< \frac{\partial h(t)}{\partial u(t')} \right>$ recover our DMFT equations from the main text.

\section{Linear Regression Path Integral Derivation}\label{app:lin_reg_path_integral}
In the linear regression example, we introduced two variables $\{\h(t) , \bm\Delta(t)\}$ which satisfy
\begin{align}
    \bm\Delta(t) = \frac{1}{\sqrt N} \bm\Psi \bm h_0(t) \ , \  \partial_t \h(t) = - \frac{1}{\alpha \sqrt{N}} \bm\Psi^\top \bm\Delta(t)
\end{align}
We introduce both of these variables into the path integral using conjugate variables $\hat{\bm\Delta}(t)$ and $\hat{\h}_1(t)$. The resulting average over a random Gaussian $\bm\Psi$ matrix gives
\begin{align}
    &\ln \left< \exp\left( - \frac{i}{\sqrt N} \text{Tr} \bm\Psi  \int dt \left[ \h_0(t) \hat{\bm\Delta}(t)^\top - \frac{1}{\alpha} \hat{\h}(t) \bm\Delta(t)^\top \right]  \right) \right> \nonumber
    \\
    &= - \frac{1}{2} \int dt dt' \left[ \hat{\bm\Delta}(t) \cdot \hat{\bm\Delta}(t)  \underbrace{\left( \frac{1}{N} \h(t) \cdot \h(t') \right)}_{C(t,t')} + \frac{1}{\alpha} \hat{\bm h}(t) \cdot \hat{\h}(t') \underbrace{\left( \frac{1}{P} \bm \Delta(t) \cdot \bm\Delta(t') \right)}_{C_\Delta(t,t')}  \right] \nonumber
    \\
    &+ \frac{1}{P}  \int dt dt' \   \underbrace{(\hat{\h}(t) \cdot \h(t') )}_{i N R_h(t',t)}  \underbrace{(  \bm\Delta(t) \cdot  \hat{\bm\Delta}(t')  ) }_{i P  R_\Delta(t,t') }
\end{align}
Introducing the correlation and response functions, we find
\begin{align}
    Z = \int \mathcal D C_h \mathcal D \hat{C}_h \mathcal D C_\Delta \mathcal D \hat{C}_\Delta \mathcal D R_h \mathcal D R_\Delta \exp\left( - N \mathcal S[C_h, \hat{C}_h, C_\Delta, \hat C_\Delta, R_h, R_\Delta] \right)
\end{align}
where the action $\mathcal S$ has the form (recall that $\alpha = P/N$)
\begin{align}
    \mathcal S = &-\frac{1}{2} \int dt dt' \left[ \hat C_h(t,t') C_h(t,t') + \alpha \hat C_\Delta(t,t') C_\Delta(t,t')   \right] 
    \\
    &-\int dt dt' R_\Delta(t',t) R_h(t,t') -\ln \mathcal Z_h - \alpha \ln \mathcal Z_\Delta 
\end{align}
where the single site stochastic moment generating functions
\begin{align}
    \mathcal Z_h = \int \mathcal D h \mathcal D\hat{h} \  &\exp\left(  i \int dt \hat h(t) \left[ \partial_t h(t) + \int dt' R_\Delta(t,t') h(t') \right]  \right) \nonumber
    \\
    &\exp\left( -\frac{1}{2} \int dt dt' \left[ \frac{1}{\alpha} C_\Delta(t,t') \hat h(t) \hat h(t') + \hat C_h(t,t') h(t) h(t')  \right] \right)
    \\
    \mathcal Z_\Delta = \int \mathcal D \Delta \mathcal D\hat{\Delta} \  &\exp\left(  i \int dt \hat \Delta(t) \left[ \Delta(t) +  \frac{1}{\alpha} 
    \int dt' R_h(t,t') h(t') \right]  \right) \nonumber
    \\
    &\exp\left( -\frac{1}{2} \int dt dt' \left[ C_h(t,t') \hat \Delta(t) \hat \Delta(t') + \hat C_\Delta(t,t') \Delta(t) \Delta(t')  \right] \right)
\end{align}
The relevant saddle point equations give the defining equations for the correlation and response 
\begin{align}
    &\frac{\partial \mathcal S}{\partial \hat C_h(t,t')} = -\frac{1}{2} C_h(t,t') + \frac{1}{2} \left< h(t) h(t') \right> = 0 \nonumber
    \\
    &\frac{\partial \mathcal S}{\partial \hat C_\Delta(t,t')} = -\frac{1}{2} C_\Delta(t,t') + \frac{1}{2} \left< \Delta(t) \Delta(t') \right> = 0 \nonumber
    \\
    &\frac{\partial \mathcal S}{\partial R_\Delta(t',t)} = - R_h(t,t') - i \left< h(t) \hat h(t')  \right> = 0  \nonumber
    \\
    &\frac{\partial \mathcal S}{\partial R_h(t',t)} = - R_\Delta(t,t') - i \left< \Delta(t) \hat \Delta(t')  \right> = 0 
\end{align}
Following the same manipulations of the last sections, we now introduce Gaussian variables $u_h(t), u_\Delta(t)$ to linearize the terms involving $\hat h(t), \hat{\Delta}(t)$ which enables us to characterize the single site stochastic processes for $h(t)$ and $\Delta(t)$
\begin{align}
    &\frac{\partial}{\partial t} h(t) = u_h(t) - \int dt' R_\Delta(t,t') h(t') \ , \ u_h(t) \sim \mathcal{GP}\left(0, \frac{1}{\alpha} C_\Delta  \right) \nonumber
    \\
    &\Delta(t) = u_\Delta(t) - \frac{1}{\alpha} \int dt' R_h(t,t') \Delta(t')  \ , \ u_\Delta(t) \sim \mathcal{GP}(0, C_h)  .
\end{align}
The correlation and response functions can then be obtained from the above equations. 

\section{Structured Random Features}\label{app:RF_path_integral}

In the structured random feature model, we decompose the dynamics into a collection of variables $\{\h_0, \h_1, \h_2, \h_3, \h_4 \}$ defined as
\begin{align}
    &\bm h_1(t) = \bm\Psi \h_0(t) \ , \ \h_2(t) =\frac{1}{P} \bm\Psi^\top \h_1(t) \nonumber
    \\
    &\h_3(t) = \bm A \h_2(t) \ , \ \h_4(t) = \frac{1}{N} \bm A^\top \h_3(t)   \nonumber
    \\
    &\frac{\partial}{\partial t} \h_0(t) = - \h_4(t) 
\end{align}
where the matrices $\bm A$ and $\bm\Psi$ are zero mean with covariance structure
\begin{align}
    \left< A_{ij} A_{k\ell} \right> = \delta_{ik} \delta_{j\ell} \ , \ \left< \Psi_{\mu k} \Psi_{\nu \ell} \right> = \delta_{\mu\nu} \lambda_k \delta_{k\ell} 
\end{align}
The averages over the 
\begin{align}
  \ln &\left<  \exp\left( - i \  \text{Tr} \bm\Psi \int dt  \left[ \h_0(t) \hat{\h}_1(t)^\top + \frac{1}{P} \hat{\h}_2(t) \h_1(t)^\top    \right] \right) \right>   \nonumber
  \\
  = &- \frac{1}{2} \int dt dt' \left[  \hat{\h}_1(t) \cdot \hat{\h}_1(t') \underbrace{\h_0(t)^\top \bm\Lambda \h_0(t')}_{C_0(t,t')} + \frac{1}{P} \hat{\h}_2(t)^\top \bm\Lambda \hat{\h}_2(t') \underbrace{\left( \frac{1}{P} \h_1(t) \cdot \h_1(t') \right)}_{C_1(t,t')} \right]  \nonumber
  \\
  &-  \int dt dt' \  \underbrace{\h_0(t) \bm\Lambda \hat{\h}_2(t')}_{i R_{0,2}(t,t')}  \ \ 
  \underbrace{\left( \frac{1}{P} \h_1(t') \cdot \hat{\h}_1(t) \right)}_{ i R_1(t,t') }
\end{align}
We thus need to introduce the correlation functions and response function order parameters highlighted above. Next, we perform the average over $\bm A$
\begin{align}
    \ln &\left< \exp\left( - i \text{Tr} \bm A \int dt  \left[ \h_2(t) \hat{\h}_3(t)^\top + \frac{1}{N}  \hat{\h}_4(t) \h_3(t)^\top \right] \right) \right> \nonumber
    \\
    &= -\frac{1}{2} \int dt dt'  \left[ \hat{\h}_3(t) \cdot \hat{\h}_3(t')  \underbrace{\left( \h_2(t) \cdot \h_2(t')  \right)}_{C_2(t,t')} + \frac{1}{N} \hat{\h}_4(t) \cdot \hat{\h}_4(t')  \underbrace{\left( \frac{1}{N} \h_3(t) \cdot \h_3(t') \right)}_{C_3(t,t')}  \right] \nonumber
    \\
    &-  \int dt dt'  \underbrace{\h_2(t) \cdot \hat{\h}_4(t')  }_{i R_{2,4}(t,t')}  \underbrace{\left( \frac{1}{N} \h_3(t') \cdot \hat \h_3(t) \right)  }_{i  R_3(t',t) }
\end{align}
The path integral now has the following form after introducing these order parameters
\begin{align}
    Z = \int \left( \prod_{\ell=0}^3 \mathcal D C_\ell \mathcal D \hat C_\ell  \right) \prod_{(j,\ell) \in \Pi} \mathcal D R_{j,\ell} \  \exp\left(  - \mathcal S[ \{C_\ell \}_{\ell \in [3]} , \{ R_{j,\ell} \}_{(j,\ell) \in \Pi} ] \right) 
\end{align}
where $\Pi = \{ (0,2), (2,4), (1,1), (3,3) \}$ are the pairings of variables involved in response functions and the action $\mathcal S$ is defined as
\begin{align}
    \mathcal S &= - \frac{1}{2} \int dt dt'  \left[ C_0(t,t') \hat C_0(t,t') + P C_1(t,t') \hat C_1(t,t') + C_2(t,t') \hat C_2(t,t')  + N C_3(t,t') \hat C_3(t,t')  \right] \nonumber
    \\
    &+ \int dt dt' \left[ R_1(t,t') R_{0,2}(t',t) + R_3(t,t') R_{2,4}(t',t) \right] - P \ln \mathcal Z_1 - N \ln \mathcal Z_3 - \sum_k \mathcal \ln \mathcal Z_{(0,2,4),k}
\end{align}
where the single site processes have the form
\begin{align}
    \mathcal Z_1 = \int \mathcal Dh_1 \mathcal D\hat h_1 &\exp\left( - \frac{1}{2} \int dt dt' \left[ \hat h_1(t) \hat h_1(t') C_0(t,t') + h_1(t) h_1(t') \hat C_1(t,t') \right]  \right) \nonumber
    \\
    &\exp\left(   i \int dt \hat h_1(t) \left [h_1(t) + \frac{1}{P} \int dt' R_{0,2}(t,t') h_1(t') \right] \right) \nonumber
    \\
    \mathcal Z_3 = \int \mathcal Dh_3 \mathcal D\hat h_3 &\exp\left( - \frac{1}{2} \int dt dt' \left[\hat h_3(t) \hat h_3(t') C_2(t,t') + h_3(t) h_3(t') \hat C_3(t,t') \right]\right) \nonumber
    \\
    &\exp\left(   i \int dt \hat h_3(t) \left [h_3(t) + \frac{1}{N} \int dt' R_{2,4}(t,t') h_3(t') \right]  \right) \nonumber
    \\
    \mathcal Z_{(0,2,4),k} = \int \prod_{j \in \{0,2,4 \} }\mathcal D h_\ell \mathcal D \hat h_\ell \  &\exp\left( - \frac{1}{2} \int dt dt' \left[ \frac{\lambda_k}{P} C_1(t,t') \hat h_2(t) \hat h_2(t') + \frac{1}{N} C_3(t,t') \hat h_4(t) \hat h_4(t')   \right] \right) \nonumber
    \\
    &\exp\left( - \frac{1}{2} \int dt dt' \left[  \lambda_k \hat C_0(t,t') h_0(t) h_0(t')  +  \hat C_2(t,t') h_2(t) h_2(t')  \right] \right) \nonumber
    \\
    &\exp\left(  i \int dt \hat h_2(t) \left[  h_2(t) + \int dt' R_1(t,t') h_0(t') \right] \right) \nonumber
    \\
    &\exp\left(  i \int dt \hat h_4(t) \left[  h_4(t) + \int dt' R_3(t,t') h_0(t') \right] \right) \nonumber
    \\
    &\exp\left(  i \int dt \hat h_0(t) \left[  \partial_t h_0(t) + h_4(t)  \right] \right)
\end{align}
At finite $N,P$ the above path integral is not perfectly dominated by the saddle point. However, the mean field approximation to the test loss is obtained by taking a saddle point over these variables. This incurs an approximation error, but that approximation error decays gracefully as $N,P$ increase \cite{bordelon2020spectrum}. The saddle point equations recover the DMFT equations provided in the main text.


\section{Free Product Dynamics}\label{app:free_product}

In this section, we work out the mean field action for the free product dynamics for both the asymmetric and symmetrized cases. 

\subsection{Asymmetric}
The path integral for the asymmetric free product ensemble has the frequency space expression
\begin{align}
    Z = \int \mathcal D \bm\chi \mathcal D \bm h \ \left< \exp\left( - \int d \omega  \ \bm\chi(\omega) \cdot \left[ i\omega \h(\omega) +  \bm O \bm B \bm O^\top \bm A \h(\omega) - \h_0 \right] \right) \right>
\end{align}
where $\h_0$ is the initial condition at $t=0$. To compute the average over $\bm O$, we introduce two vectors
\begin{align}
    \bm v(\omega) = \bm O^\top \bm A \h(\omega) \ , \ \bm u(\omega) = \bm O^\top \bm\chi(\omega)
\end{align}
As $N \to \infty$, the measure of finitely many vectors $\bm v(\omega)$ and $\bm u(\omega)$ is uniform up to constraints on the inner products, which is the main insight of low-rank HCIZ integrals \cite{potters2020first}. The inner product constraints on $\v(\omega)$ and $\u(\omega)$ are
\begin{align}
    &\frac{1}{N} \v(\omega)^\top \v(\omega') = \frac{1}{N} \bm h(\omega)^\top \A^2 \h(\omega') \equiv \Sigma_{hh}(\omega,\omega')  \nonumber
    \\
    &\frac{1}{N} \v(\omega)^\top \u(\omega') = \frac{1}{N} \bm h(\omega)^\top \A \bm\chi(\omega') \equiv \Sigma_{h\chi}(\omega,\omega')  \nonumber
    \\
    &\frac{1}{N} \u(\omega)^\top \u(\omega') = \frac{1}{N} \bm\chi(\omega) \cdot \bm\chi(\omega') \equiv \Sigma_{\chi\chi}(\omega,\omega')
\end{align}
We introduce a two-by-two matrix $\bm\Sigma(\omega,\omega')$ which has the structure
\begin{align}
    \bm\Sigma(\omega,\omega') = \begin{bmatrix}
        \Sigma_{hh}(\omega,\omega') & \Sigma_{h\chi}(\omega,\omega')
        \\
        \Sigma_{\chi h}(\omega,\omega') & \Sigma_{\chi\chi}(\omega,\omega') 
    \end{bmatrix}
\end{align}
For a given $\bm\Sigma$, we let $\mu(\v,\u,\bm\Sigma)$ represent the normalized probability distribution for $\u,\v$ induced by the random $\bm O$. We desire to compute the following integral
\begin{align}
    &\int \mathcal D\v \mathcal D\u \  \mu(\v,\u, \bm\Sigma) \exp\left( - \int d\omega \  \bm u(\omega)^\top \bm B \bm v(\omega)   \right)
\end{align}
As $N \to \infty$ the formula for $\mu$ is
\begin{align}
    \mu(\v,\u, \bm\Sigma) \propto \prod_{\omega,\omega'} &\delta\left (\v(\omega) \cdot \v(\omega') - N \Sigma_{hh}(\omega,\omega')  \right)\delta\left(\v(\omega) \cdot \u(\omega') - N \Sigma_{h\chi}(\omega,\omega') \right) \nonumber
    \\
    &\times \delta\left(\u(\omega) \cdot \v(\omega') - N \Sigma_{\chi h}(\omega,\omega') \right)  \delta\left (\u(\omega) \cdot \u(\omega') - N \Sigma_{\chi\chi}(\omega,\omega') \right)
\end{align}
where the $\propto$ represents the distribution up to normalization. To enforce the inner product constraints between $\bm v$ and $\bm u$, we introduce Lagrange multipliers $\hat{\bm\Sigma}$ (Fourier variables for a Dirac delta function integral representation), giving the following integral
\begin{align}
  \frac{ \int \mathcal D\v \mathcal D\u \  \mathcal D \hat{\bm\Sigma} \exp\left( \frac{1}{2} \int d\omega d\omega'  \text{Tr} \hat{\bm\Sigma}(\omega,\omega')^\top \left( N \bm\Sigma(\omega,\omega') - \begin{bmatrix} \bm v(\omega) \cdot \bm v(\omega') & \bm v(\omega) \cdot \bm u(\omega')
    \\
    \bm v(\omega') \cdot \bm u(\omega) & \bm u(\omega ) \cdot \bm u(\omega') 
    \end{bmatrix} \right)  - \int d\omega \  \bm u(\omega)^\top \bm B \bm v(\omega)   \right)}{\int \mathcal D\v \mathcal D\u \  \mathcal D \hat{\bm\Sigma} \exp\left( \frac{1}{2} \int d\omega d\omega'  \text{Tr} \hat{\bm\Sigma}(\omega,\omega')^\top \left( N \bm\Sigma(\omega,\omega') - \begin{bmatrix} \bm v(\omega) \cdot \bm v(\omega') & \bm v(\omega) \cdot \bm u(\omega')
    \\
    \bm v(\omega') \cdot \bm u(\omega) & \bm u(\omega ) \cdot \bm u(\omega') 
    \end{bmatrix} \right) \right) }
\end{align}
We introduce a matrix notation for the integrals over $(\omega,\omega')$ and also introduce a tensor product representation of the above Gaussian integral. We use the notation $\textbf{Tr}$ to represent a super-trace over both frequency and matrix dimensions. The integral in the numerator thus becomes
\begin{align}
    \int \mathcal D \hat{\bm\Sigma} \exp\left( \frac{N}{2} \textbf{Tr}  \hat{\bm\Sigma} \bm\Sigma - \frac{1}{2} \textbf{Tr} \log\left( \hat{\bm\Sigma} \otimes \bm I + \bm I \otimes \bm B  \right)  \right) = \int \mathcal D \hat{\bm\Sigma} \exp\left( - \frac{N}{2} \mathcal S_{B}(\hat{\bm\Sigma})  \right)
\end{align}
As $N \to \infty$, we are justified utilizing a saddle point to compute the numerator integral.
\begin{align}
    \frac{\partial \mathcal S_B}{\partial \hat{\bm\Sigma}}= - \bm\Sigma + \frac{1}{N} \textbf{Tr} \left(\hat{\bm\Sigma} \otimes \bm I + \bm I \otimes \bm B  \right)^{-1} = 0
\end{align}
This equation defines $\hat{\bm\Sigma}_\star$ which is a function of $\bm\Sigma$. The numerator integral is thus
\begin{align}
    &\int \mathcal D \hat{\bm\Sigma} \exp\left( \frac{N}{2} \textbf{Tr}  \hat{\bm\Sigma} \bm\Sigma - \frac{1}{2} \textbf{Tr} \log\left( \hat{\bm\Sigma} \otimes \bm I + \bm I \otimes \bm B  \right)  \right) \nonumber
    \\
    &\sim \exp\left( \frac{N}{2} \textbf{Tr}  \hat{\bm\Sigma}_\star \bm\Sigma - \frac{1}{2} \textbf{Tr} \log\left( \hat{\bm\Sigma}_\star \otimes \bm I + \bm I \otimes \bm B  \right)  \right)
\end{align}
Performing the same analysis for the integral arising in the denominator. For the denominator integral, the dominant contribution comes from $\hat{\bm\Sigma} = \bm\Sigma^{-1}$. Therefore our original average over the $\v$ and $\u$ fields reduces to (up to irrelevant constants)
\begin{align}
    &\int \mathcal D\v \mathcal D\u \  \mu(\v,\u, \bm\Sigma) \exp\left( - \int d\omega \  \bm u(\omega)^\top \bm B \bm v(\omega)   \right) 
    \\
    &\sim \exp\left( \frac{N}{2} \textbf{Tr}  \hat{\bm\Sigma}_\star \bm\Sigma - \frac{1}{2} \textbf{Tr} \log\left( \hat{\bm\Sigma}_\star \otimes \bm I + \bm I \otimes \bm B  \right)  - \frac{N}{2} \textbf{Tr} \log \bm\Sigma  \right)
\end{align}
Next, we must introduce a Lagrange multiplier variable to enforce the relationship between $\{ \h, \bm\chi\}$ and $\bm\Sigma$. We let this set of Lagrange multipliers be $\bm\Psi$. 
\begin{align}
    &Z = \int \mathcal D \bm\Psi \mathcal D \bm\Sigma \exp\left( -\frac{N}{2} \mathcal S[\bm\Sigma,\bm\Psi]   \right)
    \\
   &\mathcal S[\bm\Sigma,\bm\Psi]  = - \textbf{Tr} \bm\Psi \bm\Sigma - \frac{2}{N} \ln \mathcal Z_A(\bm\Psi) -  \textbf{Tr}  \hat{\bm\Sigma}_\star \bm\Sigma + \frac{1}{N} \textbf{Tr} \log\left( \hat{\bm\Sigma}_\star \otimes \bm I + \bm I \otimes \bm B  \right)  +  \textbf{Tr} \log \bm\Sigma 
\end{align}
where in the above expression $\hat{\bm\Sigma}_\star$ is implicitly a function of $\bm\Sigma$ and the function $\mathcal Z_A$ is defined as
\begin{align}
    \mathcal Z_A = \int \mathcal D\h \mathcal D \bm\chi &\exp\left( - \int d\omega \bm \chi(\omega) \left[ i\omega \bm h(\omega) + \h_0 \right]   \right)
    \\
    &\exp\left( - \frac{1}{2} \int d\omega d\omega' \text{Tr } \bm\Psi(\omega,\omega') \begin{bmatrix}
        \h(\omega)^\top \bm A^2 \h(\omega') &  \h(\omega)^\top \bm A \bm\chi(\omega') \nonumber
        \\
        \bm\chi(\omega)^\top \bm A \bm h(\omega') & \bm\chi(\omega)^\top \bm\chi(\omega')  
    \end{bmatrix} \right)
\end{align}
The saddle point equations over $\bm\Psi$ and $\bm\Sigma$ are
\begin{align}
    &\frac{\partial \mathcal S}{\partial \bm\Sigma} = - \bm\Psi - \hat{\bm\Sigma}_\star + \bm\Sigma^{-1} = \bm 0
    \\
    &\frac{\partial \mathcal S}{\partial \bm\Psi} = - \bm\Sigma + 
    \begin{bmatrix}
        \frac{1}{N} \left<\h(\omega)^\top \bm A^2 \h(\omega') \right>  & \frac{1}{N} \left<\h(\omega)^\top \bm A \bm\chi(\omega') \right>  
        \\
        \frac{1}{N} \left<\h(\omega')^\top \bm A \bm\chi(\omega) \right>  & \frac{1}{N} \left<\bm\chi(\omega) \cdot \bm\chi(\omega') \right>
    \end{bmatrix} = \bm 0
\end{align}
where the average $\left< \cdot \right>$ is over the distribution induced by $\mathcal Z_A$. We thus need to compute the mean and variance of $\h$ and $\bm\chi$. We note that $\h(\omega)$ will have nonzero mean and variance due to the term involving $\bm\chi(\omega) ( i\omega \h - \h_0 )$, even if $\bm\Psi$ is zero. However, we note that there is a physically meaningful solution where $\bm\chi$ has vanishing self-correlation, giving a $\bm\Sigma(\omega,\omega')$ matrix with vanishing lower block
\begin{align}
    \bm\Sigma(\omega,\omega') = \begin{bmatrix}
        \Sigma_{hh}(\omega,\omega') & \Sigma_{h\chi}(\omega,\omega')
        \\
        \Sigma_{\chi h}(\omega,\omega') & 0
    \end{bmatrix} .
\end{align}
\paragraph{Off diagonal blocks} As a consequence the matrices $\hat{\bm\Sigma}_\star(\omega,\omega')$ and $\bm\Psi(\omega,\omega')$ have vanishing lower block. Further, we note that the off-diagonal blocks decouple over frequencies $\Sigma_{h\chi}(\omega,\omega') = \delta(\omega-\omega') \Sigma_{h\chi}(\omega)$
\begin{align}
    &\Sigma_{h\chi}(\omega) = \text{tr} \bm A \left( i\omega + \Psi_{h\chi}(\omega) \bm A \right)^{-1} \nonumber
    \\
    &\Psi_{h\chi}(\omega) = \Sigma_{h\chi}(\omega)^{-1} - \hat{\Sigma}_{h\chi}(\omega)  \nonumber
    \\
    &\Sigma_{h\chi}(\omega) = \text{tr} \left(  \hat{\Sigma}_{h\chi}(\omega) + \bm B \right)^{-1}
\end{align}
We define $\mathcal T_{A} = \text{tr} \bm A \left( i\omega + \bm A \right)^{-1}$. Thus, the first equation gives
\begin{align}
    \Sigma_{h\chi}(\omega) \Psi_{h\chi}(\omega) = \mathcal T_A( i\omega_A ) \ , \ i\omega_A = i\omega / \Psi_{h\chi}(\omega) .
\end{align}
The second and third equations imply
\begin{align}
    \mathcal T_A( i\omega_A ) &= \Sigma_{h\chi}(\omega) \Psi_{h\chi}(\omega) = 1 - \Sigma_{h\chi}(\omega) \hat{\Sigma}_{h\chi}(\omega)  \nonumber 
    \\
    &= \text{tr}\bm B \left( \hat{\Sigma}_{h\chi}(\omega) + \bm B \right)^{-1} = \mathcal T_B( i\omega_B) \ , \ i\omega_B = \hat{\Sigma}_{h\chi}(\omega) 
\end{align}
Lastly, we note that the $\mathcal T$ transform is also identical 
\begin{align}
    \mathcal T_M(i\omega) = \text{tr} \bm A \left( i\omega + \Psi_{h\chi}(\omega) \bm A \right)^{-1} = \mathcal T_A(i\omega_{A}) = \mathcal T_B(i\omega_B) 
\end{align}
Since each of these $\mathcal T$ variables are identical when evaluated at their respective frequencies, we can simply use $\mathcal T$ without a subscript. Lastly, we can deduce a relationship between the three frequencies $\{i\omega_A, i\omega_B, i\omega\}$ 
\begin{align}
    i\omega = i\omega_A \Psi_{h\chi}(\omega) = i\omega_A \left[ \Sigma_{h\chi}(\omega)^{-1} - i\omega_B  \right] = i\omega_A i\omega_B \left[ \frac{\mathcal T}{1-\mathcal T}  \right]
\end{align}
Rearranging this equation gives the stated result in the main text
\begin{align}
    i\omega_A i\omega_B = \frac{1-\mathcal T}{\mathcal T} i\omega
\end{align}

\paragraph{Diagonal Blocks} The diagonal blocks satisfy the following equations 
\begin{align}
    &\Sigma_{hh}(\omega,\omega') = \text{tr} \bm A^2 \left( i\omega + \Psi_{h\chi}(\omega) \bm A \right)^{-1} \left[ \h_0 \h_0^\top - \Psi_{\chi\chi}(\omega,\omega') \right] \left( i\omega' + \Psi_{h\chi}(\omega') \bm A \right)^{-1}  \nonumber
    \\
    &\Psi_{\chi\chi}(\omega,\omega') = - \Sigma_{hh}(\omega,\omega') \Sigma_{h\chi}(\omega)^{-1} \Sigma_{h\chi}(\omega')^{-1} - \hat\Sigma_{\chi\chi}(\omega,\omega') \nonumber
    \\
    &\Sigma_{hh}(\omega,\omega') = - \hat\Sigma_{\chi\chi}(\omega,\omega')  \  \text{tr} \left( \hat\Sigma_{h\chi}(\omega) + \bm B  \right)^{-1} \left( \hat\Sigma_{h\chi}(\omega') + \bm B  \right)^{-1} .
\end{align}
When combined these equations yield our full equations for the two point functions. 

\subsection{Symmetrized Case}

For the symmetrized case, we instead have the following dynamics
\begin{align}
       Z = \int \mathcal D \bm\chi \mathcal D \bm h \ \left< \exp\left( - \int d \omega  \ \bm\chi(\omega) \cdot \left[ i\omega \h(\omega) +  \bm A^{1/2} \bm O \bm B \bm O^\top \bm A^{1/2} \h(\omega) - \h_0 \right] \right) \right>
\end{align}
In this case, the relevant correlations are 
\begin{align}
    \bm\Sigma(\omega,\omega') = 
    \begin{bmatrix}
        \frac{1}{N} \h(\omega)^\top \bm A \h(\omega')  & \frac{1}{N} \h(\omega)^\top \bm A \bm \chi(\omega')
        \\
        \frac{1}{N} \bm\chi(\omega)^\top \bm A \h(\omega')  & \frac{1}{N} \bm\chi(\omega)^\top \bm A \bm\chi(\omega')
    \end{bmatrix} 
\end{align}
The calculation proceeds as in the previous section. However, the difference comes in the computation of the diagonal blocks which now have the following structure
\begin{align}
    &\Sigma_{hh}(\omega,\omega') = \text{tr} \bm A \left( i\omega + \Psi_{h\chi}(\omega) \bm A \right)^{-1} \left[ \h_0 \h_0^\top - \bm A \Psi_{\chi\chi}(\omega,\omega') \right] \left( i\omega' + \Psi_{h\chi}(\omega') \bm A \right)^{-1}  \nonumber
    \\
    &\Psi_{\chi\chi}(\omega,\omega') = - \Sigma_{hh}(\omega,\omega') \Sigma_{h\chi}(\omega)^{-1} \Sigma_{h\chi}(\omega')^{-1} - \hat\Sigma_{\chi\chi}(\omega,\omega') \nonumber
    \\
    &\Sigma_{hh}(\omega,\omega') = - \hat\Sigma_{\chi\chi}(\omega,\omega')  \  \text{tr} \left( \hat\Sigma_{h\chi}(\omega) + \bm B  \right)^{-1} \left( \hat\Sigma_{h\chi}(\omega') + \bm B  \right)^{-1} .
\end{align}
This subtle change in the two point functions can generate vastly different dynamics as we demonstrate in Figure \ref{fig:two_point_correlation_dynamics}.

\pagebreak

\end{document}